\documentclass[twocolumn,tighten,numberedappendix,floatfix]{aastex631}

\usepackage{bm}
\usepackage{xspace}
\usepackage{enumitem}
\usepackage{graphicx}
\usepackage{color}
\usepackage{comment}
\usepackage{amsmath}
\usepackage[T1]{fontenc}
\usepackage{natbib}
\usepackage[capitalize]{cleveref}
\usepackage{colortbl}

\Crefmultiformat{equation}{Eqs.~#2#1#3}%
{,~#2#1#3}{, #2#1#3}{ and~#2#1#3}

\Crefformat{equation}{Eq.~#2#1#3}

\usepackage{mathtools}

\usepackage[caption=false]{subfig}

\newcommand\sref[1]{\S \ref{#1}}
\newcommand\aref[1]{Appendix \ref{#1}}

\newcommand{\newp}[1] {{#1}}

\graphicspath{{./}{graphics/}{graphics/pileup/}{graphics/pileup/Final_mass_distribution/}{graphics/CMI/}{graphics/CMI/Maps/}}

\shorttitle{Feedback-Dominated Accretion Flows}
\shortauthors{Gilbaum \& Stone}

\begin{document}

\title{Feedback-Dominated Accretion Flows}
\correspondingauthor{Shmuel Gilbaum}
\email{shmuel.gilbaum@mail.huji.ac.il}
\author[0000-0002-6462-6657]{Shmuel Gilbaum}
\affiliation{Racah Institute of Physics, The Hebrew University, Jerusalem, 91904, Israel}
\author[0000-0002-4337-9458]{Nicholas C.~Stone}
\affiliation{Racah Institute of Physics, The Hebrew University, Jerusalem, 91904, Israel}
\affiliation{Department of Astronomy, University of Maryland, Stadium Drive, College Park, MD 20742, USA}
\affiliation{Department of Physics, Department of Astronomy, and Columbia Astrophysics Laboratory, Columbia University, New York, NY 10027, USA}

\begin{abstract}
We present new two-fluid models of accretion disks in active galactic nuclei (AGNs) that aim to address the long-standing problem of Toomre instability in AGN outskirts.  In the spirit of earlier works by Sirko \& Goodman and others, we argue that Toomre instability is eventually self-regulated via feedback produced by fragmentation and its aftermath.  Unlike past semianalytic models, which (i) adopt local prescriptions to connect star formation rates to heat feedback, and (ii) assume that AGN disks self-regulate to a star-forming steady state (with Toomre parameter $Q_{\rm T}=1$), we find that feedback processes are both temporally and spatially nonlocal.  The accumulation of many stellar-mass black holes (BHs) embedded in AGN gas eventually displaces radiation, winds and supernovae from massive stars as the dominant feedback source. The nonlocality of feedback heating, in combination with the need for heat to efficiently mix throughout the gas, gives rise to steady-state AGN solutions that can have $Q_{\rm T} \gg 1$ and no ongoing star formation.  We find self-consistent steady-state solutions in much of the parameter space of AGN mass and accretion rate.  These solutions harbor large populations of embedded compact objects that may grow in mass by factors of a few over the AGN lifetime, including into the lower and upper mass gaps.  These feedback-dominated AGN disks differ significantly in structure from commonly used 1D disk models, which has broad implications for gravitational-wave-source formation inside AGNs.
\end{abstract}


\section{Introduction}
\label{sec:intro}
Active galactic nuclei (AGNs) are the brightest steady sources of electromagnetic radiation in the universe \citep{Jiang+16b, Schindler+19}, are the dominant mechanism for supermassive black hole (SMBH) growth \citep{Soltan82, Marconi+04}, and play a key role in galaxy evolution \citep{KormendyRichstone95, FerrareseMerritt00, Tremaine+02}.  Despite their central importance to multiple areas of astrophysics, and despite decades of progress in understanding their qualitative phenomenology, many aspects of AGNs remain poorly understood.  Long-standing questions exist concerning AGN variability timescales \citep{Clavel+91, Lawrence18}, the linear stability of AGN disks to thermal \citep{ShakuraSunyaev76, Piran78} and inflow \citep{LightmanEardley74} perturbations, and the geometry and physical origin of key components of AGN spectra (most notably, the broad-line region and the corona; see,  \citealt{Wilkins+16} and \citealt{Netzer20}, respectively, for recent discussions of each).  Two additional outstanding AGN puzzles, which are the primary motivations for this paper, are the well-known Toomre instability \citep{Toomre1964} manifested by AGN disks at large radii \citep{ShlosmanBegelman89, Shlosman+90}, and the mismatch between inflow timescales and typical AGN lifetime.

The Toomre instability \citep{Toomre1964} is the cylindrical, rotating analog of the classical Jeans instability and develops whenever a rotating accretion disk contains too much mass to be stable against the growth of small overdensities (instability corresponds to the dimensionless parameter $Q_{\rm T} \lesssim 1$).  The subsequent fragmentation cascade of Toomre-unstable disks can produce large, self-gravitating objects. In the context of protoplanetary disks, this is one possible origin of gas giants \citep{Boss97, Durisen+07}, but in AGN disks, it is generally thought to lead to star formation.  Circumstantial evidence exists for this outcome in the center of the Milky Way, where a flattened disk of young massive stars orbits deep inside the influence radius of the SMBH \citep{Genzel+03, LevinBeloborodov03}.  The challenge Toomre instability poses for AGN models is the question of saturation: What fraction of gas is converted into stars, and what fraction, if any, flows inward to be accreted?  Past resolutions of this problem either introduce new physics to suppress the linear development of the Toomre instability (e.g. magnetic pressure support; \citealt{BegelmanPringle07, DexterBegelman2019}), or assume that Toomre instability is a self-limiting phenomenon due to energy feedback from star formation.  In the latter class of solutions, one can find disk models that are constructed to be marginally Toomre stable: for example, the pioneering works of \citet{SirkoGoodman2003} and \citet{Thompson+2005} assume a local star formation rate that is precisely large enough to keep $Q_{\rm{T}}\sim \mathcal{O}(1)$.  Early work on $Q_{\rm{T}}=1$ AGN disks usually assumed that stellar winds and Type II supernovae (SNe) were the primary feedback mechanisms, but other papers have suggested that accretion feedback from a population of embedded compact objects can be energetically important \citep{Levin07, DittmanMiller20}.

Another problem for AGN disks is the long inflow  (or ``viscous'') timescales present in simple disk models that use a local ``$\alpha$-viscosity'' parameterization \citep{ShakuraSunyaev1973} of angular momentum transport.  Although these $\alpha$-models succeed in many areas of accretion physics, they predict outer viscous timescales many orders of magnitude longer than plausible AGN lifetimes \citep{ShlosmanBegelman89}.  This issue may be closely linked to that of Toomre instability, in that neither is a problem at small radii, but both begin to challenge Shakura--Sunyaev-type models once one moves beyond $\sim 10^{3}-10^5$ Schwarzschild radii.  Indeed, most existing resolutions to the inflow problem begin by postulating the existence of large-scale deviations from axisymmetry that are capable of exerting global torques on gas streamlines \citep{ShlosmanBegelman89, Thompson+2005}, increasing rates of angular momentum transport far above that provided by a local $\alpha$ viscosity.  These nonaxisymmetric features, such as spiral arms, can arise naturally in self-gravitating regions of a disk \citep{Gammie+01}.  However, this solution to the inflow problem is not wholly satisfactory, as there is a large parameter space of $\alpha$ models for which the zone of inflow disequilibrium is substantially larger than the zone of Toomre instability.  Whether or not spiral arms or other self-gravitating features can effectively torque gas far interior to their own radius is not clear.

More recently, the ground-breaking discoveries of LIGO/Virgo \citep{Abbott+16, Abbott+19a, Abbott+2021} have focused greater attention on the ``two-fluid'' nature of AGN disks.  Both {\it in situ} star formation \citep{Stone+17} and the capture of preexisting stars via hydrodynamic drag \citep{Bartos+17a} will build up a population of stars and compact objects that are embedded within (i.e. orbitally aligned with) the AGN disk. For brevity, the remainder of this paper will refer to these objects generally as ``embeds,'' and embedded compact objects specifically as ``ECOs.''  

ECOs have attracted much interest as sources of gravitational-wave (GW) radiation \citep{McKernan+14, Bellovary+2016,Bartos+17a, Stone+17,  McKernan+18, Yang+20b}.  Binary ECOs may be driven to rapid merger via hydrodynamic torques, producing bursts of GWs that may explain a large fraction of the LIGO/Virgo binary black hole (BH) population.  Even singleton ECOs may participate in high-frequency GW production, as they will exchange torques with disk gas and migrate through the AGN \citep{McKernan+11}.  During this process, they may capture into binaries in low relative-velocity encounters \citep{McKernan+12, Tagawa+20b}, possibly at migration traps within the AGN disk \citep{Bellovary+2016, Secunda+19, Secunda+20}.  The deep potential well of the SMBH makes the ``AGN channel'' uniquely capable of retaining merger products against GW recoil kicks, and the possibility of hydrodynamic capture enables repeated mergers and hierarchical ECO growth \citep{GerosaBerti19, Yang+19}.  This scenario may explain anomalously massive BH mergers, such as GW190521 \citep{Abbott+20b}, where one or both component masses lie in the pair-instability gap forbidden to single star evolution. As ECOs migrate inwards, they may eventually source low-frequency GWs as well, in an ``extreme mass ratio inspiral,'' or EMRI \citep{Levin07, Pan+21}.

In this paper, we build analytic and semianalytic disk models for steady-state, two-fluid AGN disks, exploring new regimes of disk structure.  Our goal is to delineate the astrophysical conditions under which heating from embeds -- typically accretion feedback from stellar-mass BHs -- is the primary heating source for the AGN gas and to explore the consequences for AGN structure and accretion in these feedback-dominated accretion flows.  In \S \ref{sec:disks}, we introduce our microphysical and ``mesophysical'' assumptions in order to present the set of disk equations we solve later on.  \S \ref{sec: Disc structure} presents the general features of these solutions and delineates the regions of parameter space over which they are self-consistent (and those in which they are not).  \S \ref{sec:implications} uses these steady-state solutions to address the original questions of Toomre instability and inflow timescales while making preliminary predictions for the novel effects of the feedback-dominated regime on the AGN channel.  We summarize in \S \ref{sec:conclusions}.

\begin{table}
\centering
\caption{\newp{Glossary}} \label{table: glossary}
\renewcommand{\arraystretch}{0.7}
\small{
\begin{tabular}{>{}l >{}l} 
  \hline  \hline
  Variable & Definition \\
\hline 
$Q_{\rm T}$ & Toomre instability parameter, if $Q_{\rm T}<1$ gas is \\
 & unstable to density perturbations and  \\
  & potentially  stars may form (\cref{eq: Toomre stability parameter} )\\ 
$M$ & SMBH mass \\
$\dot{M}_{\rm g}$ & SMBH accretion rate of the gas\\
$R$  & Distance from the center of the disk\\
$R_0$ & ISCO of the SMBH \\
$\Omega$ & Kepleration angular frequency $\left( \sqrt{GM/R^3} \right)$ \\
$T_{\rm eff}$ & Local effective temperature of the disk \\
$T_{\rm c}$ & Local central temperature of the disk (\cref{eq: temperature})\\
$c_{\rm s}$ & Local sound speed (\cref{eq: sound speed}) \\
$\Sigma$ & Local gas surface mass density ( \cref{eq: Vertical structure}) \\
$\rho$ & Local gas volume density \\
$H$ & Local disk scale height (\cref{eq: hydrostatic equilibrium}) \\
$P$ & Local gas total pressure (\cref{eq: pressure equation}) \\
$\tau$ & Local disk optical depth(\cref{eq: optical depth}) \\
$\nu$ & Local gas effective viscosity (\cref{eq: alpha viscosity}) \\
$\alpha$ & Effective viscosity unitless prefactor \\
$Q_\bullet$ & Heating rate  per unit area due to accretion \\
 & feedback from ECOs $\left( S_\bullet L_\bullet \right)$ \\
$S_\bullet$ & Local number of black holes per unit area in \\ 
 &  the continuous approximation \\
$L_\bullet$ & Local luminosity exerted due to accretion  \\ 
 & feedback from a single black hole \\
$m_\bullet$ &  Average ECO mass  \\
$\dot{M}_\bullet$ & Mass-flow rate of black holes (\cref{eq: General black hole flow rate}) \\
$\mathfrak{M}_{\varphi}$ & Azimuthal effective heat-mixing parameter, if \\ 
& $\mathfrak{M}_{\varphi}<1$ heat is dissipated  before heating the \\
 & disk effectively (\cref{Azimuthal mixing parameter}) \\
\hline
\end{tabular}}
\end{table}

\section{Two-fluid AGN Disks}
\label{sec:disks}

In this section we construct a new model for AGN accretion discs, with SMBH mass $M$ and constant gas accretion rate $\dot{M}_{\rm g}$, that is initially based on the 1D (axisymmetric with averaged vertical structure) Shakura--Sunyaev $\alpha$ model \citep{ShakuraSunyaev1973}. The primary difference from classical $\alpha$ models is that we also include a collisionless (or at most, weakly collisional) second fluid of ECOs (most importantly, stellar-mass BHs), that accrete gas and exert heat in return, stabilizing the disk structure. The primary effect of the ``second fluid'' in our model comes in the local energy conservation equation: 
\begin{equation}
 \sigma T_{\rm{eff}}^4= \frac{3GM\dot{M}_{\rm g}}{8\pi R^{3}}\left[1-\left(\frac{R_{0}}{R}\right)^{\frac{1}{2}}\right]+Q_{\bullet}
 \label{eq: Energy conservation}
\end{equation}
Here $\sigma$ is the Stefan–Boltzmann constant, $T_{\rm{eff}}$ is the local effective temperature of the disk, $G$ is the gravitational constant, $R_0$ is the inner edge (usually the innermost stable circular orbit, i.e. ISCO) of the SMBH\newp{, $M$ is the mass of the SMBH, $\dot{M}_{\rm g}$ is the accretion rate of the SMBH, and $R$ is the distance from the center of the SMBH}. The first term on the right-hand side of \cref{eq: Energy conservation} is the viscous dissipation per unit area in the disk, and the second term, $Q_\bullet$, is the heat per unit area generated by the ECOs ($Q_\bullet$ will be discussed in greater detail in \sref{sec : accretion onto embedded objects}). \cref{eq: Energy conservation} states that in thermal equilibrium, all heat generated by the disk's physics is locally radiated 
Here $c_{\rm s}$ is the local gas sound speed, $\Omega$ the orbital frequency (which we approximate as being equal to the epicyclic frequency $\kappa_\Omega$), and $\Sigma$ the gas surface mass density. Key variables for this model are summarized in \cref{table: glossary}. Many of these past works assume that the self-regulation is driven by stellar feedback (e.g. winds or radiation from massive stars) or SN explosions \citep{SirkoGoodman2003, Thompson+2005}. However, the simulations of \citet{Torrey+17} have shown that it is difficult to achieve a marginally Toomre stable disk via stellar feedback alone because of its inherently bursty nature: the small ratio of the orbital time to the stellar evolution timescale implies that too many stars form to sustain a $Q_{\rm T}=1$ steady state. Instead, stellar feedback triggers cycles of star formation, gas expulsion, and AGN quenching, followed by further gas inflow and AGN turn-on. Consequently, stellar feedback alone cannot sustain a steady-state solution, and disks without other power sources must exist in a constant limit cycle of forming and exploding new stars.away 
with an effective temperature $T_{\rm eff}$, as shown in \cref{eq: Energy conservation}. 

Past solutions to the problem of Toomre instability in AGN disks usually assume that these disks find a way to self-regulate to the point of marginal Toomre stability, i.e. $Q_{\rm T} \sim 1$, where the Toomre parameter \citep{Toomre1964} is defined as
\begin{equation}
 Q_{\rm{T}}\equiv \frac{c_s \kappa_\Omega}{\pi G \Sigma} \approx \frac{c_s \Omega}{\pi G \Sigma}.
 \label{eq: Toomre stability parameter}
\end{equation} 


However, many of the stars formed in the first generation of Toomre instability will end their lives as compact objects in the disk, with an additional source of ECOs coming from a preexisting quasi-spherical population (i.e. a nuclear star cluster) that are captured by the disk \citep{Bartos+17a}. Accretion onto these ECOs, in particular stellar-mass BHs, may in turn become the main heating source for a Toomre-stable disk ($Q_{\rm{T}}\geq 1$), halting the cycles of star formation and outflow seen in \citet{Torrey+17}. AGN stabilization via accretion onto ECOs has been considered before, first in pioneering work by \citet{Levin07} and more recently by \citet{DittmanMiller20}. However, neither of these works construct global disk models accounting for the dynamical properties of the ECOs that can self-regulate AGN disks to a $Q_{\rm T} \approx 1$ state; in particular, the accumulation, distribution, and migration of ECOs are all dynamical processes which must be treated self-consistently. Our goal in this section is to form a general set of two-fluid disk equations that can describe {\it feedback-dominated accretion flows}, or FDAFs: gaseous accretion disks where the primary heat source is energy feedback from a coplanar collisionless disk.

\subsection{Sources of Feedback} \label{sec: other sources}
Aside from accretion onto embedded BHs, we must also consider, as suggested in past work, other sources of feedback into the disk. In this subsection we show that these are either unrealistic or subdominant compared to BH accretion feedback. These include feedback from stellar winds, radiation and SNe \citep{SirkoGoodman2003, Thompson+2005}, 
and Bondi--Hoyle--Lyttleton accretion onto other types of ECOs. 
\paragraph{Stellar winds and radiation.}
Massive young stars emit energetic line-driven winds with velocities $v_{\rm{wind}}\sim 1000 \rm{~km~s^{-1}}$. These stars will, over their lifetimes, unbind of order half of their mass ($M_{\star}$) through stellar winds and therefore the time-averaged mass-loss rate is
\begin{equation} \label{stelar wind rate}
\dot{M}_{\rm{wind}} \sim \frac{1}{2} \frac{M_\star}{\tau_\star},
\end{equation} 
where $\tau_\star$ is the main-sequence lifetime. Now we can approximate the heating rate from one such star:
\begin{equation}
 L_{\rm{wind}}=\frac{1}{2}\dot{M}_{\rm{wind}} v_{\rm{wind}}^2,
 \label{Stellar winds heating rate}
\end{equation}
which we will compare to the luminosity of a stellar-mass BH (mass $m$) accreting at the Eddington limit,
\begin{equation} \label{eq: eddington luminocity}
 L_{\rm{Edd}}=\frac{4\pi Gm c }{\kappa_{\rm{es}}},
\end{equation}
where $\kappa_{\rm{es}}\approx 0.035\frac{\rm{m^2}}{\rm{kg}}$ is the electron-scattering opacity. The typical ratio of BH accretion to stellar wind luminosity is thus
\begin{equation}
 \frac{L_{\rm{Edd}}}{L_{\rm{wind}}} \sim \frac{4 \pi G c m / \kappa_{\rm{es}}}{\frac{1}{2}\dot{M}_{\rm{wind}} v_{\rm{wind}}^2} \approx 1.5\times 10^3 \left( \frac{m}{10 M_\odot}\right) \left( \frac{\dot{M}_{\rm wind}}{3 M_\odot~{\rm Myr}^{-1}} \right)^{-1},
 \label{L ratio stellar wind}
\end{equation}
where we have set $v_{\rm wind}=10^3~{\rm km~s}^{-1}$.  Stellar radiation can be a more important feedback source, however: if we assume that $\sim 10\%$ of the massive stars' H burns into He with a mass-to-energy efficiency of 0.007, the time-averaged radiative luminosity $L_{\rm rad}$ is a factor $\sim 200$ larger than the wind kinetic luminosity and thus $L_{\rm Edd}(m) \sim 5 L_{\rm rad}$.  


These comparisons show that the number of massive stars needed to supply a finite amount of energy feedback into the AGN disk over timescales of several Myr will exceed the number of BHs that perform the same task. Thus, a disk that supports itself in some type of marginally stable state via stellar feedback will cease forming stars once the youngest and most massive stars collapse into BHs.  We note that this comparison has assumed that $100\%$ of the kinetic luminosity of stellar winds, the photon luminosity of stellar radiation, and the total luminosity (radiative plus kinetic) of accretion feedback are absorbed by the AGN gas.  This assumption is reasonable for radiative accretion feedback (\S \ref{section: opacity}), so our argument is robust against it.

Of course, the total energetics of stellar wind feedback may dominate ECO accretion feedback initially, when the AGN disk is poor in ECOs. Only after the first generation of massive stars has formed and died in the disk (i.e. after the first few Myr) will ECO accretion feedback come to dominate stellar winds, although even at early times it is possible that preexisting BHs captured into the AGN through gas drag \citep{Syer+91, MiraldaEscudeKollmeier05, Bartos+17a} can contribute substantially to disk heating. 

\paragraph{SN explosions.}
Here we will approximate the rate of SNe inside the disk that is needed to heat the gas as effectively as the BHs would. 
For an explosion with average ejecta mass $M_{\rm{SN}}$, and an average velocity of $v_{\rm{SN}}$, the energy added to the disk from one supernova is $E_{\rm{SN}}=\frac{1}{2}M_{\rm{SN}} v_{\rm{SN}}^2$. By defining the time-averaged rate of supernova explosions per unit area $\dot{S}_{\rm{SN}}$ we get the total rate of heating per unit area :
\begin{equation}
 Q^+_{\rm{SN}}=\frac{1}{2} \dot{S}_{\rm{SN}} M_{\rm{SN}} v_{\rm{SN}}^2
\end{equation}
\newp{It is important to note that, similarly to the  stellar winds comparison, we are assuming that entire energy budget of the SN (the dominant term is the kinetic energy of the supernova)  is used to heat the gas without going into the specific mechanism for heating.  This is the most charitable possible assumption for SN heating.} Similarly to stellar winds, we compare this to the areal ECO accretion feedback rate, where $S_{\bullet}$ is the surface number density of ECOs in the disk:\footnote{\newp{That is the number of embedded BHs per unit area, akin to the gas surface mass density $\Sigma$.}}
\begin{equation}
 \frac{Q^+_{\rm{SN}}}{Q^+_{\bullet}}=\frac{\frac{1}{2} \dot{S}_{\rm{SN}} M_{\rm{SN}} v_{\rm{SN}}^2}{S_{\bullet}L_{\rm{Edd}}}.
\end{equation}
Here we use \cref{eq: eddington luminocity} for the Eddington luminosity and take $m=10 M_\odot$ as a typical BH mass. Typical Type II SNe have total explosion energies $E_{\rm SN} \approx 10^{51}$ erg \citep{KasenWoosley09, Rubin+16}, which we approximate by choosing $M_{\rm{SN}}=5 M_{\odot}$ and $v_{\rm{SN}}=5\times 10^3 \frac{\rm{km}}{s}$. Solving for the areal SN rate that sets $Q_{\rm SN}^+ = Q_\bullet^+$, we find\footnote{Note that the critical SN rate required to set $Q_{\rm SN}^+ = Q_\bullet^+$ (assuming an Eddington capped accretion luminosity) can also be simplified in terms of the Salpeter time $t_{\rm Sal}$ as $\dot{S}_{\rm SN} = 2(S_\bullet / t_{\rm Sal}) (m / M_{\rm SN}) (v_{\rm SN}/c)^2$. }:
\begin{equation}
 \dot{S}_{\rm{SN}}\simeq 3.7 \times 10^{-5} \frac{1}{\rm{yr}} S_{\bullet}.
 \label{eq: SN rates}
\end{equation}
If anything, this is an optimistic result for SN heating; in reality a higher rate may be needed when we consider that (i) the explosion is spherically symmetric and (usually) $v_{\rm SN} \gg \sqrt{GM/R}$, so that a major fraction (dependent on the local AGN aspect ratio $\frac{H}{R}$) of the ejecta will fly out of the disk and will fail to add its kinetic energy to the gas \citep{Grishin+2021}; and (ii) that the low frequency of SNe may require an even higher rate for the heat to mix effectively through the disk (see \sref{sec : Heat mixing} for further discussion of this point). However,  even with  this optimistic value of $\dot{S}_{\rm SN}$, \newp{we get unrealistically high SN rates\footnote{\newp{Using the results discussed in \sref{sec: pileup solution} together with \cref{eq: SN rates} (see \aref{app: additional reults}).}}. Furthermore, this} shows that if an SN-heated disk can achieve a $Q_{\rm T}=1$ steady state, the heating rate will only remain dominated by SNe for the first generation of star formation (i.e. $\sim 3\times 10^6$ yr). For every SN explosion, there is an $\mathcal{O}(1)$ number of stellar-mass BH remnants created in failed SNe, so after $\sim3$ Myr, enough BHs will have been generated to completely replace SN heating (and perhaps even over stabilize the disk). 
\paragraph{Stellar accretion.} 
Consider a star with mass $M_\star$ and radius $R_\star$.
In principle, accretion onto this star could also help stabilize the disk against further fragmentation, and one might speculate that this type of accretion feedback could outcompete BH accretion feedback due to the greater number of stars. However, the accretion efficiency onto BHs is much higher. Per accretor, the luminosity ratio at fixed $\dot{M}_{\rm g}$ is $L_\star / L_\bullet \sim R_{\rm g}(m) / R_\star \ll 1$, where $R_{\rm g}=GM/c^2$ is the gravitational radius. This estimate assumes accretion onto neither object is Eddington limited (i.e. small gas density $\rho$); if both are Eddington limited (large gas density $\rho$), the luminosity ratio is $L_\star / L_\bullet \sim M_\star / m$. In an intermediate regime of $\rho$, accretion onto BHs will be Eddington capped but the (less efficient) stellar accretion will not be. We now consider the relative accretion efficiency of BHs in comparison to main-sequence stars, white dwarfs (WDs) and neutron stars (NSs). 

For a single generation of star formation, the ratio of main-sequence stars to stellar-mass BHs will be given by the initial mass function (IMF), the age of the population $t_{\rm age}$, and the transition mass(es) between NS and BH formation. From both theoretical arguments and observations of the Milky Way nuclear stellar disk, disk-mode star formation is likely to have a top-heavy IMF. We take the observational estimates of \citet{Lu+2013} $dN_{\star}/dM_{\star}\propto M^{-1.7}$ as fiducial, and leave the uncertain minimum (maximum) mass of disk-mode star formation as a free parameter $M_{\rm min}$ ($M_{\rm max}$). We also assume a single zero-age main-sequence mass $M_{\rm trans}$ that separates NS from BH formation. The best-case scenario for stellar accretion feedback is when both BH and stellar accretion are Eddington limited. In this limit, the ratio of total stellar to BH accretion luminosity in a zone of fixed gas properties is
\begin{equation}
 \frac{L_{\rm MS}}{L_\bullet} = \frac{\int_{\rm M_{\rm min}}^{\rm M(t_{\rm age})} M M^{-1.7}dM}{\int_{\rm M_{\rm trans}}^{\rm M_{\rm max}} m(M) M^{-1.7}dM} \sim 1.
\end{equation}
Here we have assumed that $M_{\rm trans} > M(t_{\rm age})$, the turnoff mass from the main-sequence given the system age. Assuming that the BH initial--final-mass relation gives $m(M) \sim M$, $M_{\rm trans} \approx 20 M_\odot$, and $M(t_{\rm age}) \approx 10 M_\odot$ gives $L_{\rm MS} \sim L_\bullet$. However, this estimate is generally far too generous for stellar accretion efficiency, because gas densities in the $Q_{\rm T} \lesssim 1$ regions of AGN disk models are almost never high enough for stellar accretion to be Eddington limited (we verify this against our detailed models later). 

NS feedback is always subdominant to BH feedback because the number of NSs is comparable, but their Eddington limit is an order of magnitude lower. WDs do not have time to form in typical AGN episodes, although a small population could be captured by gas drag. For these reasons, we ignore all accretion feedback apart from BHs but note that even in the limiting regime where stellar accretion feedback is maximally effective, it represents an $\mathcal{O}(1)$ correction to the BH feedback.

\subsection{Disk Microphysics} \label{sec:microphysics}

In this subsection we discuss two key ``microphysical'' components of the gas dynamics in our model: angular momentum transport and opacity. For simplicity we assume an ideal gas equation of state with and adiabatic index $\gamma = 5/3$.

\subsubsection{Angular Momentum Transport}

The classic Shakura--Sunyaev model employs an $\alpha$ prescription for effective viscosity in accretion disks. In this picture, angular momentum (AM) is transferred through the disk via turbulence, which can be approximated with an effective kinematic viscosity $\nu$. AM transport is fundamentally local, is sourced by the magnetorotational instability (MRI), and is limited by the plausible motion of turbulent eddies, i.e. by the local disk scale height $H$ and the local sound speed $c_{\rm{s}}$. The Shakura--Sunyaev ansatz states $\nu = \alpha c_{\rm s} H$, with the dimensionless parameter $\alpha \leq 1$. Using this model for an AGN, one finds that the inflow timescales (``viscous timescales'') become much higher than the AGN lifetime beyond a modest radius, $R_{\rm{visc}} \sim 10^{3-5} R_{\rm{g}}$. Nonetheless, observed AGN gas accretion rates, $\dot{M}_{\rm g}$, imply a total mass accreted $\Delta M = \dot{M}_{\rm g} t_{\rm{AGN}}$ that is much greater than the mass enclosed inside $R_{\rm{visc}}$ \citep{ShlosmanBegelman89}.

This classical problem with AGN disks, which we will refer to as the ``inflow problem,'' has motivated theorists to consider stronger mechanisms of angular momentum transport that are usually nonlocal in nature. The most common of such an approach is to assume angular momentum is transported by global torques produced by low-order nonaxisymmetric features, which may arise naturally in Toomre-unstable accretion disks \citep{HopkinsQuataert11}. An additional possibility is that the combined effect of many small linear density waves (launched by individual ECOs) may also transport angular momentum at enhanced rates, either locally or globally. This possibility is discussed in greater detail in \sref{sec : Migration}, although we find it to be unlikely to dominate local effective viscosity from the MRI.

As we will demonstrate later in the paper, the inflow problem is significantly alleviated by considering two-fluid disks that are substantially heated by accretion feedback (these hotter disks have larger aspect ratios and thus shorter viscous times). In our model, we therefore use the local AM transport $\alpha$ prescription, and we will show post hoc that the viscous timescales are not as long as in the Shakura--Sunyaev model (\cref{fig: pileup visc time}). For specificity, we take $\alpha=0.1$ in our numerical disk solutions.

\subsubsection{Opacity} \label{section: opacity}
We construct our model using realistic tabulated results for the Rosseland mean opacity $\kappa$. We use the updated OPAL tables \citep{IglesiasRogers1996} for high temperature $\left( T>10^4\rm{K} \right)$ and opacities from \citet{Semenov+2003} for lower temperatures. At temperatures immediately below $\sim 10^4 \rm{K}$ most of the gas is made of neutral molecules and therefore the opacity becomes very low, but at around $1500$ $\rm{K}$ dust particles start to form and become the dominant opacity source. The drop in opacity for the range of temperatures from $10^3 -10^4 \rm{K}$ is known as the ``opacity gap'' and has major implications for disk structure and Toomre stability \citep{Thompson+2005}. Diskontinuous jumps in opacity at low temperatures due to specific dust sublimation edges (e.g. sublimation temperatures for silicates or water) will create discontinuities in our local disk structure. These should be taken with a grain of salt, as in reality the turbulent motion of fluid within the disk will advect some dust from cold regions into hotter annuli with average temperatures slightly above the dust sublimation threshold, increasing the opacity there (the reverse process will also reduce the opacity of annuli slightly colder than the dust sublimation threshold). Turbulent mixing will thus preclude the development of razor-sharp discontinuities in disk properties, but we defer a detailed investigation of this for future work.

The standard Shakura--Sunyaev model assumes an optically thick disk, but the presence of the opacity gap means that in our solution we need to include the optically thin limit as well. We bridge the gap between these two limits by modeling radiative cooling rates with an effective temperature $T_{\rm eff}$ as in \citep{SirkoGoodman2003, Thompson+2005}: 
\begin{equation} \label{eq: temperature}
 T_{\rm{eff}}^4=T_{\rm{c}}^4 \left( \frac{3}{8} \tau +\frac{1}{2} +\frac{1}{4\tau} \right)^{-1}. 
\end{equation}
Here $T_{\rm c}$ is the midplane temperature.

Even if the AGN disk is optically thin to its own radiation, it may not be optically thin to all radiation. The miniature accretion disks around embedded BHs put out most of their energy in $\sim0.1 - 1$ keV photons. When these encounter quasi-neutral gas (as in the opacity gap and larger radii), they experience huge bound--free opacities from metals (carbon, oxygen, etc.). 
Using the fitting functions for frequency-dependent X-ray absorption cross sections $\sigma_{\nu}^{\rm bf}$ in \citet{Verner+1996}, we approximate the opacity experienced by high-energy photons as $\kappa_\nu = \sum \kappa_{i, \nu}$, where $\kappa_{i, \nu} = \sigma_{i, \nu}^{\rm bf} Z_i / (\mu m_{\rm p})$ is the opacity of an individual species with index $i$. Considering neutral\footnote{We assume mass fractions $Z_i$ of H I, He I, C I, and O I that are 0.74, 0.24, 0.01, and 0.01, respectively. We note that this may be a conservative choice if the AGN disk has been substantially metal enriched due the deaths of massive stars formed {\it in situ} \citep{Cantiello+21, Dittmann+21}.} H, He, C, and O, using the thin-disk approximation, $\tau_\nu \approx \kappa_\nu \Sigma/2 $, and noting that in our results the surface mass density $\Sigma\gtrsim 5\times 10^{-1} \ \rm{kg} \ \rm{m}^{-2} $ in essentially all regions of interest (see \cref{fig: pileup Sigma}), we find that the AGN disk is always opaque (and usually very optically thick) to high-energy photons. For example, at 0.1 keV (1 keV), we find $\kappa_\nu \approx 4\times 10^3~{\rm m}^2~{\rm kg}^{-1}$ ($\kappa_\nu \approx 5 ~{\rm m}^2~{\rm kg}^{-1}$). Thus, even in regions that are optically thin in the Rosseland-mean sense (i.e. to their own thermal photons), the disk can be heated via feedback from accretion onto ECOs.

\subsection{Mesophysics of Embedded Objects}
\label{sec:mesophysics}
In this section, we describe in greater detail the interactions between the AGN gas and its ECO population. These interactions are referred to as ``mesophysical'' as they concern scales in between gas microphysics and the global properties of the two-fluid disk as a whole.
\subsubsection{Migration of Embedded Objects } \label{sec : Migration}
Consider an ECO of mass $m$ in orbit inside the disk. There are several mechanisms that result in its migration through the disk. In this section, we generally make the approximation that all ECO orbits are circular, deferring the more general case of nonzero eccentricity to future work. However, we note when this assumption could qualitatively impact our results.

\paragraph{Type I Migration.}
Type I migration is the (usually) inward movement of an embedded body in a disk of gas due to the exchange of torques between the body and the linear density waves it creates in the gas. In general, linear tidal perturbations from a single ECO will create two spiral arms that exert torques of opposite sign on the ECO; the residual difference between these two torques is the Type I torque $\Gamma_I$ \citep{GoldreichTremaine79, GoldreichTremaine80, Artymowicz93, Tanaka+02}. Note that although the net torque exchange on one ECO is relatively small (as it is a second-order residual of the difference between each one-arm torque), the absolute magnitude of the one-armed torques can be much larger than the net type I torque, and for a very large number of ECOs, this could become a mechanism for global torque transfer. However, if one considers a continuous ECO approximation (as we do) then the torque from inward spiral arms passing through a specific annulus will be canceled, to leading order, by the torque from the outer spiral arms also passing through this annulus. 

Alternatively, the density waves that produce type I torques can provide a source of local angular momentum transport in regions where the continuous embed approximation holds via the shock dissipation of the gravitational wakes \citep{GoodmanRafikov2001}. These shocks exert local torques that can be expressed in the effective $\alpha$-viscosity prescription with a local $\alpha$ parameter. This is an alternative (or additional) source of local effective viscosity (beyond the usual MRI picture). Using our BH surface number density $S_\bullet$ and Eqs. 8 and 35 from \citet{GoodmanRafikov2001}, we obtain the approximate value of this local effective viscosity, $\alpha_{\rm eff}$:
\begin{equation}
 \alpha_{\rm eff} \approx 14\left(\frac{\gamma+1}{7/5+1}\right)^{-2/5}S_{\bullet}R^{2}\left(\frac{H}{R}\right)^{-14/5}\left(\frac{m_{\bullet}}{M}\right)^{8/5}
\end{equation} 
In our model, in the regions where the continuum approximation holds, we get $\alpha_{\rm eff}\sim 10^{-2}-10^{-1}$ which is comparable to expected effective viscosity values from MRI turbulence, both those seen in simulations \citep{Hawley+2013,Penna+2013} and also inferred from observations \citep{King+2007}. Note that in our numerical results, we always take $\alpha = 0.1$. 

For ECOs in AGN disks, we find that type I torques are usually the dominant form of migration. Sufficiently strong tidal torques could also clear a gap between the gas and the ECO, placing it in the type II migration regime \citep{LinPapaloizou93}, but ECOs are generally not massive enough to open gaps (\citealt{Stone+17}, unless they grow into intermediate mass BHs; \citealt{GoodmanTan04, McKernan+14}). In our model, we assume that the two-fluid disk never enters the Type II regime, which we confirm in {\it post hoc} (see \aref{app: additional reults}) analysis using the gap opening criterion of \citet[][see our \cref{eq: gap equation}]{Crida+06}.

We therefore use the type I migration torque \citep{Paardekooper+10}: 
\begin{equation}
 \Gamma_I=-C_{I}\left(\frac{2m}{M}\right)^{2}\left(\frac{H}{R}\right)^{-2}\Sigma R^{4}\Omega^{2}_k,
 \label{eq: Type I torque}
\end{equation}
where 
$C_I$ is a factor $\sim \mathcal{O}(1)$ dependent on the temperature and surface density power laws. Assuming $T_{\rm c} \propto R^{-\xi}$ and $\Sigma \propto R^{-\delta}$, this factor becomes \citep{Paardekooper+10}:
\begin{equation}
 C_I=0.8+ 1.0\xi +0.9\delta. 
 \label{eq: C_I equation}
\end{equation}
If temperature and surface density are declining functions of $R$, the migration will always be inwards. However, if the surface density and/or temperature profiles increase with radius $R$ quickly enough, the net torque can actually lead to outspiral. If there is a transition in the disk from negative to positive $C_I$ we get a classical ``migration trap'' \citep{Bellovary+2016}.
Using \cref{eq: Type I torque} and the Keplerian angular velocity, and assuming a steady state in the disk, we can calculate the migration rate for a circular orbit of semimajor axis $R$:
\begin{equation}
 \frac{da}{dt}\bigg|_{\rm{I}} = -8C_{\rm{I}}G^{\frac{1}{2}} m M^{-\frac{3}{2}} R^{\frac{3}{2}} \Sigma \left(\frac{H}{R}\right)^{-2}.
 \label{eq: type I migration rate}
\end{equation}

\paragraph{Gravitational Wave Emission}
If the orbiting ECO is close enough to the SMBH, all interactions with the gas become negligible and the main cause of migration is direct GW emission. Using the post Newtonian approximation, the average migration rate for a circular orbit is \citep{Peters1964}:
\begin{equation} \label{eq: GW migration rate}
 \frac{da}{dt} \bigg|_{\text{GW}} = - \frac{64}{5} \frac{G^3 M m (M+m)}{c^5 R^3}
\end{equation}
where $c$ is the speed of light.
For the full migration calculation we use the sum of both migration types mentioned above.

\subsubsection{Accretion onto Embedded Objects} \label{sec : accretion onto embedded objects}
Consider an embed (with mass $m$) moving in a uniform gaseous medium. Fluid elements will become gravitationally bound and accrete onto the object if at infinity their impact parameters are smaller than the Bondi--Hoyle--Lyttleton radius:
\begin{equation}
 R_{\rm{BHL}}=\frac{2Gm}{c_s^2+v_{\rm{rel}}^2} 
 \label{eq: Bondi-Hoyle radius}
\end{equation}
where $v_{\rm{rel}}$ is the relative velocity between the gas and the massive body. The accretion will happen at the rate \citep{Bondi1952,HoyleLyttleton1939,Shima+1985}:
\begin{equation}
 \dot{m}_{\rm{BHL}}=\pi R_{\rm{BHL}}^2 \left(c_s^2+ v_{\rm{rel}}^2 \right)^{1/2} \rho=\frac{4 \pi G^2 m^2 \rho}{\left(c_s^2+ v_{\rm{rel}}^2 \right)^{3/2}}
 \label{eq: BH accretion rate}
\end{equation}
where $\rho$ is the gas mass density, and $\pi R_{\rm{BHL}}^2$ is the effective cross section for the accreted gas. If we change the geometry to an embedded mass orbiting an SMBH (with mass $M$) inside a disk of gas, with a finite scale height $H$, the cross section will be limited by the height of the disk, and also by the gravitational pull of the SMBH \citep{Stone+17, Dittmann+21}. Using the Hill radius \citep{MurrayDermott1999} $R_{\rm{H}}=R \left(\frac{m}{3M} \right)^{1/3}$, where $R$ is the semi-major axis of the embed, we can determine the maximum impact parameter needed for a particle to become gravitationally bound to the embedded mass. Specifically, we compute the radius where a freefalling fluid element with impact parameter $b$ and velocity at infinity $v_{\rm rel}$ crosses the symmetry axis of the problem and require this to be within the Hill sphere of the accretor, yielding a maximum impact parameter:
\begin{equation}
 b_{\rm{H}}=\sqrt{R_{\rm{BHL}} R_{\rm{H}}}.
 \label{eq: Hills impact parameter}
\end{equation}
We can now formulate a ``reduced Bondi--Hoyle--Lyttleton'' (RBHL) accretion rate, with a new effective elliptic cross section:
\begin{equation}
 \dot{m}_{\rm{RBHL}}=\pi A_{\rm{RBHL}}B_{\rm{RBHL}} \left(c_s^2+ v_{\rm{rel}}^2 \right)^{1/2},
 \label{eq: RBH accretion rate}
\end{equation} 
where
\begin{align*}
 A_{\rm{RBHL}} &\equiv \min (R_{\rm{BHL}},b_{\rm{H}}) \\
 B_{\rm{RBHL}} &\equiv \min (R_{\rm{BHL}},b_{\rm{H}},H),
\end{align*}
though we use the approximate expressions:
\begin{align}
 A_{\rm{RBHL}} &\approx \left(\frac{1}{R_{\rm{BHL}}}+\frac{1}{b_{\rm{H}}} \right)^{-1} \label{eq: A-cross section} \\
 B_{\rm{RBHL}} &\approx \left(\frac{1}{R_{\rm{BHL}}}+\frac{1}{b_{\rm{H}}} +\frac{1}{H} \right)^{-1}
 \label{eq: B-cross section}
\end{align}
in our disk models. After being gravitationally captured, the AGN gas forms an accretion mini disk around the embedded object, producing radiation with a luminosity:
\begin{equation}
 L_{\rm RBHL}=\eta c^2 \dot{m}_{\rm RBHL},
 \label{eq: Luminocity from accretion} 
\end{equation}
where $\eta$ is the dimensionless radiative efficiency. For thin, prograde accretion disks \newp{of an ECO}, $\eta$ ranges from 0.075 (the Schwarzschild limit) to $\sim 0.32$ (the efficiency of a $\chi=0.998$ BH, the maximum spin achievable through thin-disk accretion; \citealt{Thorne1974}). 

Even this reduced Bondi--Hoyle--Lyttleton accretion rate generally translates into highly super-Eddington accretion onto the ECOs, with attendant uncertainties. Theoretical radiation-hydrodynamical simulations of super-Eddington accretion disks have historically found different results on the time-averaged accretion rate $\langle \dot{m} \rangle$, with some numerical radiation transport schemes showing $\langle \dot{m} \rangle \sim \dot{m}_{\rm Edd}$, and others showing $\langle \dot{m} \rangle \gg \dot{m}_{\rm Edd}$. Likewise, different numerical methods find different emergent luminosities, both in radiation and in the kinetic luminosity of subrelativistic outflows \citep[see, e.g., Fig. 5 in][]{Inayoshi+20}.


Unfortunately, the aforementioned small-scale super-Eddington accretion disk simulations are currently unable to resolve radii comparable to $R_{\rm BHL}$, and the full picture of time-dependent, super-Eddington Bondi--Hoyle--Lyttleton accretion with feedback remains incomplete.  However, if it were possible for an ECO to accrete at highly super-Eddington time-averaged rates, it would swiftly grow into an intermediate-mass BH large enough to open gaps in the AGN disk (see \aref{app: additional reults}; also \citealt{GoodmanTan04, Stone+17}).  \newp{Likewise, if it were possible for a {\it population} of ECOs to accrete and grow at highly super-Eddington rates, they would consume $\gg 100\%$ of the total $\dot{M}_{\rm g}$ flowing through the AGN disk, starve the central SMBH, and quickly deactivate the AGN, as we show later in \S \ref{sec: pileup solution}.} Inside the opacity gap, $\dot{m}_{\rm RBHL}$ is generally super-Eddington by many orders of magnitude, and the presence of even a small number of ECOs would gravely disrupt the disk structure (both by eventual gap opening but more immediately simply by starving the central SMBH of gas), making it challenging to feed the central SMBH at observed accretion rates.

For these reasons, we follow most past work (e.g. \citealt{Park+20,Tagawa+20b}) and cap the time-averaged accretion rate onto ECOs at the Eddington luminosity:
\begin{equation*}
 L_{\bullet}=\min \left(L_{\rm{RBHL}},L_{\rm{Edd}} \right)
\end{equation*}
or approximately :
\begin{equation}
 L_{\bullet}\approx \left( \frac{1}{L_{\rm{RBHL}}}+\frac{1}{L_{\rm{Edd}}} \right)^{-1}
 \label{eq: capped Luminocity}
\end{equation}
Although the radiative efficiency of near-Eddington accretion flows is debated, we take $\eta=0.32 $ (corresponding to thin-disk accretion with BH spin $\chi=0.998$), as we expect most ECOs to be spinning near-extremally (see \sref{sec: mass growth} for a discussion on this). In practice, the choice of $\eta$ has a limited effect on our results, because at most radii, accretion is Eddington limited.

\subsubsection{Heat Mixing} \label{sec : Heat mixing}
One limitation of past 1D models for marginally Toomre-stable disks is the implicit assumption of efficient heat mixing. Disk models that manually set 
$Q_{\rm T}=1$ (in regions where $Q_{\rm T}$ would otherwise be smaller) and work out the level of ``feedback'' required to sustain marginal stability, assume by construction that sources of feedback are smoothly distributed through the disk. However, this is not at all obvious given the spatially and sometimes temporally discrete nature of feedback sources. Feedback associated with embedded stars/compact objects will have an intrinsic level of spatial discreteness based on the fact that the stellar/ECO ``fluid'' is not truly continuous, and feedback from SN explosions will also be discrete in time because of the huge energy output of individual SN explosions (see e.g. \cref{eq: SN rates}). In this subsection, we formulate an approximate set of criteria for determining whether embedded energy sources can efficiently mix enough heat through the disk to prevent fragmentation.

Consider an AGN accretion disk filled with embedded stellar-mass BHs, in the continuum approximation, with surface number density $S_{\bullet}$ and mass $m_{\bullet}$ that accrete mass and radiate out according to \cref{eq: capped Luminocity}. 
In order for this heating source to have any effect on the disk structure, we need to make sure that most of the injected heat will not escape immediately. 

\begin{figure} 
 \centering
 \includegraphics[width=0.45\textwidth]{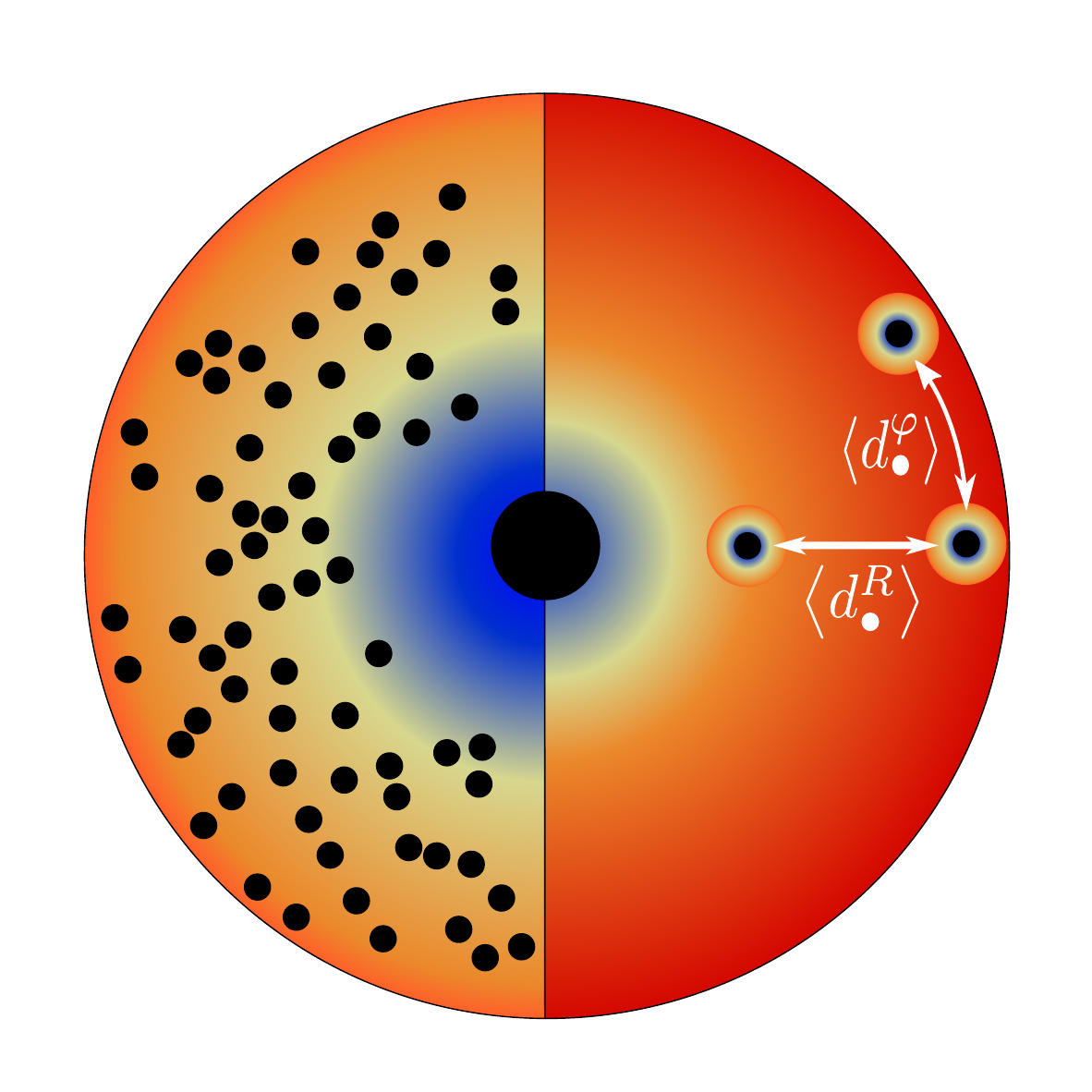}
 \caption{Schematic picture of heat mixing and its importance in a two-fluid AGN disk. On the right half of the disk, we show a highly discrete ``fluid'' of ECOs that fail to mix heat efficiently due to the large azimuthal ($\langle d_\bullet^\varphi \rangle$) and radial ($\langle d_\bullet^R \rangle$) average distances between embedded BHs. As a result, accretion feedback heats localized hot spots but fails to change global disk properties. On the left half of the disk, we show a more continuous ECO distribution that effectively mixes heat from accretion feedback. The difference between these two limits is quantified by the heat-mixing parameters $\mathfrak{M}_R$ and $\mathfrak{M}_\varphi$ (see \sref{sec : Heat mixing}).}
 \label{fig: Heat mixing }
\end{figure}

The (radial) linear BH number density, $dN_\bullet/dR$ is, in the continuum limit, $2 \pi R S_{\bullet}$, where $R$ is the semi-major axis of the orbit. The average radial distance between one BH orbit and its radially nearest neighbor is thus $\sim (2 \pi R S_{\bullet})^{-1}$. If we consider a small typical eccentricity $e\ll 1$, embedded objects will carve out radial annuli of width $eR$, thus the average interannular distance will be
\begin{equation}
 \left< d_{\bullet}^R \right>=\frac{1}{2 \pi R S_{\bullet} }-2Re.
 \label{radial distance of black holes}
\end{equation}
Photons emitted by ECOs will engage in a quasi-isotropic random walk until they reach the disk photosphere\footnote{Note that as long as the motion of turbulent eddies is quasi-isotropic in the local frame, similar considerations would apply to advective heat transport by turbulence.} at a height $z \sim H$. Therefore, there will be large radial heating gaps if $\langle d_\bullet^R\rangle \gg H(R)$: radial zones are unable to absorb heat from ECOs at different radii (see \cref{fig: Heat mixing } for a sketch of this). 

We therefore set the condition $\langle d_{\bullet}^R \rangle \lesssim H(R)$ as an approximate criterion for efficient radial heat mixing. We now can define the dimensionless radial heat-mixing parameter (by reordering the terms in the radial heat-mixing condition inequality) :
\begin{equation}
 \mathfrak{M}_R \equiv 2 \pi R H S_{\bullet} \left(1+2e\frac{R}{H} \right) \gtrsim 1.
 \label{Radial mixing parameter}
\end{equation}
If $\mathfrak{M}_R < 1$, most of the heat generated by the accreting embedded objects will diffuse outside the disk before covering the radial distance between the orbits of nearby ECOs. As a result, large annular zones will remain vulnerable to Toomre instability and further fragmentation. A region of the disk which satisfies $Q_{\rm T}=1$ {\it in the continuum limit of} $S_\bullet$ may thus continue to be Toomre unstable, due to discreteness effects, if $\mathfrak{M}_{\rm R} < 1$.

Similarly to the radial case, the average azimuthal distance between two BHs can be written as
\begin{equation*}
 \left< d_{\bullet}^{\varphi} \right>=\frac{2 \pi a}{\Delta N_{\bullet} }
\end{equation*}
where $\Delta N_{\bullet}$ is the number of BHs per annulus (each annulus is assumed here to be thin, of width $\Delta R$):
\begin{equation}
 \Delta N_{\bullet}=2 \pi a \Delta R S_{\bullet} 
\end{equation}
As with radial heat mixing, we assume that heat transport by photon diffusion is highly suppressed over distance scales greater than $H$, so we set $\Delta R \approx H$ when considering the width of annular zones of interest. Thus,
\begin{equation}
 \left< d_{\bullet}^{\varphi} \right>=\frac{1}{HS_{\bullet}}
 \label{Azimuthal distance of black holes},
\end{equation}
as is illustrated in \cref{fig: Heat mixing }. 

However, azimuthal heat mixing can be complicated by the nonzero relative speed between gas and ECOs. Three \newp{potentially} important sources of relative bulk velocity exist. 
First, we consider the sub-Keplerian rotation rate of gas, which in a thin disk moves with an average speed $\mathbf{v}_{\rm{g}} = v_k\sqrt{1-n c_s^2/v_k^2} \hat{\varphi} {  \approx v_k \left(1-\frac{1}{2} H^2/R^2\right) \hat{\varphi}}$ (\citealt{Frank+2002}; here $v_{\rm k}$ is the Keplerian velocity for circular motion and $n$ is the power-law index for the pressure, $P\propto R^{-n}$, \newp{which we approximate to be $n \sim 1$}). The second source we consider is shearing motion due to the nonzero gradient $dv_{\rm k}/dR$. The average velocity difference within an annulus of width $\sim H$ is $\Delta v_{\rm shear} \simeq 0.25 v_{\rm k} \left( H/R \right)$ \newp{to leading order in $H/R$}. \newp{As the sub-Keplerian velocity is only a second-order effect (in $H/R$) we will ignore it in the remainder of this work.}  While sub Keplerian gas rotation and shear velocities create relative bulk velocity between AGN gas and ECOs on circular orbits, this relative velocity will be modulated, and sometimes greatly increased, for ECOs on orbits with eccentricity $e>0$, which move at a speed $\mathbf{v}=v_k \frac{1}{\sqrt{1-e^{2}}} \left[e(\sin f )\hat{r}+\left(1+e\cos f\right)\hat{\varphi}\right]$ (here $f$ is the true anomaly).  \newp{However, a proper treatment of ECO eccentricity evolution is beyond the scope of this paper, so we also ignore relative bulk velocities due to finite eccentricity, and in the discussion that follows, only consider shear.}

Without relative bulk velocity, an individual ECO would only effectively heat gas a distance $\sim H$ from it in the $\hat{\varphi}$ direction. With relative bulk velocities, the continuum approximation for feedback will be roughly valid (in the azimuthal direction) as long as the time an individual gas parcel waits between heating events (i.e. passage of an ECO within an azimuthal distance $\sim H$) is less than the thermal time, $t_{\rm{th}}\sim \Sigma c_s^2/Q_-$ (here $Q_-$ is the areal heat loss rate, which in our models is due to radiative cooling). \newp{Setting $\Delta v_{\rm bulk} = \Delta v_{\rm shear}$,} we get an effective azimuthal heating distance due to relative velocity:
\begin{equation} \label{eq : delta d}
 \Delta d^\varphi_v = t_{\rm{th}} \Delta v_{\rm bulk}.
\end{equation}
If relative velocities are sufficiently low or the thermal time is sufficiently short,\footnote{In practice, we find that extra heat mixing due to bulk relative velocities can usually be ignored in our numerical solutions, due to the very short thermal time (less than the dynamical time) in outer regions of AGN disks supported by feedback sources \citep{Thompson+2005}.} then $\Delta d^\varphi_v$ will be less than $H$ and can be ignored. To consider both methods of heat mixing, we say that the total azimuthal distance that can be effectively heated by one ECO is $\approx H + \Delta d^\varphi_v$. Heat will efficiently mix in the azimuthal direction so long as the average inter-ECO distance $ \langle d_{\bullet}^{\varphi} \rangle < H + \Delta d^\varphi_v$. 
In terms of a dimensionless heat-mixing parameter, we can say that heat mixes azimuthally as long as the inequality
\begin{equation}
 \mathfrak{M} _{\varphi} \equiv S_{\bullet} H^2 \left( 1+\Delta d^\varphi_v/H \right) \gtrsim 1
 \label{Azimuthal mixing parameter}
\end{equation}
is satisfied. If $\mathfrak{M} _{\varphi} \ll 1$, large azimuthal sectors of each radial annulus will be effectively unheated by ECO feedback, {\it even if} the continuum limit of $S_\bullet$ predicts a marginally stable, $Q_{\rm T}=1$ annulus. 

To summarize, if $\mathfrak{M}_R<1$ or $\mathfrak{M}_{\varphi}<1$, accretion feedback from the embedded objects will fail to efficiently mix through the AGN disk. These dimensionless heat-mixing parameters offer a way to quantify the discreteness of the ECO distribution. While we explore discreteness effects more fully in later sections, we note here that over much of AGN parameter space, $Q_{\rm T}=1$ equilibria will fail to satisfy these heat-mixing criteria, and thus would not represent astrophysically stable solutions. Such disks would likely continue to form stars (in unheated radial annuli or azimuthal sectors) until $S_\bullet$ had increased to the point where both $\mathfrak{M}_R\ge 1$ and $\mathfrak{M}_\varphi \ge 1$, producing a marginally stable equilibrium with $Q_{\rm T} \gg 1$. In such a case, we can relate the Toomre parameter to the relevant heat-mixing parameter ($\mathfrak{M}_\varphi$, as this is always smaller than $\mathfrak{M}_R$) and disk parameters:\footnote{Here we combine \cref{eq: Toomre stability parameter} with \cref{eq: 8 SS equation} in the steady-state limit, using their reduced form in \aref{app: reduced equations }. We also assume the limit where bulk velocities are negligible, so that ($ \mathfrak{M} _{\varphi} = S_{\bullet} H^2$. In the relevant regions, the thermal timescales for our case are very short, as we will discuss in \sref{section: pileup}).}
\begin{equation}
 Q_{\rm T }=3G^{-1}\mathfrak{M}_{\varphi}^{\frac{3}{2}}S_{\bullet}^{-\frac{3}{2}}\Omega^{3}\cdot\left[\alpha^{-1}\dot{M}_{{\rm g}}\left(1-\left[\frac{R_{0}}{R}\right]^{1/2}\right)\right]^{-1}.
 \label{eq: Q-m_phi relation}
\end{equation} 
Although we have not quantified temporal discreteness in variable sources of heating, such as SN explosions, it is likely that this will pose further challenges for stability against fragmentation in AGN disks.

As noted previously, we take the $e=0$ limit for two-fluid disk solutions presented in this paper, although we have presented a more general discussion of heat-mixing considerations here. When $e=0$, $\mathfrak{M}_{\varphi}<\mathfrak{M} _R$ always, so the condition for effective heat mixing reduces to just $\mathfrak{M}_{\varphi}>1$.
\subsection{Feedback-dominated Accretion Flows}
\label{sec:Two-Fluid Disk Equations}
We now are at the point where we can incorporate the above assumptions regarding disk microphysics and embed mesophysics into a solvable system of equations. Our starting point is the classic Shakura--Sunyaev family of models \citep{ShakuraSunyaev1973}, which we follow by assuming a thin ($H\ll R$), axisymmetric accretion disk. As in the Shakura--Sunyaev picture, we assume a 1D disk (i.e. vertically averaged structure) together with quasi-viscous angular momentum transport and local thermal equilibrium. Our model for the gas disk differs from standard disk theory by including new heating sources in the energy equation, representing feedback from embedded objects. This includes both accretion feedback from ECOs, with mass $m_\bullet$, accretion luminosity $L_\bullet$, a continuum surface number density $S_{\bullet}$, and also feedback from young massive stars, represented with an areal heating term $Q_\star$. All together, we have seven algebraic equations and one partial differential equation describing the time-dependent gas-disk structure:
\begin{subequations} \label{eq: 8 SS equation}
\begin{align}
 & \Sigma=2 \rho H \label{eq: Vertical structure} \\
 & c_s= \left( \frac{GM}{R^3} \right)^{\frac{1}{2}} H=\Omega H & \label{eq: hydrostatic equilibrium} \\
 & c_s^2=\gamma \frac{P}{\rho} \label{eq: sound speed} \\
 & P=\frac{k_B}{\mu m_p} \rho T_c+\frac{\tau \sigma}{2 c} T_{\rm{c}}^4 \left( \frac{3}{8} \tau +\frac{1}{2} +\frac{1}{4\tau} \right)^{-1} \label{eq: pressure equation} \\
 & \frac{\sigma T_{\rm{c}}^4} {\frac{3}{8} \tau +\frac{1}{2} +\frac{1}{4\tau} } = \frac{9GM}{8R^{3}}\nu\Sigma+ S_{\bullet} L_{\bullet} +Q_\star \label{eq: full energy conservation} \\
 & \tau = \frac{\kappa_R \Sigma}{2} \label{eq: optical depth}\\
 & \nu =\alpha c_s H \label{eq: alpha viscosity} \\
 & \frac{\partial \Sigma}{\partial t} = \frac{3}{R}\frac{\partial}{\partial R} \left[R^{1/2} \frac{\partial}{\partial R}\left(R^{1/2}\nu \Sigma \right) \right] + \Xi \label{eq:viscousPDE} .
\end{align}
\end{subequations}
Here $k_{\rm B}$ is the Boltzmann constant, $\mu$ is the gas mean molecular weight, and $m_{\rm p}$ is a mass of a proton. 
Into the diffusion equation for gas surface mass density $\Sigma$, we have also added $\Xi$, an areal source/sink term for the gas disk, which would typically represent the rate at which gas is converted into stars or the rate at which stellar matter is returned to the disk via winds or SNe. By introducing a new variable ($S_{\bullet}$), we have also created the need for at least one new equation to close the problem properly. In the general context of a 1D, two-fluid disk model, the extra equation is a continuity equation of the form:
\begin{equation}
 \frac{\partial S_\bullet}{\partial t} = -\frac{\partial}{\partial r} (\dot{a}_\bullet S_\bullet) + X_\bullet.
\end{equation}
Here $\dot{a}_\bullet$ represents the summed contribution of all migration types, and $X_\bullet$ is an areal source term accounting for ECO formation\footnote{In principle, if these equations are coupled to a model for a quasi-spherical background star cluster, an additional term should be added to account for the capture of preexisting star/compact objects through gas drag \citep{Bartos+17a}.}. In the simplest possible model, $X_\bullet$ could be equated directly to $-\Xi/m_\bullet$, or alternatively, we could account for a ``third fluid'' of stars in the disk in a similar way (i.e. with surface number density $S_\star$ and mass $m_\star$): 
\begin{equation}
 \frac{\partial S_\star}{\partial t} = -\frac{\partial}{\partial r} (\dot{a}_\star S_\star) + X_\star-X_\bullet.
\end{equation} 
The latter approach is advantageous because it can more self-consistently account for the time delay between star formation via Toomre instability and ECO formation. Similarly to the ECOs, $\dot{a}_\star$ is the migration rate for embedded stars, and $X_\star$ is the stellar number formation rate per unit area. We account for the fact that over time stars will evolve and convert into compact objects by using the present-day mass function (PDMF) $\frac{dN}{dm_\star}=f_{\rm PDMF}(m_\star,R)$ and the initial mass function $f_{\rm{IMF}}(m_\star)$. The rate of change in the PDMF is: 
\begin{equation}
 \newp{\frac{\partial f_{\rm PDMF}}{\partial t} \left(m_\star,R \right) = X_\star f_{\rm{IMF}}(m_\star)-\frac{f_{\rm PDMF} \left(m_\star,R \right)}{T_\star\left(m_\star\right)}}
\end{equation}
where $T_\star \left(m_\star,R \right)$ is the expected lifetime of a star with mass $m_\star$. Using the above we approximate 
\begin{equation}
 \Xi =X_\bullet \left( \langle m_\star \rangle-\langle m_\bullet \rangle \right)- X_\star \langle m_\star \rangle 
\end{equation}
where $\langle m_\star\rangle$ is the first moment of the stellar PDMF.

The above formulation provides a general set of multifluid disk equations to describe the stabilization of AGN disks by feedback from embeds. However, the full solution of this set of PDEs and algebraic equations is beyond the scope of this preliminary paper, though we will explore its full time-dependent dynamics in the future. In the remainder of this paper, we will  \newp{ explore the steady-state limits of these two-fluid equations}, i.e. limits where $X_\bullet = 0$, $X_\star = 0$, $\partial S_\bullet / \partial t = 0$, $\partial S_\star / \partial t = 0$, $\Xi = 0$, and $\partial \Sigma / \partial t = 0$. As we have shown in \sref{sec: other sources}, stellar feedback is negligible when a modest number of ECOs exist, and therefore, we approximate $Q_\star=0$ (and neglect the stellar population). In these limits, the two remaining PDEs presented above reduce to 
\begin{align}
 & \nu \Sigma = \frac{\dot{M}_{\rm g}}{3 \pi } \left( 1- \left[\frac{R_0}{R}\right]^{1/2} \right) \label{eq: steady viscous transport} \\
 & 2\pi R \left| \dot{a}_\bullet \right| S_\bullet = \dot{N}_\bullet, \label{eq: steady BH transport}
\end{align}
where $\dot{N}_\bullet$ is the radial number flux of ECOs. We now present the two limiting, steady-state cases of this model.

\subsubsection{Pileup Regime} \label{section: pileup}
\begin{figure}
 \centering
 \subfloat[Pileup \label{fig:Pileup schematic}]{
 \centering
 \includegraphics[width=\linewidth]{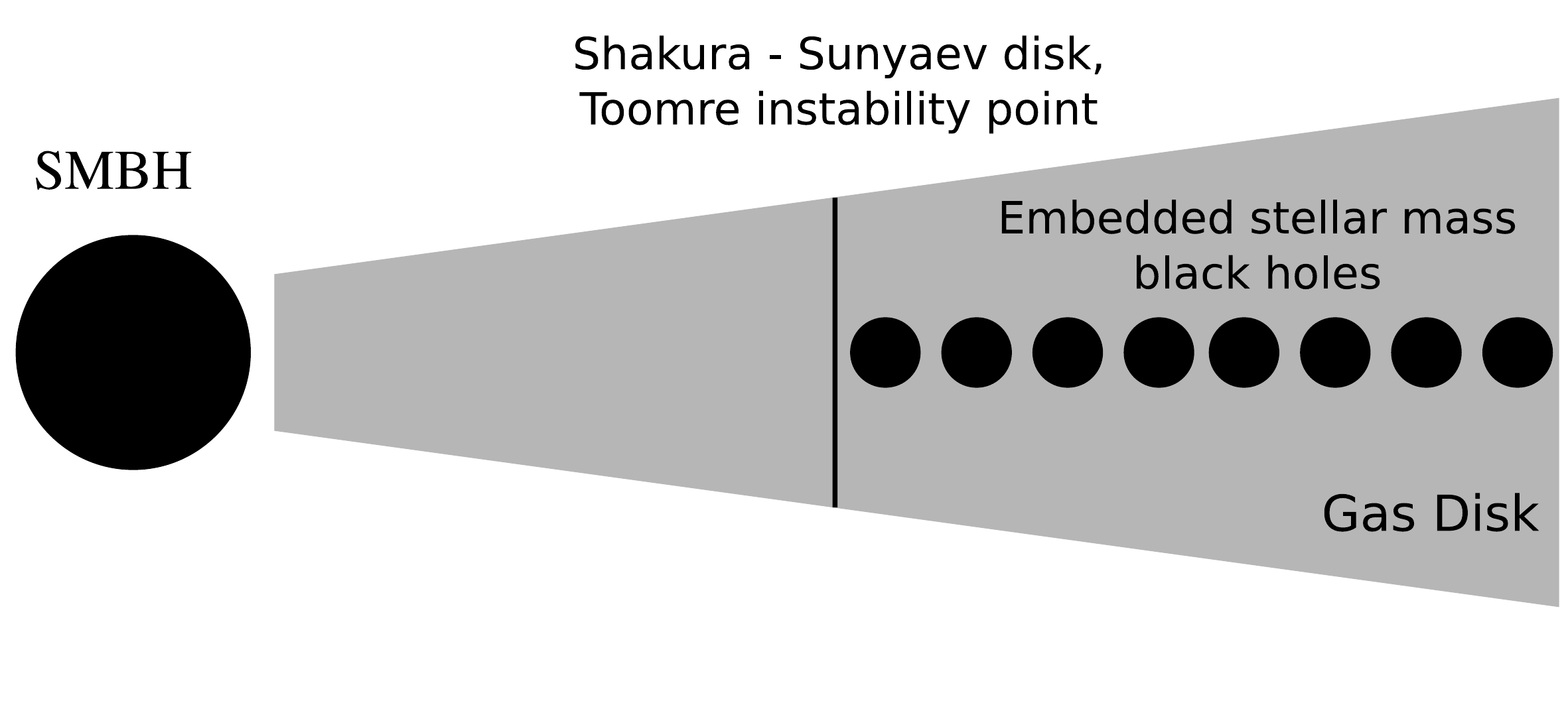}
 
 } 
 
 \subfloat[Constant mass influx \label{fig:CMI schematic}]{
 \centering
 \includegraphics[width=\linewidth]{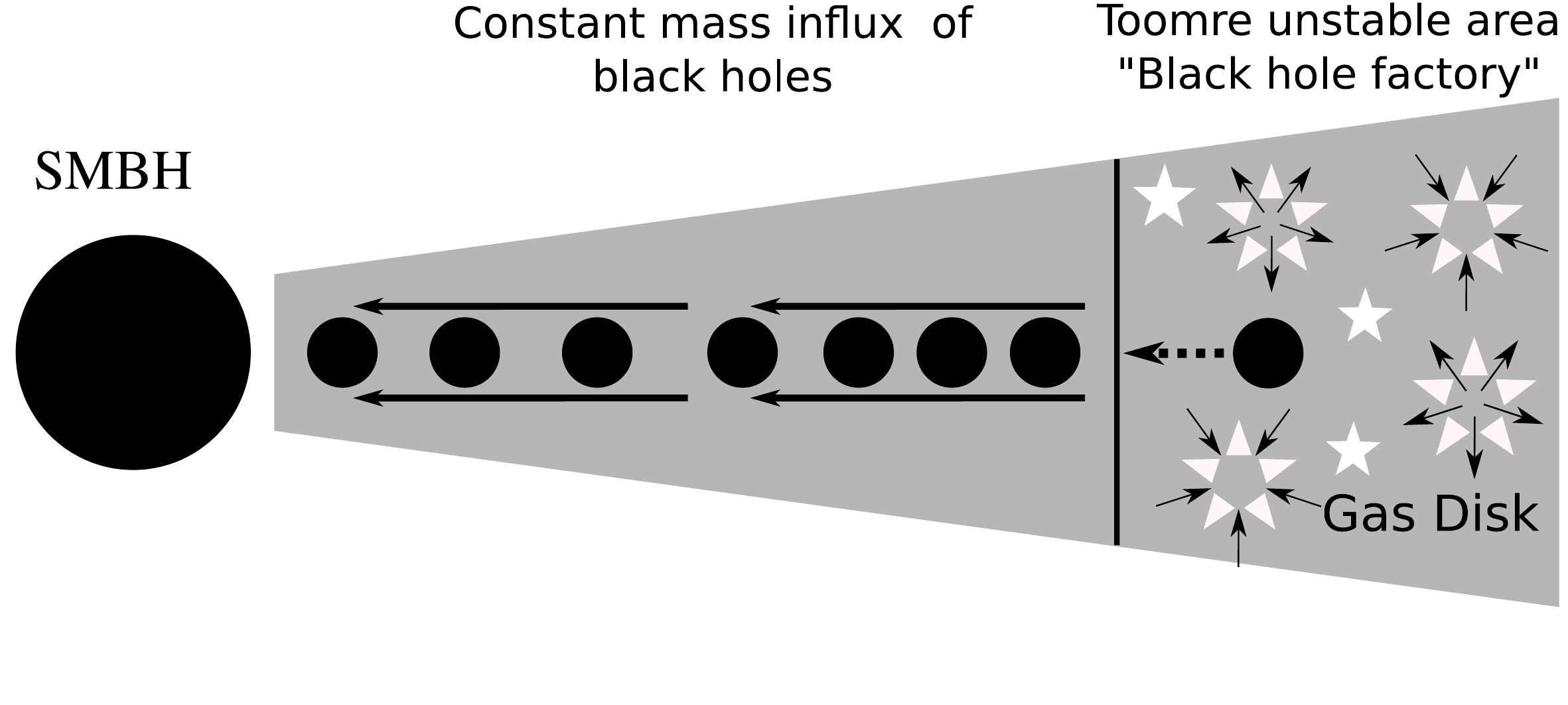}
 
 }
 
 \subfloat[Combined pileup and CMI \label{fig:combined schematic}]{
 \centering
 \includegraphics[width=\linewidth]{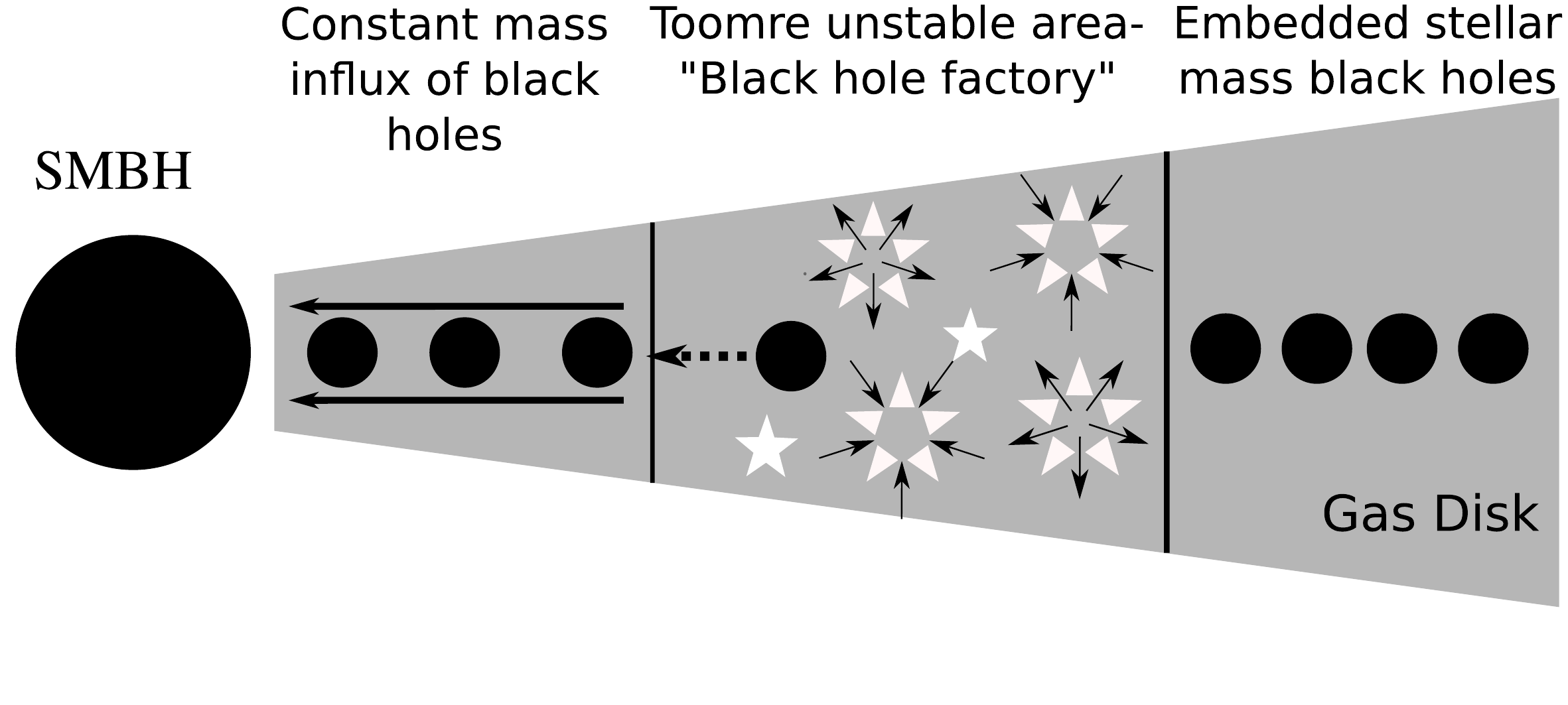}
 
 } 
 \caption{Schematics for the different regimes in our model, in which the SMBH is located on the left. In the top panel \textbf{(a)}, we present the \textbf{pileup} regime, where at small radii the SMBH is surrounded by a gas disk in the Shakura--Sunyaev regime, but at $R>R_{\rm Q}$, BHs stabilize the disk and do not migrate. In the middle panel \textbf{(b)} we present the \textbf{CMI} regime. The disk is Toomre unstable at larger radii and therefore is fragmenting and forming stars (inward arrow stars) which will eventually explode (outward arrow stars) and form new BHs that migrate inwards with a constant mass flux. And in the bottom panel \textbf{(c)}, we present a more realistic combination of the two regimes based on our results in \sref{sec: Disk structure}.}
 \label{fig: schematics}
\end{figure}

The simplest type of steady-state solution is one where feedback is local: migratory torques are too weak to operate (i.e. $t_{\rm mig} > t_{\rm AGN}$), and ECOs reside at the same locations where they formed (or were captured into the disk). For brevity, we call this zero-migration, gaseous steady-state limit the ``pileup regime.'' This is sketched out in \cref{fig:Pileup schematic}.
We expect the pileup regime to emerge in the outer regions of AGN disks after the first few Myr of an AGN lifetime, once gas fragmentation in the Toomre instability zone ($R>R_{\rm Q}$ where $R_{\rm Q}$ is the initial Toomre instability point for a Shakura--Sunyaev disk) of an ECO-less disk forms new stars, which will collapse into compact objects that will then pile up until the BHs produce enough feedback from accretion to make the disk marginally Toomre stable. 
However, achieving $Q_{\rm T}=1$ does not necessarily mean the end of star formation. If in this case the azimuthal heat-mixing parameter is not sufficient for the compact objects to actually heat the disk $\left( \mathfrak{M}_{\varphi}<1 \right)$, then the number of compact objects at a given radius will continue growing until they actually heat the disk in all azimuthal sectors. This will ``overshoot'' the Toomre stability parameter, i.e. star formation only truly shuts off when the azimuthally averaged $Q_{\rm{T}} \gg 1$. 

We note that this picture could be complicated by alterations to stellar evolution in the dense environments of AGN disks.  In particular, recent work has indicated that in cold, high-density regions of these disks, rapid accretion onto embedded main-sequence stars can extend their lives \citep{Cantiello+21, Dittmann+21}, perhaps by sufficiently long periods of time to prevent {\it in situ} BH formation during the AGN lifetime.  The details of this ``runaway growth'' regime depend on the competition between (reduced) Bondi--Hoyle--Lyttleton accretion and the highly uncertain mass-loss rates of stars radiating near the Eddington limit, so we do not attempt to model its impact here, although this is an important topic for future investigation.  A second uncertainty, as previously mentioned, is that {\it in situ} formation is not the only way to embed compact objects into AGN disks; gas drag may grind down the orbits of preexisting BHs in the nuclear star cluster, which we investigate briefly in \S \ref{sec: Disk structure}.  In general, the steady-state pileup regime we investigate here is agnostic as to the origins of the ECO population that sustains it.

Because we assume no migration in this regime, \cref{eq: steady BH transport} is replaced by
\begin{equation} \label{eq: pileup equation}
\begin{cases}
 Q_{\rm{T}} =1 & \mathfrak{M}_{\varphi}>1\\
 \mathfrak{M}_{\varphi}=1 & \text{otherwise}.
\end{cases} 
\end{equation}
We expect this regime to be relevant only at larger radii, where (i) models without feedback are Toomre-unstable, and (ii) migration times are the longest. 

\subsubsection{Constant Mass Influx Regime} \label{section: Constant Mass Flux Inflow }

A more complicated kind of two-fluid steady-state exists when migration times $t_{\rm mig} < t_{\rm AGN}$.
In analogy to the steady state gas inflow rate $\dot{M}_{\rm g}$, we introduce the BH mass inflow rate $\dot{M}_{\bullet}$, and set this equal to a constant across a wide range of radii. This is illustrated in cartoon form in \cref{fig:CMI schematic}.
The mass inflow rate can be rewritten as:
\begin{equation}
 \dot{M}_{\bullet}=\dot{N}_\bullet \langle m_\bullet\rangle=-2 \pi R S_{\bullet} \langle m_{\bullet}\rangle \frac{da}{dt},
 \label{eq: General black hole flow rate}
\end{equation}
where $\langle m_{\bullet} \rangle$ is the average mass of embedded BHs (hereafter $m_\bullet$) and $\frac{da}{dt}$ is the total BH migration rate, given by \cref{eq: GW migration rate,eq: type I migration rate}. \cref{eq: General black hole flow rate} is thus the ninth equation of the Constant mass influx (CMI) regime, closing the system.

We expect the CMI regime to emerge at small radii, where migration times are relatively short. Note that the existence of a small-$R$ CMI zone implies that migrating BHs are continually being replenished from larger radii, most likely an intermediate zone between the CMI and pileup regimes. The properties of this intermediate zone likely require a solution of the full, time-dependent two-fluid (or three-fluid) equations to be understood, as the substantial time lags between star formation and the onset of ECO feedback make limit cycles likely. Furthermore, as we will see in later sections, there is never a direct, self-consistent transition between the inner CMI regime and the outer pileup regime (i.e. there is generally an order of magnitude in radius $R$ that is not self-consistently described by either regime for any combination of parameters). A cartoon illustrating this more general case is shown in \cref{fig:combined schematic}. 

\subsection{\newp{Parameter Choices}} \label{sec: numercial method}
\begin{figure} 
 \centering
 \includegraphics[width=1 \linewidth]{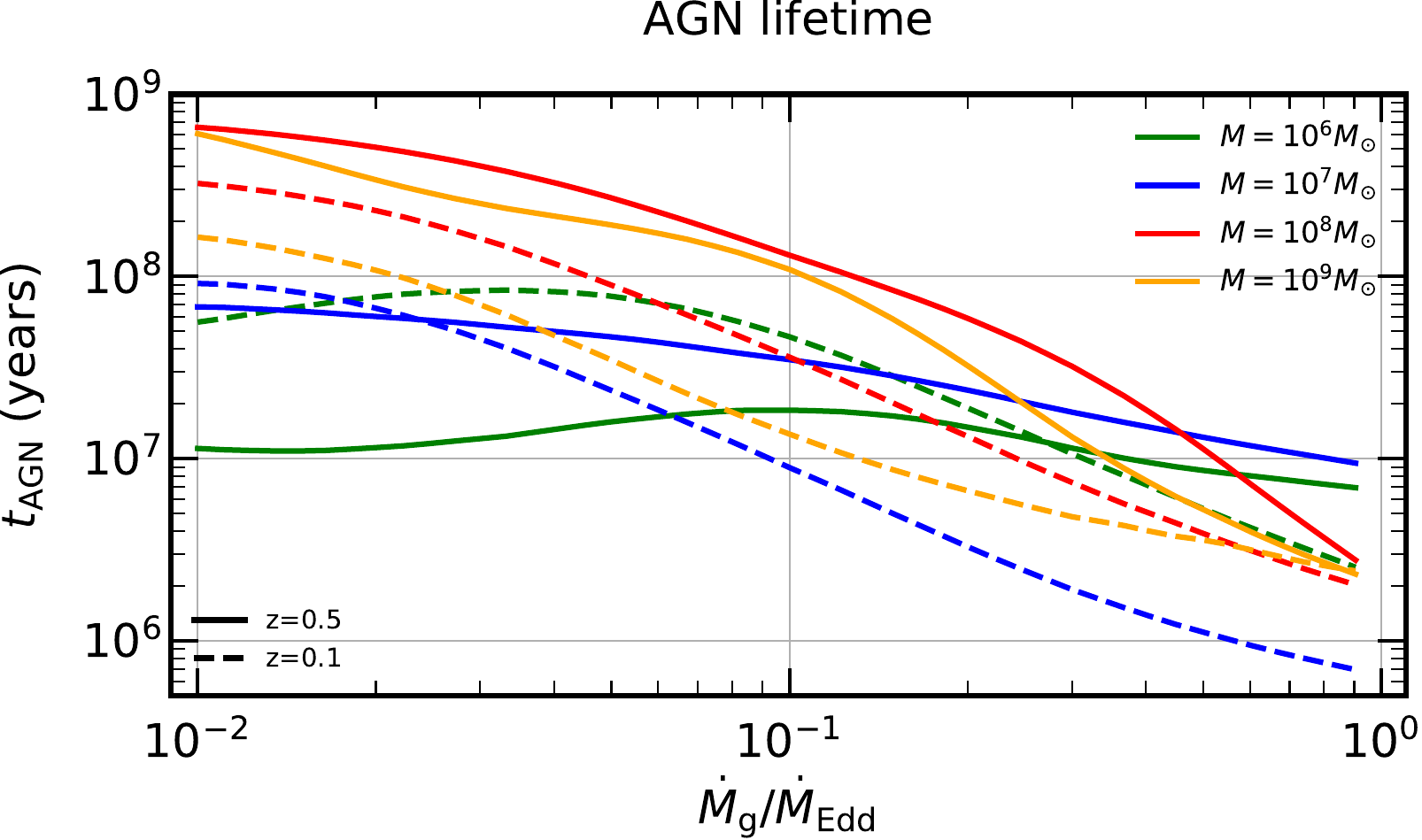}
 \caption{Effective AGN lifetime $t_{\rm AGN}$ as a function of dimensionless accretion rate $\dot{M}_{\rm g}/\dot{M}_{\rm Edd}$, with four different SMBH masses (colors listed in the panel) and two different redshifts (solid lines are $z=0.5$, dashed lines are $z=0.1$). At the very highest accretion rates ($\dot{M}_{\rm g} \sim \dot{M}_{\rm Edd}$), the AGN lifetimes are usually too short for a first generation of star formation to produce ECOs, but at more typical accretion rates ($10^{-2} \lesssim \dot{M}_{\rm g}/\dot{M}_{\rm Edd} \lesssim 10^{-1}$) AGN can last long enough for a large population of ECOs to be produced by Toomre instability.} 
 \label{fig:AGN_lifetimes}
\end{figure}

For both steady-state scenarios, we algebraically reduce Eqs. \ref{eq: 8 SS equation} (see \aref{app: reduced equations }
) in combination with either \cref{eq: pileup equation} or \cref{eq: General black hole flow rate}. In both cases, we reduce the problem to two nonlinear equations with two variables. The solutions are dependent on a few free parameters: the mass of the SMBH ($M$), the dimensionless viscosity ($\alpha$), the accretion rate of the gas ($\dot{M}_{\rm g}$), and for the CMI regime the BH mass inflow rate ($\dot{M}_{\bullet}$). These determine the other parameters we use: 
\paragraph{AGN lifetime} 
Combining empirical luminosity data from \citet{Aird+2018} and the AGN formalism from \citet{HopkinsHernquist2009} in the ``light-bulb'' limit, we compute effective AGN lifetimes as a function of (i) redshift - $z$, (ii) mass of SMBH,\footnote{The data in \citet{Aird+2018} gives luminosity distributions as a function of stellar mass rather than SMBH mass $M$, but we use scaling relations to convert total stellar mass to bulge mass \citep{Stone+18, vanVelzen2018} and from there to SMBH mass \citet{KormendyHo2013}.} and (iii) Eddington ratio. Example effective AGN lifetimes are shown at different masses, Eddington ratios, and redshifts in \cref{fig:AGN_lifetimes}. In general, most AGNs live long enough for the first generation of star formation to produce ECOs \newp{(i.e. long enough for the most massive stars, with main-sequence lifetimes of $\sim3-10$ Myr, to evolve and produce stellar-mass BHs)}, although this is not always the case at the highest Eddington ratios \newp{and at the lowest redshifts.  Even in this limit, however, ECOs may be captured via gas drag from the preexisting star cluster over shorter timescales, a process we briefly explore in \S \ref{sec: Disk structure}}.  We caution, however, that the effective AGN lifetime here represents a rough upper limit on the life of an individual AGN episode (as it is possible that the duty cycle for AGNs of certain characteristics may be broken up into individual accretion episodes of shorter duration).
\paragraph{Compact object mass}
We start with a $dN_{\star}/dM_{\star}\propto M_\star^{-1.7}$ IMF, choosing this top-heavy power law based on observations of the young stellar disk in the center of the Milky Way \citep{Lu+2013}. We remove all the stars that do not end their life before the AGN lifetime by using the MIST project evolutionary tracks \citep{,Paxton+2011,Paxton+2013,Paxton+2015,Paxton+2018,Choi+2016,Dotter2016}. Using this truncated mass distribution to identify stars that will turn into com-act objects before the end of the AGN, together with the tabulated initial--final mass relationships from \citet{SperaMapelli2017}, we use Monte Carlo sampling to calculate the mean BH mass for each specific $\{M, \dot{M}_{\rm g}, z \}$ combination. We also considered the accretion feedback from neutron stars using a constant mass of $1.4 M_{\odot}$ and a neutron star--BH ratio calculated from the distributions mentioned above but found that neutron star feedback contributes negligibly to our disk solution (usually $\lesssim 5 \% $).
\paragraph{Boundaries of the disk}
 We set our inner boundary to be at the innermost stable circular orbit (ISCO) \citep{Bardeen+72} of a Throne limit ($\chi=0.998$) Kerr SMBH: $R_0=R_{\rm{ISCO}}\approx 1.23 R_{\rm g}(M)$. Our results are insensitive to this choice, as we are mostly interested in the larger radii where the effect of the inner boundary is negligible. The outer radius where we truncate our solutions\footnote{In principle, our solutions could be straightforwardly extended to larger radii by including the non-Keplerian aspects of the galactic potential. However, the empirical diversity \citep{Lauer+05, GeorgievBoker14} of galactic nuclear potentials would add many additional free parameters to the model, so in this paper we limit ourselves to the radii inside the SMBH influence radius.} is defined by calculating the radius of influence of the SMBH, where the galactic velocity dispersion $\sigma_{\rm{v}}$ equals the Keplerian velocity for circular orbits:
\begin{equation} \label{eq: radius of influence}
 R_{\rm{infl}}=\frac{GM}{\sigma_{\rm{v}}^2}.
\end{equation}
For each solution, we compute the galactic velocity dispersion using the $M-\sigma_{\rm v}$ relation from \citet{KormendyHo2013}.
It is important to note that, because we are considering steady-state limits, our solutions are local.


\section{Disk Structure}
\label{sec: Disc structure}
 In this section we survey the parameter space for numerical solutions in the two steady-state regimes we expect to exist: the zero-migration pileup regime, relevant for large radii, and the CMI regime, relevant for small radii.
 
 \begin{figure} 
 \includegraphics[width=\linewidth]{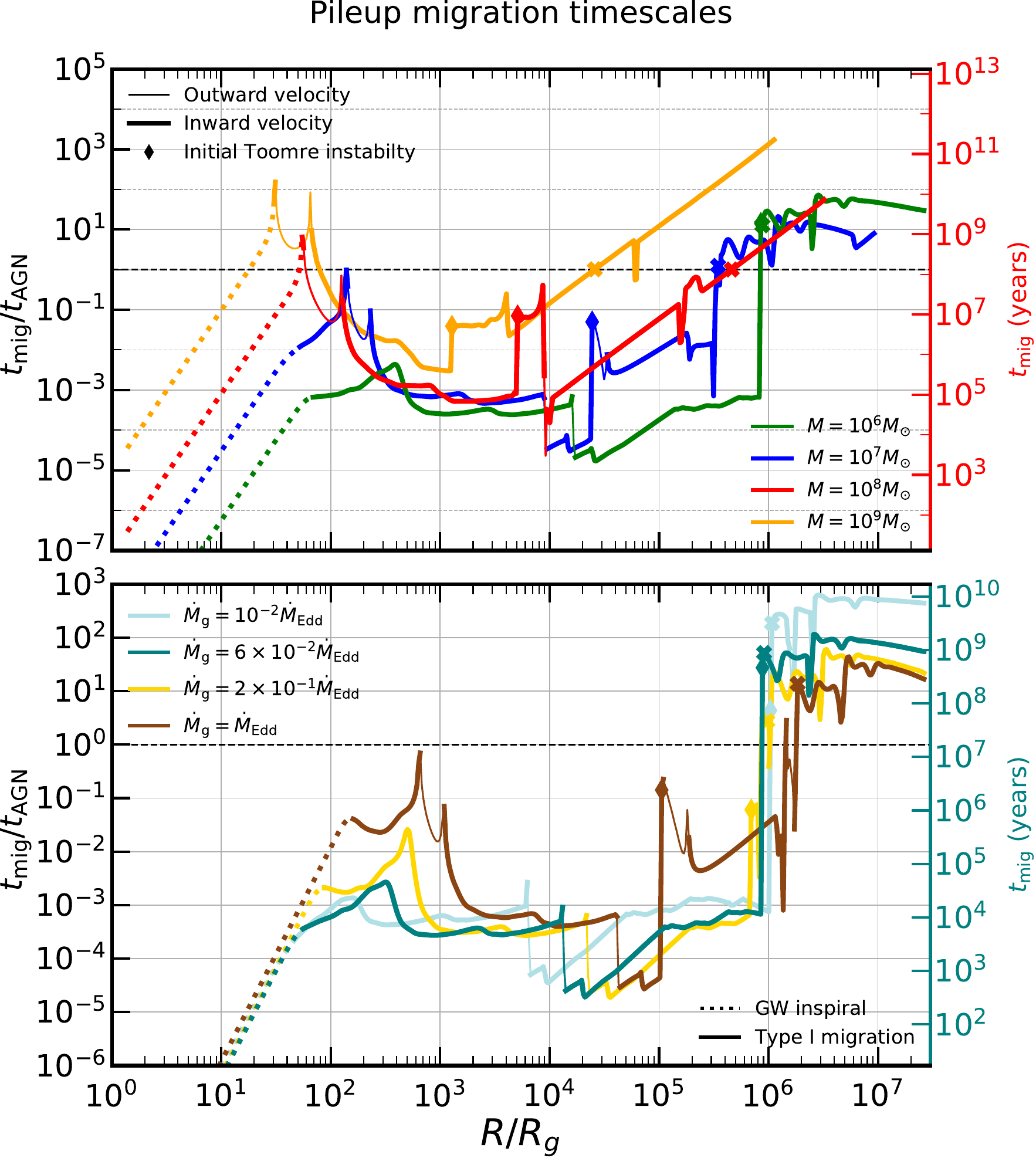}
 \caption{ECO migration timescales $t_{\rm mig} = R / (da/dt)$, normalized by AGN lifetimes $t_{\rm AGN}$ at redshift $z=0.5$, plotted against dimensionless radius $R/R_{\rm g}$. The {\it top panel} shows different SMBH masses (color-coded in the panel) at $\dot{M}_{\rm g} = 10^{-1}\dot{M}_{\rm Edd}$.  Its secondary (red) y-axis shows $t_{\rm mig}$ in physical units for the case of a $10^8 M_\odot$ SMBH. The {\it bottom panel} shows different accretion rates (color-coded in the panel) for a $10^6 M_{\odot}$ SMBH.  Its second (teal) y-axis shows $t_{\rm mig}$ in physical units for the case of $6 \times 10^{-2} \dot{M}_{\rm Edd}$ accretion rate. The various line styles indicate different dominating migration torques. Thick (thin) lines indicate inward (outward) migration. In general, the pileup (i.e. zero-migration) regime is self-consistent at large radii only. The innermost radius of validity for the pileup regime $R_{\rm pile}$ is marked by an ``x'' symbol on each curve, which ranges from $\sim 10^4 - 10^7 R_{\rm g}$, depending weakly on gas accretion rate $\dot{M}_{\rm g}$ and more strongly on SMBH mass $M$. } \label{fig: pileup t_mig}
\end{figure}
 In general, the models are not fully self-consistent across all parameter space. There is most notably an intermediate zone \newp{(see the ``factory'' zone in \cref{fig:combined schematic} and discussion in  \sref{section: Constant Mass Flux Inflow })} which cannot be in steady state \newp{given a constant gas accretion rate, as the two limiting steady-state solutions are internally inconsistent (applying the pileup regime to these radii gives $t_{\rm mig} < t_{\rm AGN}$; applying CMI to these radii gives $Q_{\rm T} < 1$).  We discuss this intermediate zone at greater length in \S \ref{sec:intermediate}.}
 
 Some general conclusions that apply in all regions where the steady-state limits are self-consistent are the following:
 \begin{enumerate}
 \item Two-fluid accretion disks are optically thick except in the opacity gap, where they can be optically thin to their own photon field. Even in the opacity gap, however, the frequency-dependent optical depth for X-ray photons, $\tau_{\rm keV} \gg 1$, implying that accretion feedback from ECOs will effectively heat the disk. 
 \item The accretion rates onto embedded BHs are always super-Eddington at small radii, but for high-mass SMBHs, in the pileup regime in the outer radii of the disk, embedded BHs accretion rates can fall below Eddington. For example, for a $10^9 M_\odot$ SMBH, this happens at $R \sim 3\times 10^5 R_{\rm g}$. The accretion rates onto embedded main-sequence stars are mostly sub-Eddington in the self-consistent regions, justifying the neglect of accretion heating from the stellar population.
 \item ECOs almost always fail to open gaps in the gaseous accretion disk according to the gap-opening criterion of \cref{eq: gap equation} \citep{Crida+06}. Gaps never open in the pileup regime characteristic of large radii. For the smallest SMBH masses ($M_\bullet \lesssim 10^6 M_\odot$) and lowest accretion rates ($\dot{M}_{\rm g}/\dot{M}_{\rm Edd} \lesssim 0.1$), gaps can open in the constant mass flux solutions in the opacity gap (see \aref{app: additional reults}). This result appears to be in tension with the implementation of the \citet{Thompson+2005} model in \citet{Tagawa+20b}, where a gap-opening on scales of $R\sim 0.01$ pc is found. This difference originates from the different gas-disk structures in our work and that of \citet{Thompson+2005}, likely due to either (i) the choice of local versus non-local effective viscosity, or (ii) the additional role of feedback in the disk energy equation.
 \end{enumerate}

\subsection{``Pileup'' solution} \label{sec: pileup solution}

\begin{figure} 
 \includegraphics[width=\linewidth]{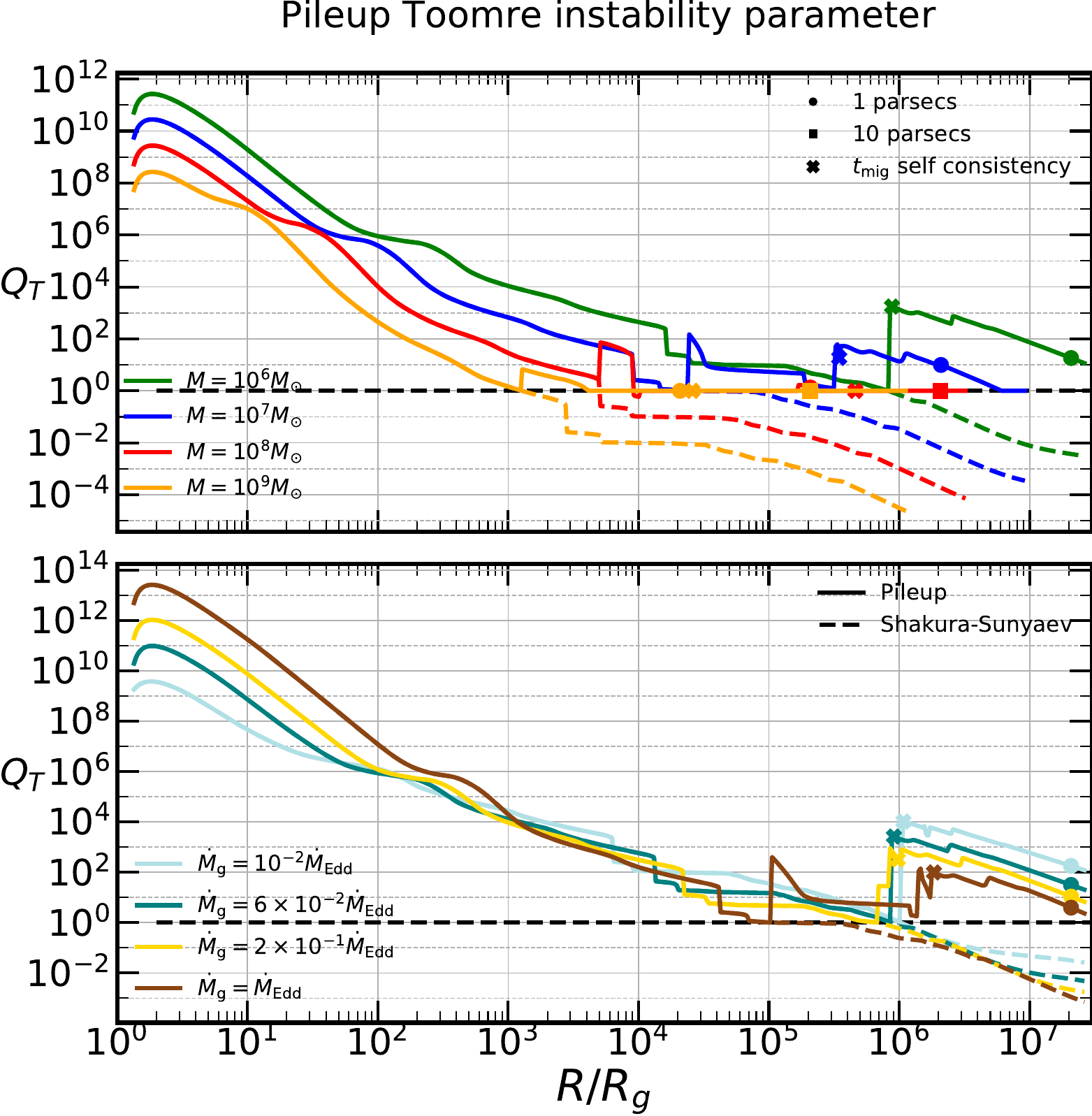}
 \caption{Toomre instability parameter $Q_{\rm{T}}$ as a function of dimensionless radius $R/R_{\rm g}$. Color/panel layouts are the same as in \cref{fig: pileup t_mig} (top panel is fixed Eddington ratio and different SMBH masses, bottom panel fixed SMBH mass and different Eddington ratios, only self-consistent at large radii) Solid lines show the pileup solutions , while the dashed lines show a Shakura--Sunyaev-type solution. Notably, the pileup regime often achieves a steady-state $Q_{\rm T} \gg 1$ because of the need to evenly mix heat throughout the disk. An azimuthally averaged $Q_{\rm T}=1$ equilibrium would suffer from severe discreteness effects, i.e. large azimuthal sectors of the disk would be Toomre unstable for times longer than the local thermal time, leading to further star formation (see \sref{sec : Heat mixing}). Toomre fragmentation will only stop when the density of sources of feedback is high enough to evenly mix their thermal energy across all azimuthal angles.} \label{fig: pileup Q_T}
\end{figure} 

One important phenomenon to mention is that the thermal time $t_{\rm th}$ (defined in \sref{sec : Heat mixing}) is much shorter than the dynamical time in regions where heating is dominated by feedback from ECOs, similar to the results in \citet{Thompson+2005}. In general, the cold branch of the pileup regime always satisfies the following timescale hierarchy: $t_{\rm diff} < t_{\rm th} \ll t_{\rm dyn} \ll t_{\rm visc}$ (here we define the photon diffusion time $t_{\rm diff} = \tau H / c$ and the dynamical time $t_{\rm dyn}= \Omega^{-1}$). The cold branch is the only one that is consistently thermally stable to linear perturbations \citep{Piran78} throughout all radii and parameter choices as is also seen in the disk models of \citet{Thompson+2005}, which is an extremely important condition for short thermal timescales. The cold branch also requires far fewer ECOs than the hot branch solutions (which are sometimes stable in a narrow range of radii, but not for all of it). As a result, we believe that the cold branch is the realistic solution for the astrophysical AGN disks - even in the radial ranges where hot branches can be thermally stable, reaching them requires the accumulation of orders of magnitude more ECOs to a disk that is no longer star-forming.
 
The key assumption in the steady-state pileup limit is that migration timescales are longer than the AGN lifetime. We check the validity of this assumption in a post hoc way to define ranges of radii for which the pileup regime is self-consistent. As we see in \cref{fig: pileup t_mig}, ECOs fail to migrate ($t_{\rm mig}\equiv R/ \frac{da}{dt} > t_{\rm AGN}$) only at larger radii (generally, the zero-migration limit only applies to regions beyond the initial Shakura--Sunyaev Toomre instability point $R>R_{\rm Q}$). The radius where the pileup regime begins, $R_{\rm pile}$, is a strong function of $M$.  $R_{\rm pile}$ can depend strongly on $\dot{M}_{\rm gas}$, but becomes a weak function of $\dot{M}_{\rm gas}$ for small SMBHs. 
Type I migration is the dominant migration torque at large radii\footnote{In these calculations, we calculated $C_I$ using \cref{eq: C_I equation}, after getting the results for all the relevant disk variables.}. From this point onwards, most of our pileup analysis will be on the larger, self-consistent radii.

\cref{fig: pileup Q_T} shows that the Toomre stability parameter can be much greater than the marginally stable $Q_{\rm T}=1$ limit. As discussed earlier, this arises because of discreteness in the ECO surface number profile. Within the self-consistent radii of the pileup regime, enforcement of the heat-mixing condition $\mathfrak{M}_\phi \ge 1$ often results in $Q_{\rm T} \gg 1$ for the lower SMBH masses (see \cref{eq: Q-m_phi relation}; e.g. $M= 10^6 M_\odot$ and $10^7M_\odot$ in \cref{fig: pileup Q_T}). At higher SMBH masses, discreteness effects are less important and the pileup regime is usually in the $Q_{\rm T}=1$ limit. These results show the importance of considering heat-mixing effects in models of feedback-dominated accretion flows: AGN disks surrounding smaller SMBHs will possess large azimuthal sectors that receive no heat over a thermal time even if their azimuthally averaged properties satisfy $Q_{\rm T}=1$.
\begin{figure*}
\centering
 \subfloat[\label{fig: pileup Tc}]{
 \centering
 \includegraphics[width=0.45\linewidth]{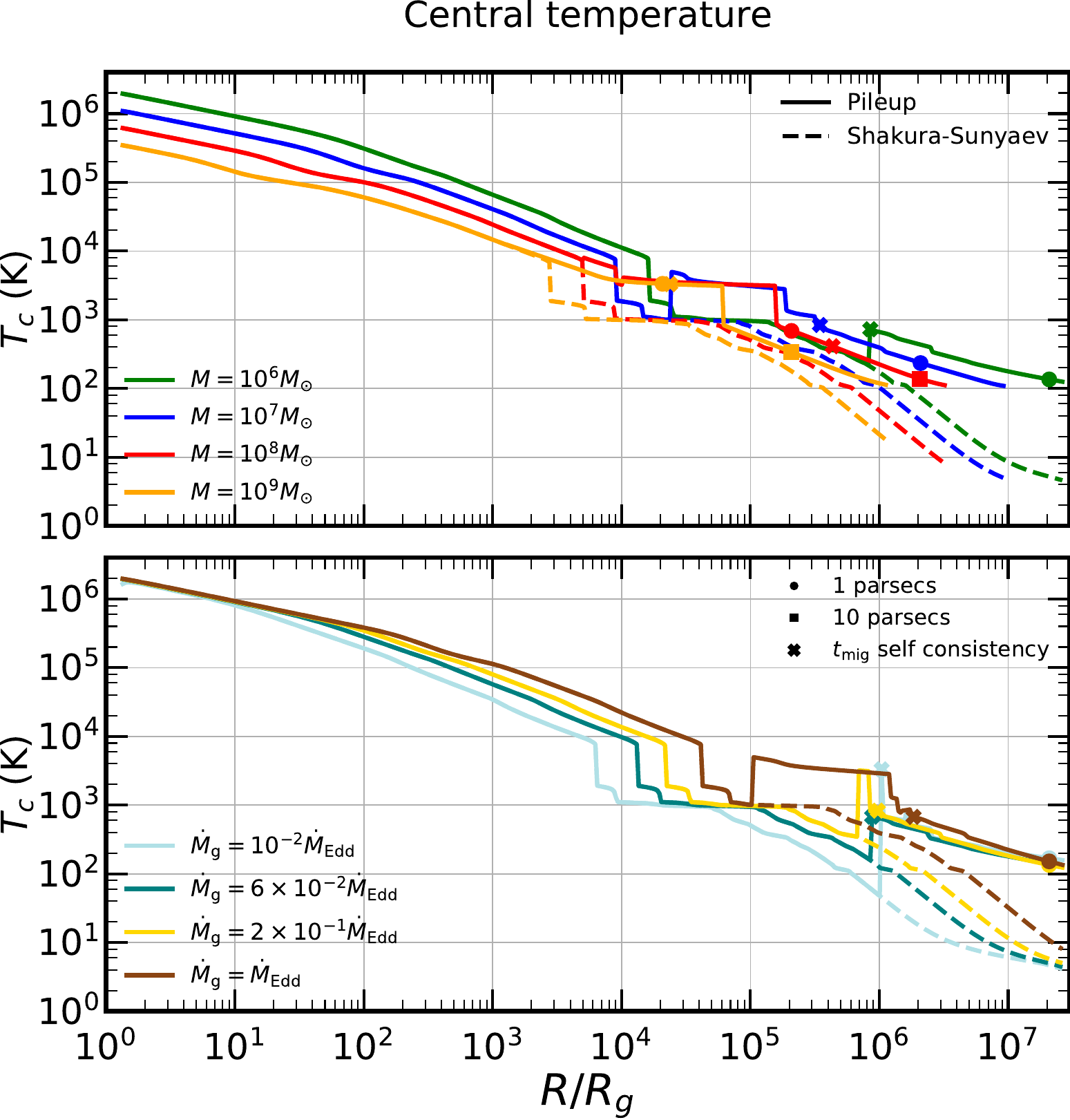}
 
 }
 \subfloat[\label{fig: pileup_hor}]{
 \centering
 \includegraphics[width=0.46\linewidth]{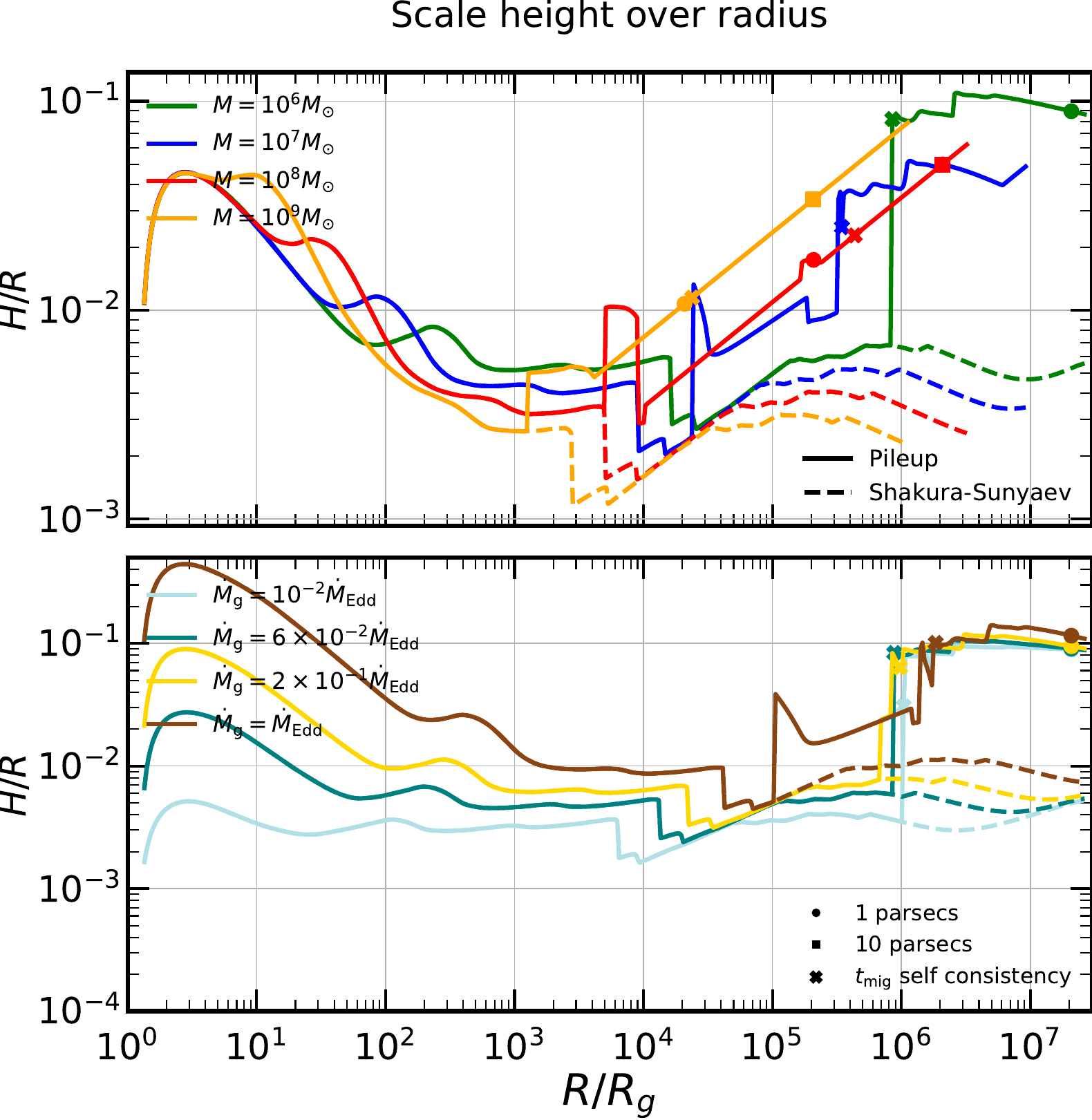}
 
 } 
 
 \subfloat[\label{subfig: pileup rho}]{
 \centering
 \includegraphics[width=0.465\linewidth]{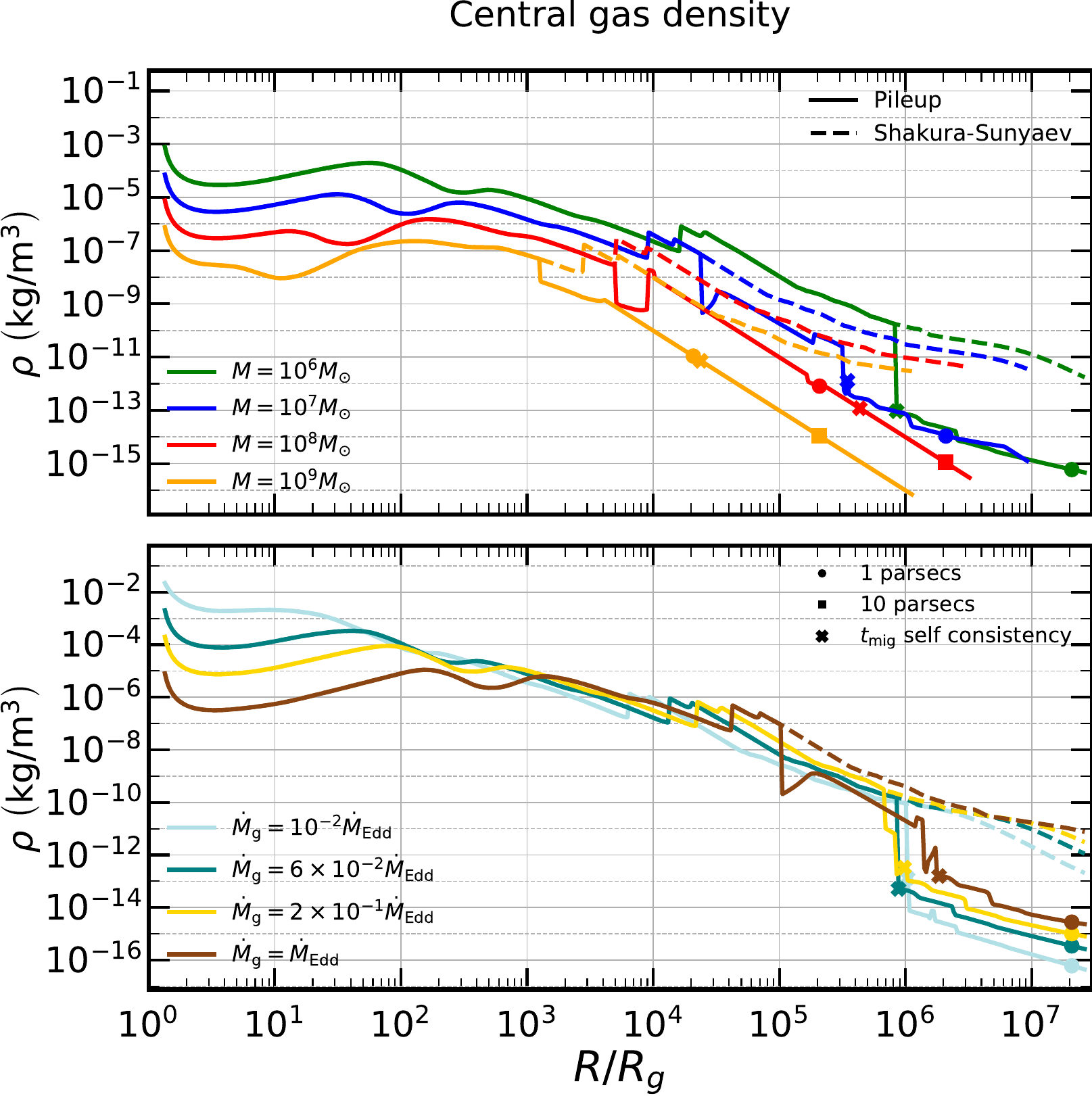}
 
 }
 \subfloat[\label{fig: pileup Sigma}]{
 \centering
 \includegraphics[width=0.45\linewidth]{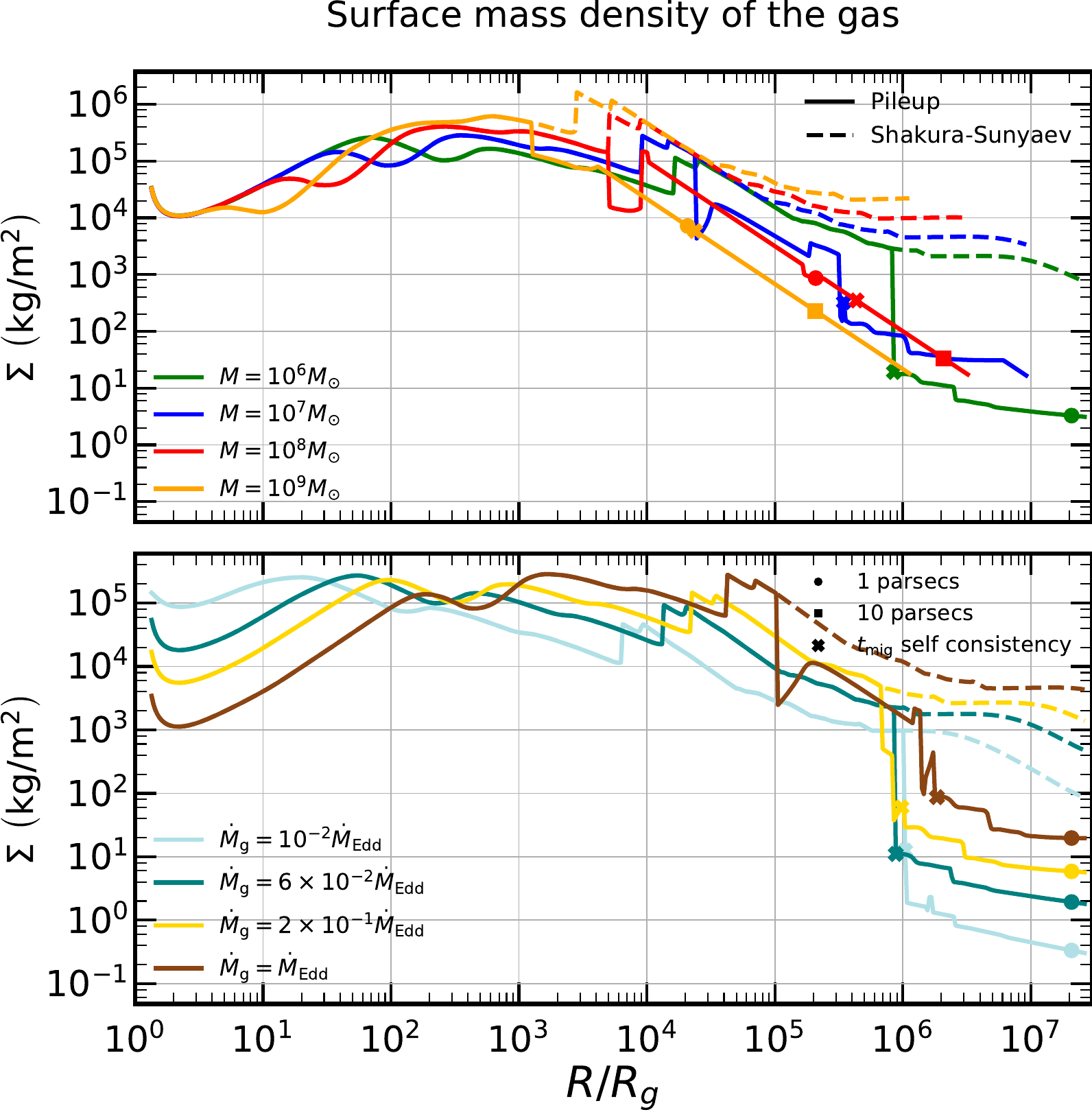}
 
 }

\caption{Radial profiles of different gas-disk variables: central temperature $T_{\rm c}$ (\textit{top left}), aspect ratio $H/R$ (\textit{top right}), 3D density $\rho$ (\textit{bottom left}) and surface density $\Sigma$ (\textit{bottom right}). All these variables are plotted against dimensionless radius $R/R_{\rm g}$ for both the pileup solution (solid) and the Shakura--Sunyaev model (dashed). \newp{We see that in general, at the outer radii, the pileups solution is hotter, less dense with a larger scale height.} The pileup solution's self-consistency radius $R_{\rm pile}$ is marked by an ``x'' symbol on each curve. In each column, the top subpanels shows results for different masses, for $\dot{M}_{\rm g}=0.1\dot{M}_{\rm Edd}$. The bottom subpanels show results for four different accretion rates, for $M=10^6 M_\odot$.} \label{fig: pileup profile}
\end{figure*}

\begin{figure}

 \centering
 \includegraphics[width=\linewidth]{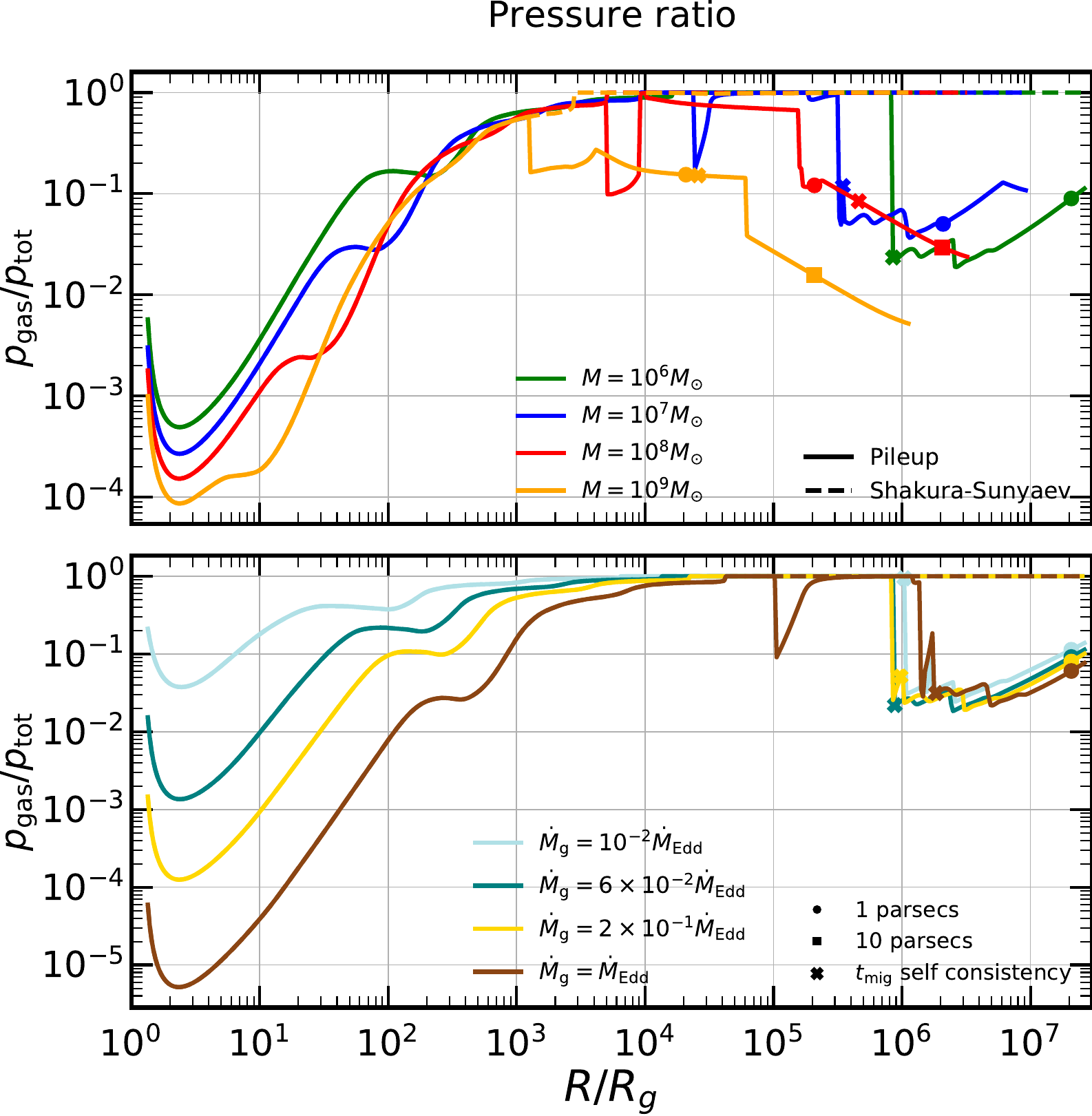}
 
\caption{Gas to total pressure ratio plotted against dimensionless radius $R/R_{\rm g}$ for both the pileup solution (solid) and the Shakura--Sunyaev model (dashed). We see that unlike Shakura--Sunyaev the outer radii are dominated by radiation pressure. The pileup solution's self-consistency radius $R_{\rm pile}$ is marked by an ``x'' symbol on each curve. In each panel, the top subpanel shows results for different masses, for $\dot{M}_{\rm g}=0.1\dot{M}_{\rm Edd}$. The bottom subpanel show results for four different accretion rates, for $M=10^6 M_\odot$. }\label{fig: pileup p_ratio}
\end{figure}
The pileup (i.e. zero-migration) steady-state solutions, which are valid at large radii (usually $R\gtrsim 10^4R_{\rm g}$), result in up to three branches of solutions for the marginally Toomre stable limit $Q_{\rm T}=1$ (this multibranch behavior is also seen in the \citealt{Thompson+2005} and \citealt{SirkoGoodman2003} models). This expands to up to five branches of solutions in the full pileup regime, where $Q_{\rm T}$ may be much greater than unity due to heat-mixing considerations (the exact number of possible branches depends on the mass and accretion rate of the SMBH; for example, for $M=10^6M_\odot$ and $\dot{M}_{\rm g}=0.1\dot{M}_{\rm Edd}$, there are only three possible solutions). In general there is one ``cold'' branch (that reaches a few $\times 10^2 \rm K$ at the radius of influence) and any others are much hotter (with minimum temperatures of a few $\times 10^3 \rm K$). 
 
\begin{figure}
\centering
 \includegraphics[width=\linewidth]{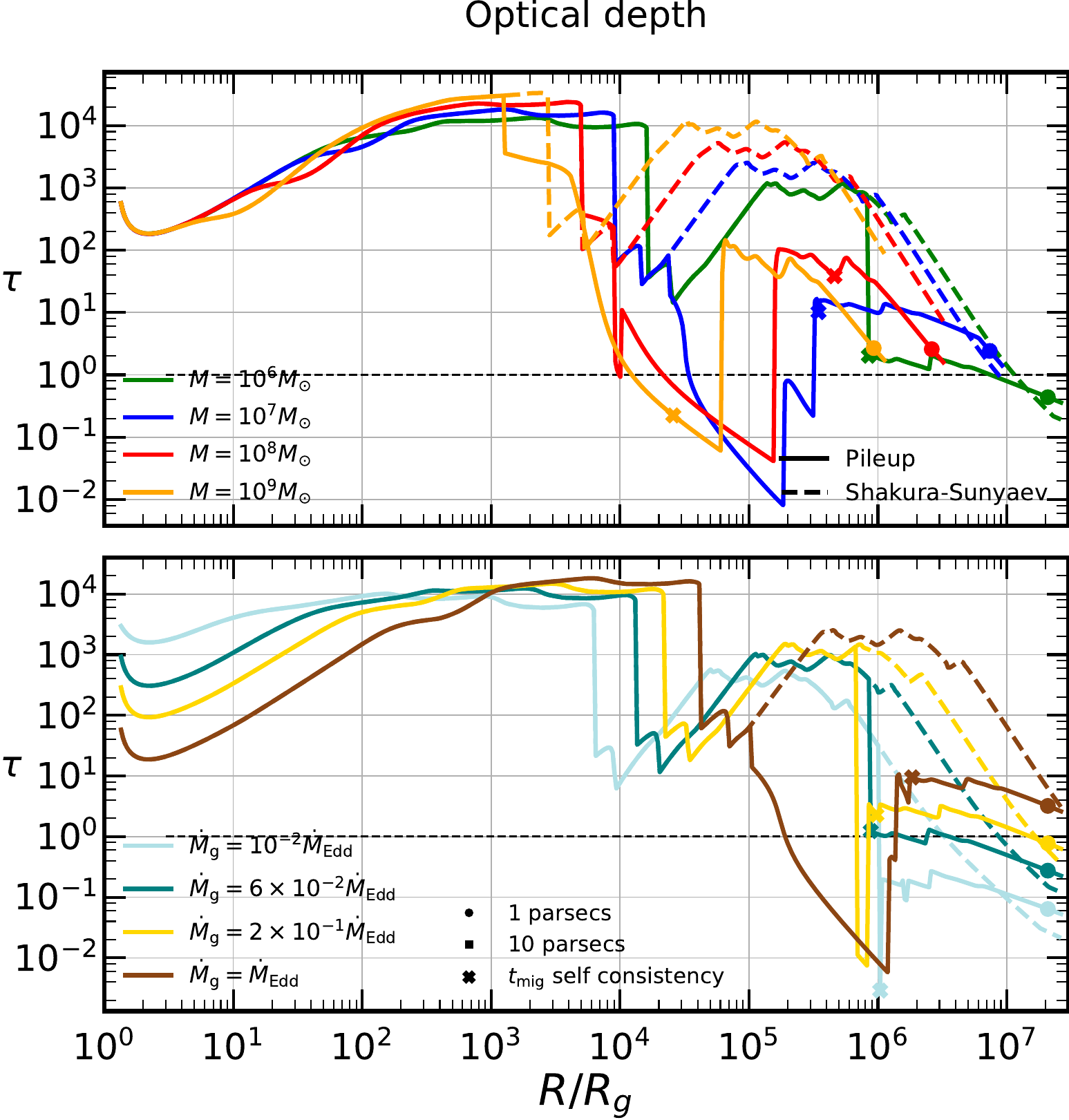}
\caption{Rosseland mean optical depth  plotted against dimensionless radius $R/R_{\rm g}$ for both the pileup solution (solid) and the Shakura--Sunyaev model (dashed). We see that in the pileup solution the disk is optically thinner in the outer radii.  The pileup solution's self-consistency radius $R_{\rm pile}$ is marked by an ``x'' symbol on each curve. In each panel, the top subpanel shows results for different masses, for $\dot{M}_{\rm g}=0.1\dot{M}_{\rm Edd}$. The bottom subpanel show results for four different accretion rates, for $M=10^6 M_\odot$. }\label{fig: pileup tau}
\end{figure}
\newp{In \cref{fig: pileup profile}, we show the radial profiles of different gas-disk variables in comparison with a simple Shakura--Sunyaev-type profile.  The general result is that, in the self-consistent regions of the pileup solution, is that  the pileup solution is hotter (higher $T_{\rm c}$), thicker (larger $H/R$), and has lower density $\rho$ (also $\Sigma$) than in an otherwise equivalent Shakura--Sunyaev-type solution. As is the case with Shakura--Sunyaev-type disks, larger $\dot{M}_{\rm g}/\dot{M}_{\rm Edd}$ ratios also increase $T_{\rm c}$, $H/R$, $\rho$ and $\Sigma$. Across all SMBH masses and accretion rates, we find that the pileup solutions remain in the thin-disk regime, with $H/R \lesssim 0.1$. }

In \cref{fig: pileup p_ratio} we  plot the ratio between gas pressure and total pressure. Similarly to \citet{SirkoGoodman2003,Thompson+2005}, we see that  in the self-consistent p ileup region, radiation pressure is the dominant form of pressure. In \cref{fig: pileup tau} we show the Rosseland mean optical depth profile, which is generally $>1$ in the pileup regime, but can become optically thin for small radii in high-$M$ systems (i.e. the opacity gap) and in the outer radii for low-$M$ systems, particularly with low accretion rates. As discussed in \S \ref{sec:microphysics}, even these regions (which are optically thin to the disk's thermal photon field) will be optically thick to the $\sim 0.1-1$ keV photons emitted by ECOs.

\begin{figure} 
 \includegraphics[width=\linewidth]{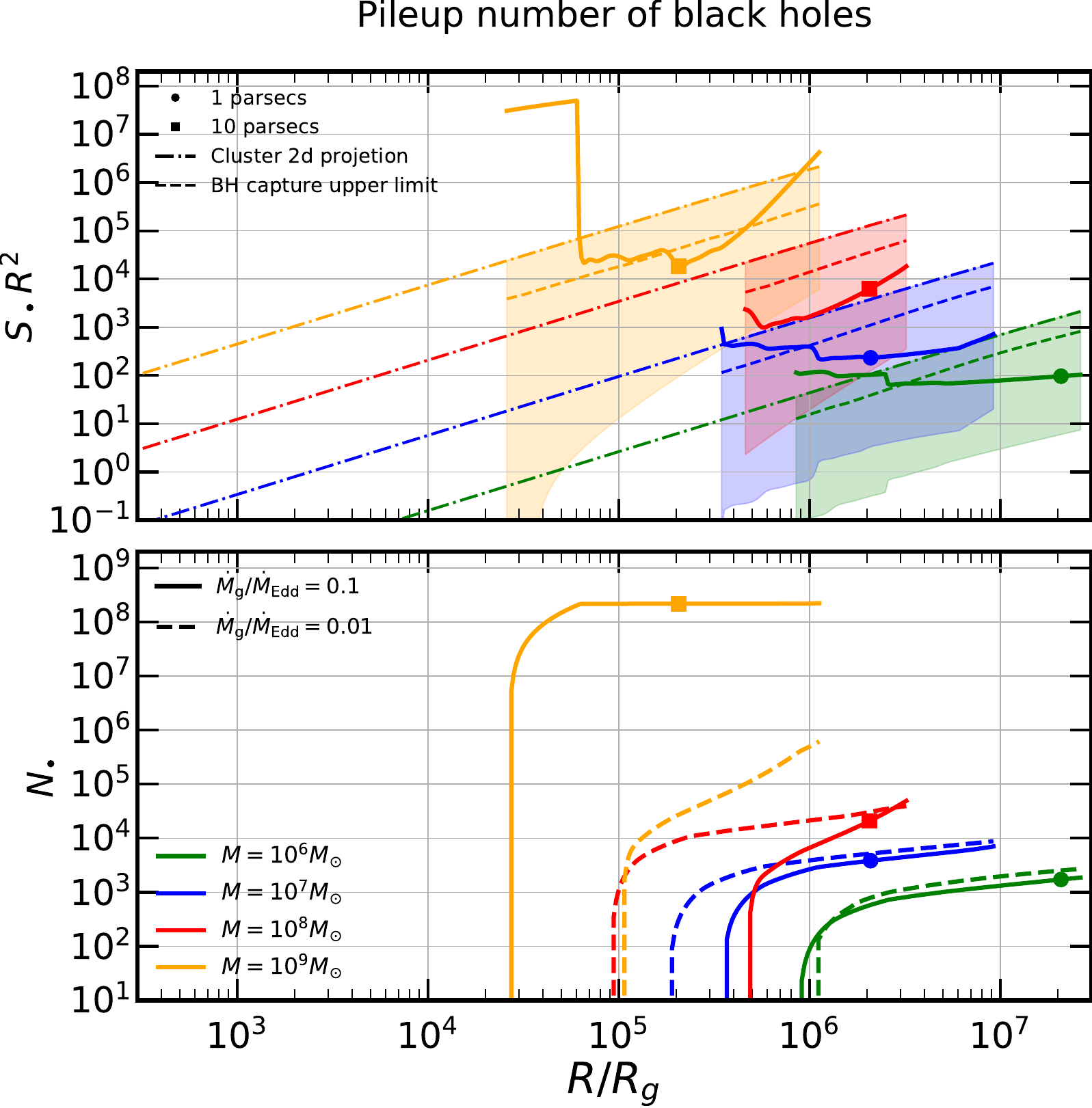}
 \caption{In the {\it top panel} we see the surface number density multiplied by $R^2$ within the self-consistent areas of the pileup regime for $\dot{M}_{\rm g}/\dot{M}_{\rm Edd}=0.1$ \newp{(solid lines)}. The dotted-dashed line shows the projected number density of BHs within a preexisting spherical nuclear stellar cluster (i.e the maximum number of BHs that the disk could absorb from such a cluster). The lower limit of the shaded area is the surface number density of BHs in the preexisting cluster that coincidentally lie within the disk of current scale height $H$, and the dashed line represents the upper limit on the surface number density of BHs that are captured within the AGN lifetime, using the capture timescale of \citet{Bartos+17a} and a Shakura--Sunyaev disk model.  By choosing a Shakura--Sunyaev-type disk, we have maximized the plausible $\Sigma$ value, thereby obtaining a probable upper limit on the captured BH population.  We see that it is possible to reach the pileup steady state partially with BHs from the cluster instead of only BHs formed {\it in situ} in the disk if the  hydrodynamic alignment of orbits \citep{Syer+91} is efficient.
 In the {\it bottom panel} the cumulative number of embedded BHs, $N_\bullet$, is plotted against the  dimensionless radius $R/R_{\rm g}$, within the self-consistent areas of the pileup regime for $\dot{M}_{\rm g}/\dot{M}_{\rm Edd}=0.1$ (solid) and $\dot{M}_{\rm g}/\dot{M}_{\rm Edd}=0.01$ (dashed). The total mass of ECOs can become comparable to the SMBH mass for large Eddington fractions {\it and} large SMBHs.}
 \label{fig: pileup self consistent NBH}
\end{figure}

 In \cref{fig: pileup self consistent NBH} we plot the cumulative number of BHs, $N_\bullet(R)$, and also the surface number density $S_\bullet$; both are plotted only within self-consistent regions of the pileup regime (i.e. exterior to at $R_{\rm pile}$, where the migration timescale exceeds the AGN lifetime).
In the top panel of \cref{fig: pileup self consistent NBH}, we show the the 2D projected surface number density (using an Abel transformation; \citealt{Abel1826}) of a preexisting spherical nuclear star cluster assumed to follow the density profile from \citet{BahcallWolf1976}.    \footnote{The Bachall-Wolf 3D number density profile scales as $n \propto r^{-7/4}$, which we normalize by setting the mass enclosed at the influence radius to be $M$ (assuming the mean stellar mass is $0.3 M_\odot$) with the number fraction of stellar-mass BHs of $10^{-3}$, as is characteristic for the Salpeter IMF. Note that the BH number fraction would be higher if most nuclear stars formed from a top-heavy disk-mode IMF. This scenario is unlikely for low-mass galaxies but may be correct for higher-mass late-type galaxies \citep{Neumayer+20}.} We see that interesting numbers of ECOs could potentially be embedded in the disk not \newp{only} via star formation and evolution but \newp{also} via absorption of preexisting BHs, which could speed the process of reaching the pileup solution. 

More specifically, \cref{fig: pileup self consistent NBH} illustrates three general trends.  First, assuming a spherical nuclear star cluster, the number of preexisting BHs embedded {\it by chance} in the AGN disk (i.e. the fraction with orbits coplanar with the AGN disk to within an angle $\approx H/R$) {\it is always insufficient} to stabilize the disk against further star formation.  Second, if {\it all} the preexisting BHs could be aligned into the disk, the resulting population of ECOs would almost always stabilize the disk against further star formation\footnote{The only exception is for the highest masses and accretion rates - roughly, $M\gtrsim 10^9 M_\odot$ and $\dot{M}_{\rm g}/\dot{M}_{\rm Edd} \gtrsim 0.1$.} 
Third, we have calculated a rough upper limit on the number of preexisting BHs that will be torqued into alignment with the disk via hydrodynamic effects.
Specifically, we calculate the BH alignment time for circular, prograde orbits\footnote{The $\approx 50\%$ of retrograde orbits will have a far longer alignment timescale and are not considered in this calculation.  We also ignore the effects of eccentricity here.} as a function of the initial inclination angle $i$ as in \citet{Bartos+17a}:
\begin{equation}
\tau_{{\rm align}} \sim \frac{M^2}{4 R^2 \Omega m_\bullet \Sigma} \sin i\left(2-2\cos i\right)^{3/2}.
\end{equation}
We equate $\tau_{\rm align} = t_{\rm AGN}/2$ and solve this equation numerically to find the critical inclination angle $i$ above which alignment does not occur over most of the AGN lifetime.  The resulting populations of captured BHs are large enough to fully stabilize the disk against star formation in parts of parameter space (e.g. $M=10^8 M_\odot$, $\dot{M}_{\rm g}/\dot{M}_{\rm Edd} =0.1$), too small to have much effect in others, (e.g. $M=10^9 M_\odot$, $\dot{M}_{\rm g}/\dot{M}_{\rm Edd} =0.1$), but for most SMBH masses and accretion rates, the picture is more nuanced, with some radii being stabilized and others not.  However, we note that this result should be seen as a rough upper limit on the number of BHs that can be ground down into the AGN disk, because we have set the gas surface density $\Sigma$ equal to its Shakura--Sunyaev value, which is higher than alternative models where $\Sigma$ is decreased by more effective angular momentum transport (as is the case for the local effective viscosity in this work because of the larger aspect ratio, and also for non-local angular momentum transport mechanisms as in \citealt{Thompson+2005}).  

\newp{Simple calculations using Fig. \ref{fig: pileup self consistent NBH} provide {\it post hoc} justification for our assumption that ECO accretion rates must be capped at the Eddington limit, at least in a time-averaged sense.  As an experiment, we have removed the Eddington cap and computed the total accretion rate onto all ECOs represented by the solid lines in this figure; this results in the ECOs consuming $100 - 290\%$ of the total $\dot{M}_{\rm g}$ passing through the disk.  However, this calculation is done for $m_\bullet \approx 10 M_\odot$, and sustained super-Eddington mass growth would quickly increase the reduced-Bondi-Hoyle accretion rate to the point where the SMBH would be completely starved of gas, deactivating the AGN.}


Note that for the $M=10^9 M_{\odot}$ SMBH, $\dot{M}_{\rm g}/\dot{M}_{\rm Edd}=0.1$ solution, the cumulative mass of the embedded BHs\footnote{\newp{This inconsistency in our pileup solutions only emerges for the largest SMBH masses and Eddington ratios.  The picture is qualitatively unchanged if one also includes the mass of main-sequence embeds, as for the top-heavy mass function we use, these only dominate BHs in mass by a factor of a few.}  In these calculations we take the cumulative mass of embedded BHs to be $\sim 10 M_{\odot} \times N_{\bullet}$.} is higher than the SMBH mass, which breaks the approximation of a Keplerian potential. Although this result could be straightforwardly recalculated with a more general potential, the bigger problem is that this result seems in tension with dynamical observations of the influence of low-redshift SMBHs, which show that for $M\sim 10^9 M_{\odot}$, $R_{\rm{infl}}\approx 50 \rm{~pc}$ \citep{Lauer+05}, and not $1-10 \rm{~pc}$, as the high-mass pileup regime would predict. This $R_{\rm infl}$ problem arises only for the largest SMBHs for high accretion rate. As the largest SMBHs experienced most of their AGN episodes at high redshift $z$ (``cosmic downsizing''; \citealt{Barger+05, Hasinger+05}), it is possible that high-$M$ SMBHs do indeed have anomalously small influence radii, but that $R_{\rm infl}$ expands significantly between their final major AGN episode and $z\approx 0$ (the only range of redshifts/times where these SMBHs are close enough to permit dynamical $R_{\rm infl}$ measurements\newp{)}. Core scouring due to satellite galaxy infall \citep{MilosavljevicMerritt01} is a plausible explanation for this influence radius expansion (and is especially relevant here, given the circumstantial evidence for greater core formation in the most massive elliptical galaxies; e.g. \citealt{Lauer+05, Thomas+14}), but a full exploration of this hypothesis is beyond the scope of this paper. 
Finally, we note that there is some variation in this ``ECO overproduction problem'' with accretion rate and SMBH mass. In our numerical results, for high-mass SMBHs with sufficiently low accretion rates, we find a more self-consistent number of BHs (i.e. they do not dominate the total potential at typical low-$z$ values of $R_{\rm infl}$), for example, for $\dot{M}/\dot{M}_{\rm g}=10^{-2}$ we get $N_\bullet\sim 10^6$.

\subsection{CMI solution} \label{sec: CMI solution}
\begin{figure}[b]
\centering
 \includegraphics[width=\linewidth]{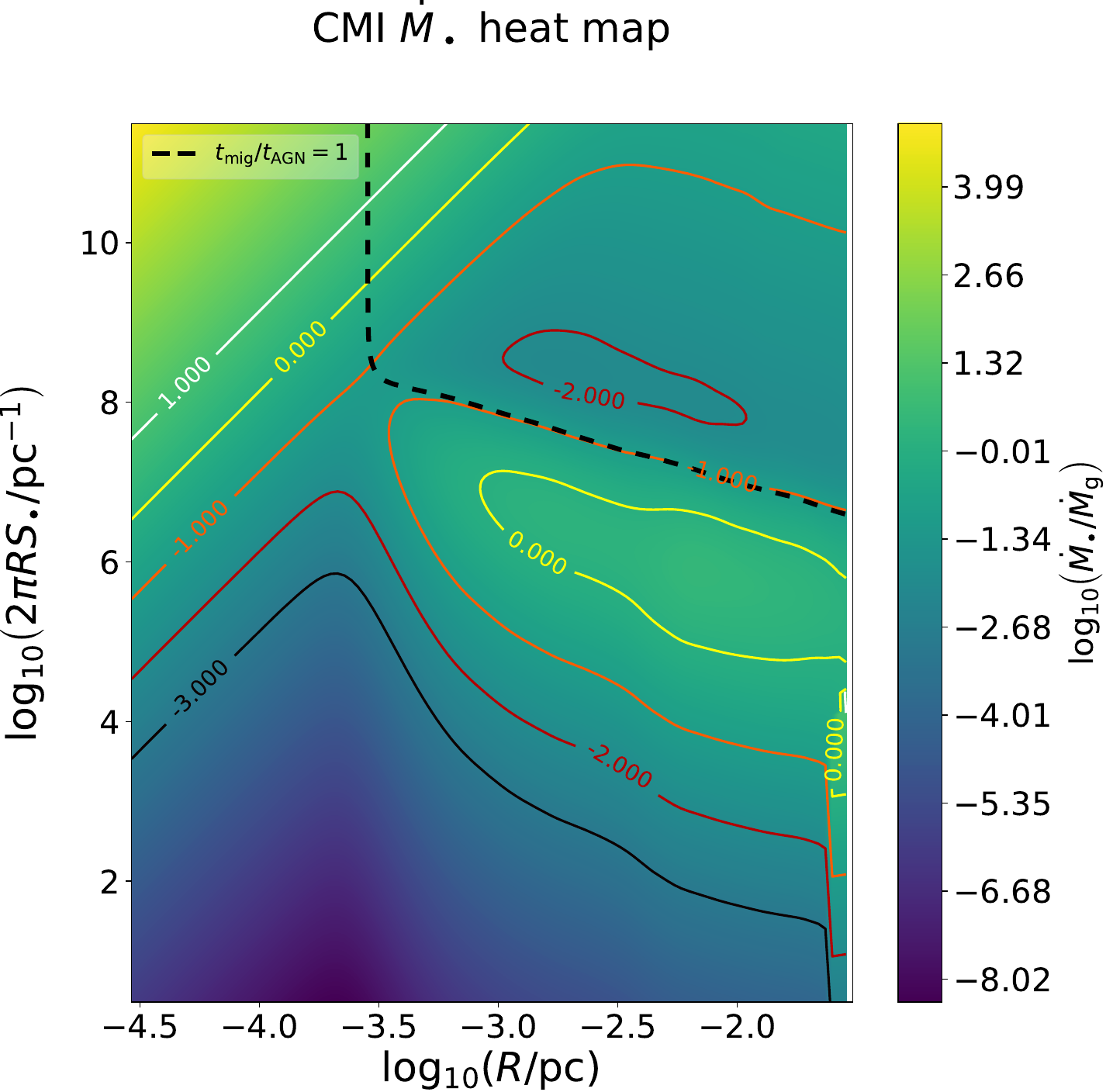}
 \caption{A heat map of the ratio between ECO mass influx ($\dot{M}_\bullet$) and gas mass influx ($\dot{M}_{\rm g}$), shown in the parameter space of radius $R$ (in parsec) and ECO linear number density $2\pi R S_\bullet$ (in number per par sec). This plot illustrates how, in a CMI steady-state, specification of the mass influx ratio $\dot{M}_\bullet/\dot{M}_{\rm g}$ determines the nature of the disk solutions. We hold constant SMBH mass $M=10^8 M_\odot$ and gas accretion ratio $\dot{M}_{\rm g}=10^{-1} \dot{M}_{\rm Edd}$. Solid contours show curves of constant mass influx ratio, which can be multivalued, indicating different possible solutions. The dashed black line separates regions of parameter space where ECOs can migrate (left of the dashed black line) from the zero-migration limit (at $z=0.5$). At low $\dot{M}_\bullet/\dot{M}_{\rm g} \lesssim 10^{-1}$, solutions are unique and aligned with the Shakura--Sunyaev limit, but at higher $\dot{M}_\bullet/\dot{M}_{\rm g} \gtrsim 10^{-1}$, an additional ``upper branch'' emerges with a much higher ECO density and inflated gas aspect ratio. Note that in this plot $t_{\rm mig}$ is calculated in a simplified way, by setting $C_I=2$.} 
 \label{fig: CMI constant S map}
\end{figure}
In the constant mass influx regime, we set $\dot{M}_\bullet$ equal to a constant along with $\dot{M}_{\rm g}$. This type of steady state solution is most likely achievable at small radii in AGN disks, though we will check its self-consistency post hoc. In this section, we will present results in terms of the dimensionless compact object mass influx $\dot{M}_\bullet / \dot{M}_{\rm g}$. Solving this system of equations, we find multiple possible branches in some parts of the parameter space, as can be seen in \cref{fig: CMI constant S map},which shows a contour map showing the mass flux ratio $\dot{M}_\bullet / \dot{M}_{\rm g}$ in the space of ECO linear density $2\pi R S_\bullet$ and radius $R$. Although \cref{fig: CMI constant S map} is plotted for the specific case of $M=10^8M_\odot, \dot{M}_{\rm g}=0.1 \dot{M}_{\rm Edd}$, we find similar results for different $\{M, \dot{M}_{\rm g}\}$ values, albeit with $\dot{M}_\bullet / \dot{M}_{\rm g}$ rescaled by the changes in the independent variables. We plot a larger parameter space exploration of $\dot{M}_\bullet / \dot{M}_{\rm g}$ in \aref{app: CMI profile}. 

In general, we find significant degeneracy between the three parameters $M, \dot{M}_{\rm g}/\dot{M}_{\rm Edd},\dot{M}_\bullet/\dot{M}_{\rm g}$. Assuming that the migration is dominated by type I torques using \cref{eq: type I migration rate,eq: General black hole flow rate}, and that the BH heating source is given by \cref{eq: full energy conservation} (assuming that the accretion onto the ECOs is at the Eddington limit) we get that the heating rate from feedback is\footnote{Note that at the relevant radii, $\Sigma$ and $H/R$ scale up weakly with both $M$ and $\dot{M}_{\rm g}/\dot{M}_{\rm Edd}$.} :
\begin{equation}
 Q_\bullet \propto \left(\frac{R}{R_{\rm g}} \right)^{-3/2} \Sigma \left(\frac{H}{R} \right)^2 \left[ \left(\frac{\dot{M}_\bullet}{\dot{M}_{\rm g}}\right) \times \left( \frac{\dot{M}_{\rm g}}{\dot{M}_{\rm Edd}}\right) \times M\right]
\end{equation}

\begin{figure} 
 \centering
 \includegraphics[width=\linewidth]{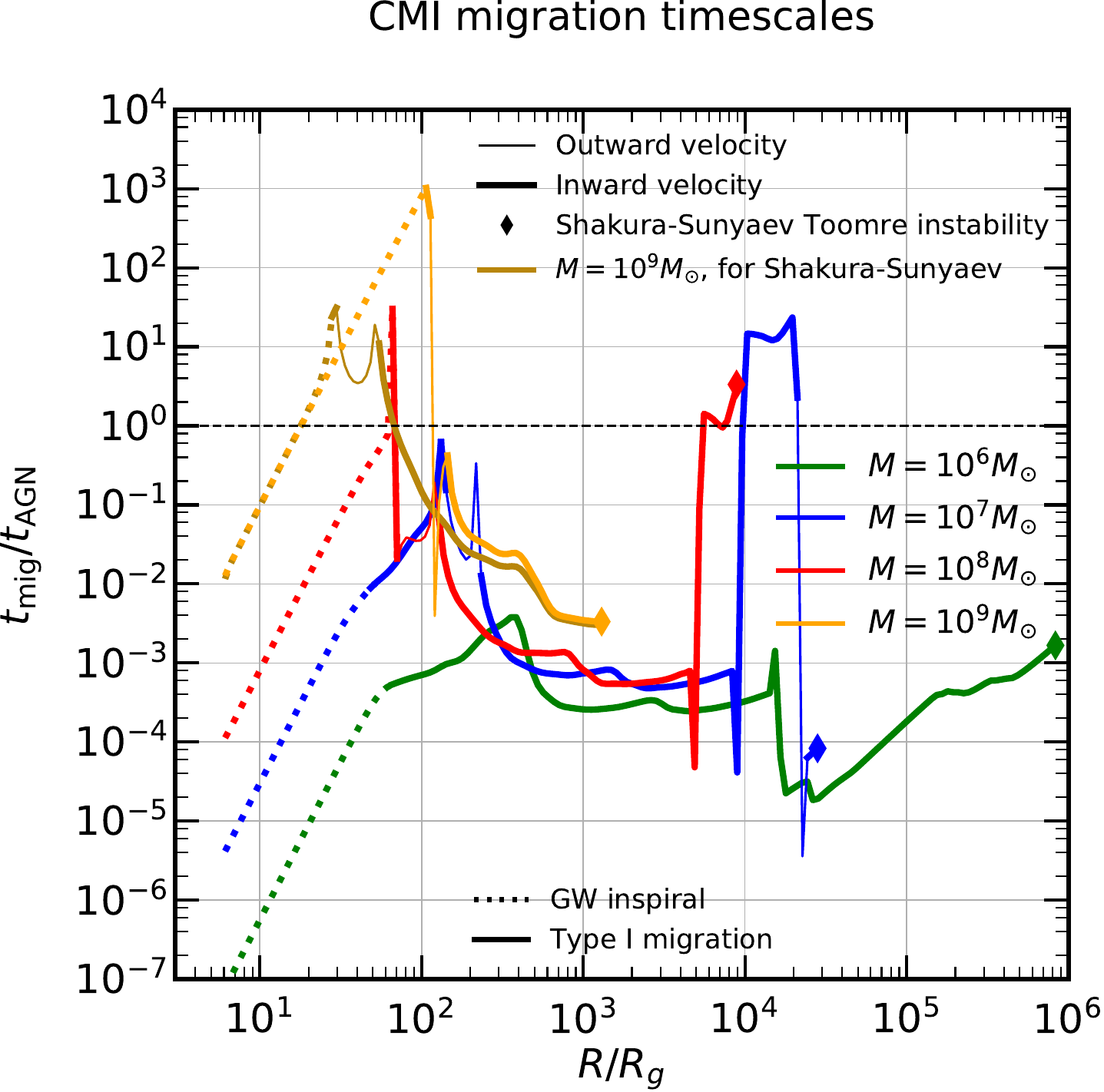}
 \caption{The ECO migration time $t_{\rm mig}$ (post hoc calculation of the type I migration prefactor) normalized to the effective AGN lifetime $t_{\rm AGN}$ at redshift $z=0.5$, plotted against dimensionless radius $R/R_{\rm g}$. Colors correspond to different SMBH masses as in the top panel of \cref{fig: pileup t_mig}, all curves are computed for $\dot{M}_{\rm g} = 10^{-1} \dot{M}_{\rm Edd}$, and we take $\dot{M}_\bullet / \dot{M}_{\rm g} = 10^{-1}$. Regions with $t_{\rm mig}/t_{\rm AGN} \lesssim 1$ are consistent with a constant mass influx steady state, but regions with $t_{\rm mig}/t_{\rm AGN} \gtrsim 1$ are not.  } 
 \label{fig: CMI mig_time}
\end{figure}

\begin{figure} 
 \centering
 \includegraphics[width=\linewidth]{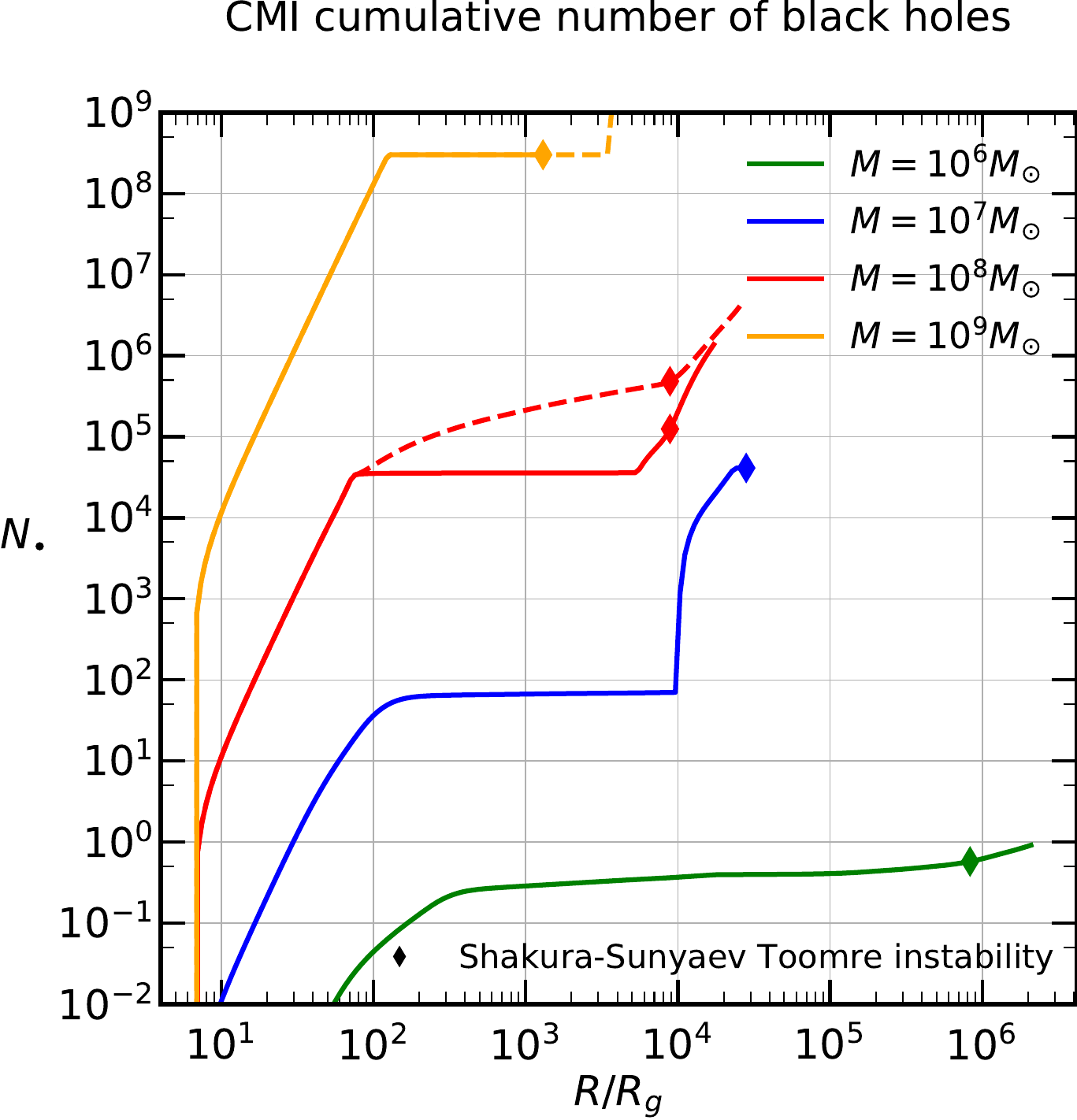}
 \caption{Number of enclosed BHs $N_\bullet$ in the CMI regime solutions, shown as a function of dimensionless radius $R/R_{\rm g}$. Line colors and styles are the same as in \cref{fig: CMI mig_time}. Notably, the lower branch of solutions (solid lines) is similar to Shakura--Sunyaev-type disks for gas properties (except for the  $10^9M_\odot$  SMBH); for small SMBHs, the number of enclosed BHs is so small they cannot be treated statistically; for large SMBHs, the ECOs do form a semicontinuous fluid. Large SMBHs also have an upper branch of solutions with  more ECOs.}
 \label{fig: CMI NBH}
\end{figure}
This degeneracy manifests in two types of solutions: 
(i) for lower values of $Q_\bullet$ (i.e, small values for one or all of the parameters mentioned above), which is usually the realistic range of variables, our disk profile converges to the Shakura--Sunyaev solution, and the presence of ECOs does not affect the disk profile (see the $\dot{M}_{\bullet}/\dot{M}_{\rm g}=10^{-2},10^{-3}$ contour line in \cref{fig: CMI constant S map}); (ii) for higher values of $Q_\bullet$, each specific $\dot{M}_\bullet/\dot{M}_{\rm g}$ ratio appears to have two to three branches of solutions. These scale with $Q_\bullet$ so that the lowest branch, similarly to (i), converges to Shakura--Sunyaev, the ``middle'' branch is below (or slightly above) the migration timescale self-consistency limit, and the ``upper'' branch is mostly in the ``no-migration'' limit, which is not self-consistent with our CMI assumption. The type (ii) solutions, which result in higher number of ECOs, usually contain a discontinuity in the $\dot{M}_\bullet/\dot{M}_{\rm g}$ space. For an example of the second kind, see the $\dot{M}_{\bullet}=10^{-1}\dot{M}_{\rm g}$ case in \cref{fig: CMI constant S map}.

All of the different types of CMI solutions converge to Shakura--Sunyaev at small radii (the precise radius of convergence depends on the parameters). Note that for some parameters (high $M_\bullet$ and high $\dot{M}_\bullet / \dot{M}_{\rm g}$) there exists a narrow region at small radii with no ``lower branch,'' as we see in \cref{fig: CMI mig_time} for $M=10^9 M_{\odot}$. When this situation arises, the result is generally not self-consistent, with $t_{\rm mig} > t_{\rm AGN}$. This self-consistency problem arises from the exceedingly large number of BHs necessary at small distances\footnote{Specifically, distances too far for GW migration to be efficient, but deep enough inside the radiation-pressure-dominated zone that gas is dilute and Type I migration also highly inefficient.} in order to attain the required $\dot{M}_\bullet$ value. This can be seen in \cref{fig: CMI NBH}, which also shows that for lower masses ($M \lesssim 10^8M_{\odot}$), the (Shakura--Sunyaev-like) CMI solutions involve so few ECOs that the continuum approximation does not hold. However, this does not change the basic result for lower-mass SMBHs that any ECOs in this region will fail to significantly affect the disk structure and will migrate quickly into the SMBH. In contrast, for $M=10^8M_{\odot}$, we see that even for the lower branch there is a large number of ECOs. In general, the lower branch always satisfies the timescale hierarchy $t_{\rm diff} < t_{\rm dyn} \ll t_{\rm th} \ll t_{\rm visc}$.
We truncate these solutions at the radius where $Q_{\rm T}=1$, as the assumption of constant mass flux will stop being valid in star-forming regions.

In summary, we conclude that small radii (i.e. those inside the original Toomre instability point $R<R_{\rm Q}$) in AGN disks can usually be described well, ignoring ECO feedback. Even large ECO inflow rates $\dot{M}_\bullet$ usually fail to produce significant changes in the small-scale gaseous disk structure. The sole exception to this conclusion is for the largest SMBHs ($M\gtrsim 10^8 M_\odot$) and highest ECO inflow rates ($\dot{M}_\bullet \gtrsim 0.1 \dot{M}_{\rm g}$), where ``middle'' and ``upper'' branches of solutions emerge that feature much larger $H/R$ than standard Shakura--Sunyaev-type solutions. The upper branches are strongly inconsistent ($t_{\rm mig} \gg t_{\rm AGN}$) with a steady-state inflow assumption, indicating that they are not astrophysically relevant solutions. The middle branches are marginally self-consistent/inconsistent ($t_{\rm mig} \sim t_{\rm AGN}$) with a steady state, indicating that in some circumstances they may be achievable. 

\subsection{Intermediate regime}
\label{sec:intermediate}

As we have shown in the previous two subsections, the pileup solution is mostly self-consistent in the outer radii of AGN disks, and the CMI solution is generally self-consistent in smaller radii (though we note that {\it usually}, the CMI regime defaults to a Shakura--Sunyaev-like disk). Notably, however, there is a substantial portion of parameter space -- typically $\approx 1$ order of magnitude in radius -- in between the domain of validity of each of these limiting regimes. In this intermediate regime, which covers $R_{\rm Q} < R < R_{\rm pile}$, a naive Shakura--Sunyaev-type model has $Q_{\rm T}<1$, but migration timescales are generally $\ll t_{\rm AGN}$, even if one adds so many ECOs to the disk that it is stabilized against fragmentation (see \cref{fig: pileup t_mig}). It is possible that this intermediate regime -- which was schematically described as the ECO ``factory'' in our earlier cartoon (\cref{fig: schematics}) -- features a more complex steady-state solution, with $\dot{M}_{\rm g} = \dot{M}_{\rm g}(R)$ (in analogy to the model of \citealt{Thompson+2005}), and steady conversion of gas into embeds, which migrate inwards and set the constant $\dot{M}_\bullet$ influx rate of the inner CMI zones. Alternatively, it is also possible that the time lag between star formation and ECO formation prevents a steady state from being achieved in these intermediate zones and instead limit cycles of fragmentation, ECO formation and disk stabilization, and ECO depletion via migration proceed periodically. Clearly, it will be necessary to investigate this issue with a time-dependent solution of our full two- (or three-) fluid disk equations (\sref{sec:Two-Fluid Disk Equations}), so we defer a full study for future work. 
\begin{figure}[b]
 \centering
 \includegraphics[width=\linewidth]{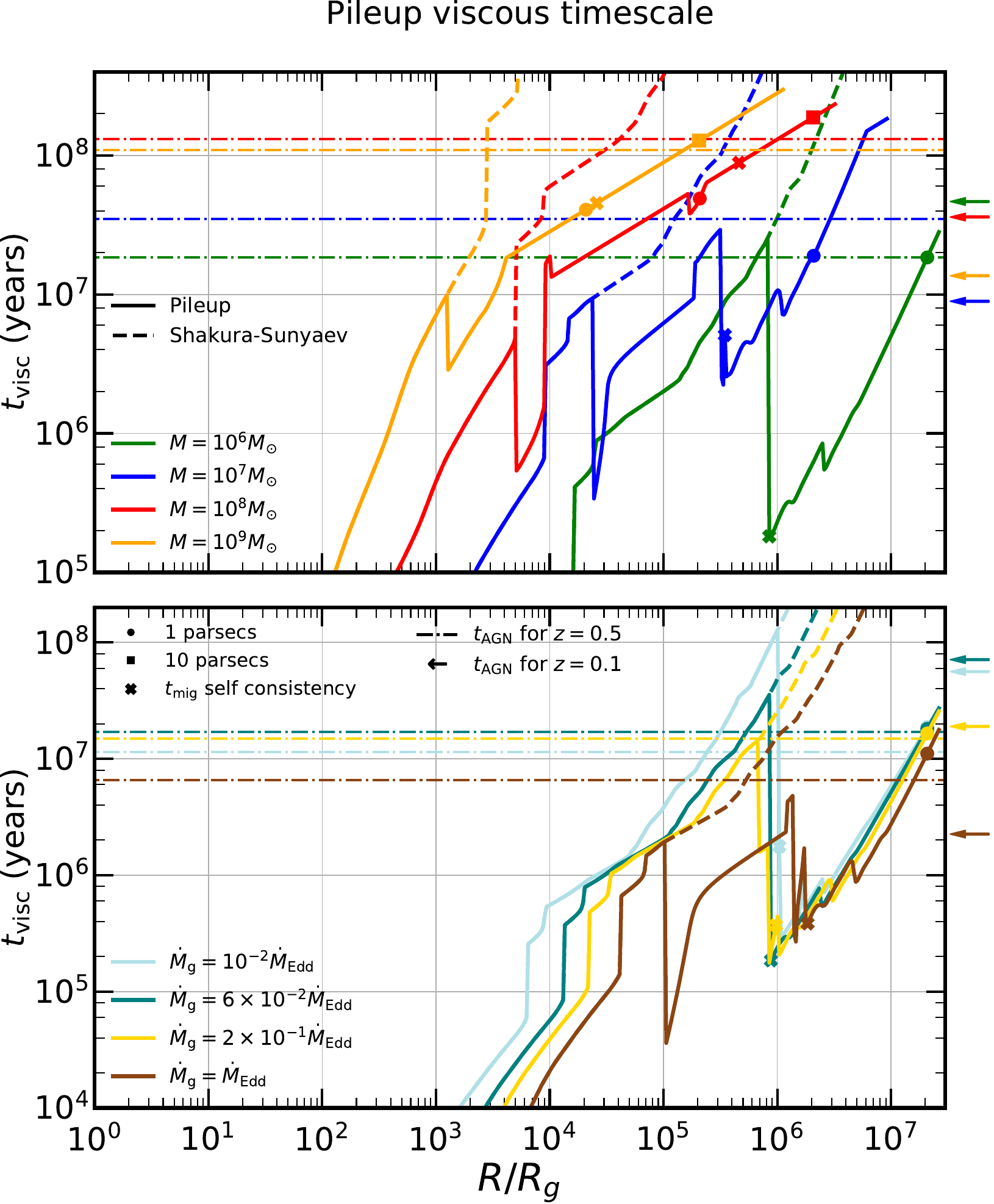}
 \caption{Viscous (or inflow) time $t_{\rm visc}$ , shown against dimensionless radius $R/R_{\rm g}$ for the pileup solution. Color/panel layouts are the same as in \cref{fig: pileup t_mig} (top panel is fixed Eddington ratio and different SMBH masses, bottom panel fixed SMBH mass and different Eddington ratios, only self-consistent at large radii) Solid lines show the pileup solutions , while the dashed lines show a Shakura--Sunyaev-type solution. The dashed--dotted lines (arrows) indicates the approximated AGN lifetime (see \cref{fig:AGN_lifetimes}) for redshift $z=0.5$ ($z=0.1)$. In general, $t_{\rm visc}$ drops by orders of magnitude when comparing the pileup solutions to Shakura--Sunyaev. The radius of inflow equilibrium $R_{\rm visc}$ increases by one to two orders of magnitude in all cases. This increase is usually larger at high $z$, but  remains very large at low $z$ for small $M$.}
 \label{fig: pileup visc time}
\end{figure}%
\section{Astrophysical Features of FDAFs}
\label{sec:implications}

In this section, we discuss the most astrophysically relevant features of feedback-dominated accretion flows in AGN.

\subsection{Inflow Equilibrium} \label{sec:viscous}

As mentioned in \sref{sec:intro}, the timescales for local angular momentum transport (i.e. the viscous timescale $t_{\rm visc}$) in a Shakura--Sunyaev disk around an SMBH become much longer than the expected AGN lifetime beyond radii of $R_{\rm visc} \sim 0.01-0.1$ pc. The enclosed gas mass within this radius, $M_{\rm g}(R_{\rm visc})$, is $\ll M$, making it clear that local angular momentum transport mechanisms alone cannot explain AGN episodes that last long enough to contribute to SMBH growth. Generally, this mass and timescale mismatch has motivated the assumption of global angular momentum transport mechanisms (e.g., nonaxisymmetric bar modes, as in \citealt{HopkinsQuataert11}) that can transport mass from larger radii down to the SMBH on shorter times. An alternative solution is to invoke dynamically important magnetic fields that inflate the AGN aspect ratio at large radii, shortening $t_{\rm visc} \propto (H/R)^{-2}$ \citep{DexterBegelman2019}. 

In analogy with models of magnetically supported accretion disks, our model for FDAFs finds large increases in $H/R$ and correspondingly large reductions in the viscous times associated with local angular momentum transport. 
In the CMI zone (where our model generally defaults to Shakura--Sunyaev-like solutions), the viscous timescales 
$t_{\rm visc} \ll t_{\rm AGN}$ generally. At the larger radii of the pileup regime, where simple disk models run into the aforementioned inflow time problems, we see substantially shorter viscous timescales for all choices of $M$ and $\dot{M}_{\rm g}$, as is shown in \cref{fig: pileup visc time}. In the pileup regime of an FDAF disk, the viscous timescale is not necessarily shorter than the AGN lifetime everywhere, but $R_{\rm visc}$ is generally extended by one to two orders of magnitude, increasing the domain of validity for $\alpha$-viscosity prescriptions and reducing (but not eliminating) the need for global torques.

\subsection{Migration Traps}
\label{sec:traps}

In protoplanetary disks, planets and planetesimals accumulate in migration traps: narrow regions where the sign of the migratory torque flips from positive to negative \citep{Masset+06, HasegawaPudritz11}. These traps have long been thought to play an important role in the growth of planets. More recently, interest has emerged in the existence of migration traps in AGN disks \citep{Bellovary+2016}. If such traps exist, they will likewise serve as accumulation points for stellar-mass embeds; ECOs interacting in such traps will be able to pair up into binaries and merge, emitting bursts of high-frequency gravitational radiation. Migration traps are one of the main hypothesized sources for binary BHs in AGN disks and could plausibly be the dominant contributor GW emission rates in the AGN channel. If these traps sit deep enough in the potential well of the SMBH, GW recoil kicks following merger will be insufficient to dislodge ECOs in the traps, enabling hierarchical merger and growth of stellar-mass BHs, possibly into the pair-instability mass gap and beyond.

An analogous possibility arises in the semianalytic two-fluid disk model of \citet{Tagawa+20b}.  Although this model finds that BH--BH mergers are not usually due to standard Type I migration traps, BHs instead accumulate in gaps that are opened within the disk and merge in these sites.  As mentioned earlier, we do not find that stellar-mass BHs are able to open gaps in our disk models, although it almost occurs for the lowest SMBH masses and accretion rates (Fig. \ref{fig: gap opening}).  The qualitative difference between these conclusions is likely due to the different gas-disk models used in the two works (most notably for the radii of interest, the opacities and angular momentum transport prescriptions differ).    

In our earlier exploration of migration timescales, we solved for the two-fluid steady state in the CMI regime assuming a simplified and constant prefactor $C_I$ to the type I migration torque (see \cref{eq: Type I torque}). From this exercise, we concluded that usually, the $R < R_{\rm Q}$ CMI regime defaults to a Shakura--Sunyaev-like structure.  Thus, the full migration profile (with a variable $C_I$) can be read off the inner regions of \cref{fig: pileup t_mig} for the solutions that converge to the Shakura--Sunyaev solution, or equivalently off \cref{fig: CMI mig_time}, where we calculated the migration of ECOs using the full migration expression (the sum of \cref{eq: type I migration rate} and \cref{eq: GW migration rate}). 
By comparing \cref{fig: CMI mig_time} to the constant $C_{\rm I}$ case we see that the main differences arise for the $M=10^7M_{\odot},10^8M_{\odot}$ and $10^9 M_{\odot}$ curves at $R\sim 10^2 R_{\rm g}$. In the $M=10^7M_\odot$ and $10^8 M_\odot$ cases, the full type I torque prescription produces migration traps that are similar to the results of \citet{Bellovary+2016} for the \citet{SirkoGoodman2003} model (which, like our CMI regime, also converges to the Shakura--Sunyaev model at small radii). 

\citet{Secunda+19} simulated BH populations in AGNs (Sirko-Goodman disk model) with a migration trap and found that BHs that migrate toward the migration traps will accumulate there in a chain of mean motion resonances, potentially forming BH-BH binaries. These migration traps might cut off the supply of migrating ECOs and therefore have two potentially important effects. The first effect concerns our models: in the high-$M$, high-$\dot{M}_\bullet$ parts of parameter space where CMI solutions deviate from Shakura--Sunyaev, these solutions may be untrustworthy interior of migration traps (which, by cutting off the supply of ECOs to the smallest radii, reduce $\dot{M}_\bullet$ there). The long-term evolution of a large population of ECOs interacting in a single migration trap is unclear, and it is possible these traps could become ``leaky'' due to strong scatterings if a large enough ECO population builds up.

Even if this speculation is correct, however, it is interesting to note that in the $M=10^9 M_\odot$ case, an additional and qualitatively new type of migration trap would emerge (we see it both for the Shakura Sunyaev case and for the $\dot{M}_\bullet = 0.1\dot{M}_{\rm g}$ ECO influx rate considered in \cref{fig: CMI mig_time}). This trap -- which would only arise if strong scatterings push some ECOs through the classic type I migration trap at larger radii-- is associated not with a reversal in the sign of the migratory torque but rather with a finite radial zone of high ECO density $S_\bullet$, large aspect ratio $H/R$, and correspondingly long migration times. We call this possible type of migration trap a ``Zeno trap,'' for brevity\footnote{An ECO approaching such a trap will not be physically repulsed due to a flip in migration torque (as is the case at a classical trap) but will instead continue migrating inwards at ever decreasing rates, akin to Zeno's paradox.}, and note that it may be a feature of embed populations in AGN disks surrounding the largest SMBHs.

The second effect, which is of significant astrophysical interest, concerns quasi-circular extreme mass ratio inspirals (EMRIs) sourced from AGN disks. EMRIs are  LISA-band GW signals that normally are thought to be sourced by two-body scatterings and thus involve a stellar-mass compact object approaching an SMBH on a high-$e$ orbit. However, a qualitatively different gravitational waveform will be generated by EMRIs that approach following quasi-circular migration in AGN disks \citep{Levin07, Pan+21}. This alternate mode of EMRI formation appears promising, but may be averted if migration traps exist at small radii. 
Finally, we note that the existence, size, and location of the migration traps are highly dependent on the different disk parameters ($M, \dot{M}_{\rm g},\dot{M}_\bullet$).

\subsection{ECO Mass Growth} \label{sec: mass growth}

\begin{figure*} 
 \centering
 \subfloat[\label{subfig: final mass zams 1e8 spin up}]{
 \centering
 \includegraphics[width= 0.305\linewidth]{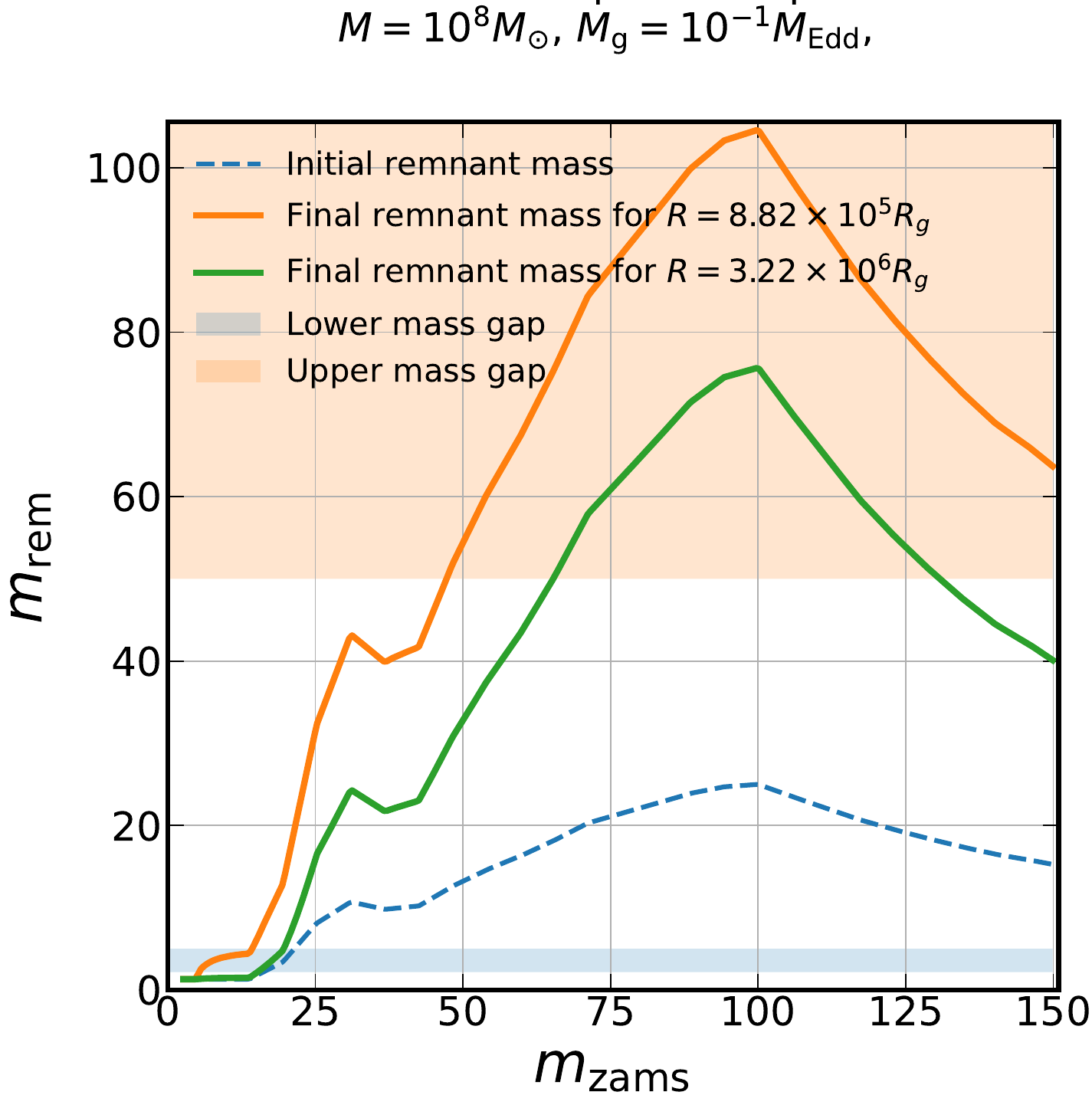}
 
 }
 \subfloat[\label{subfig: final mass zams 1e8 max spin}]{
 \centering
 \includegraphics[width= 0.305\linewidth]{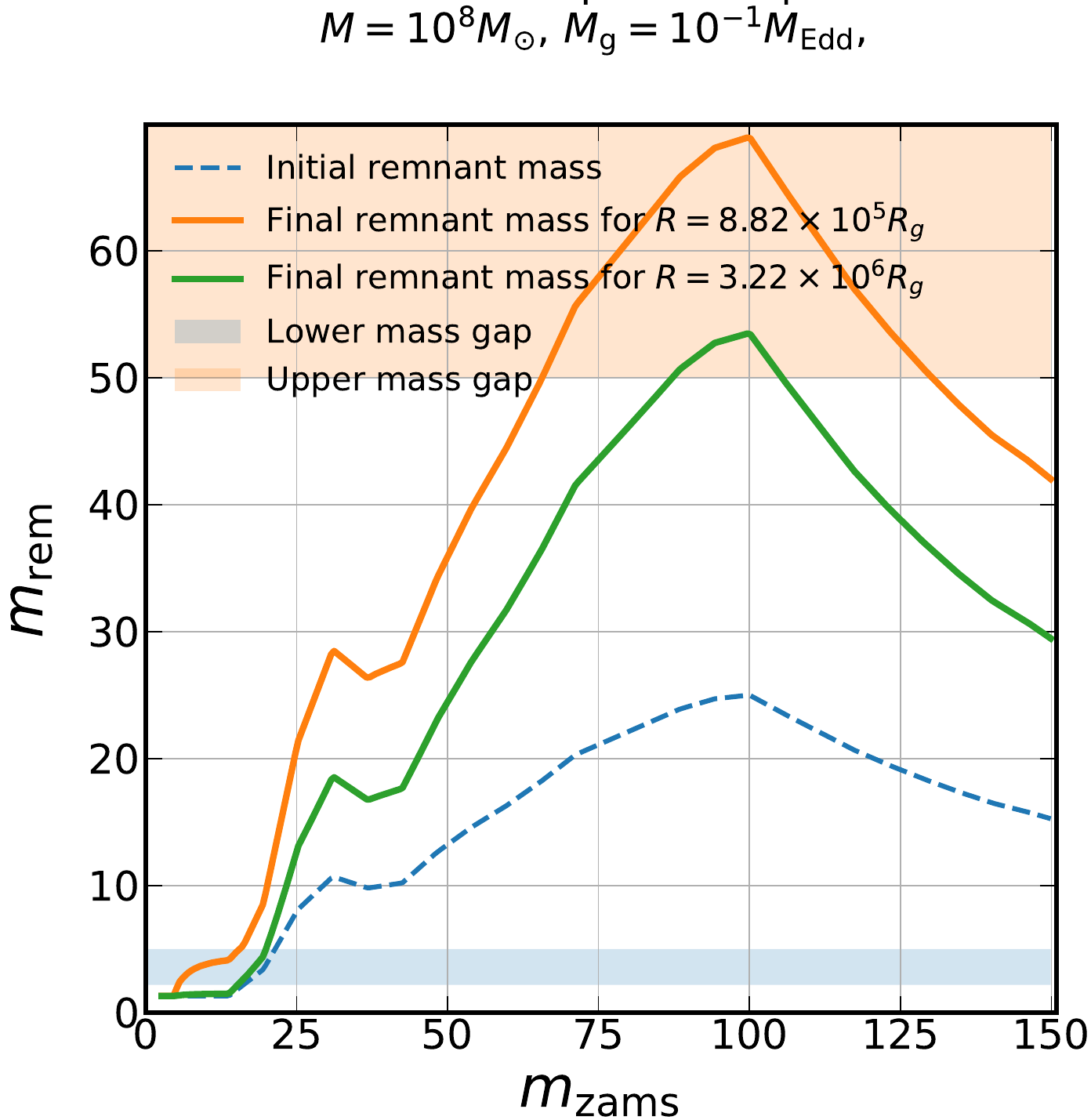}
 
 }
 \subfloat[\label{subfig: final mass zams 1e7}]{
 \centering
 \includegraphics[width= 0.305\linewidth]{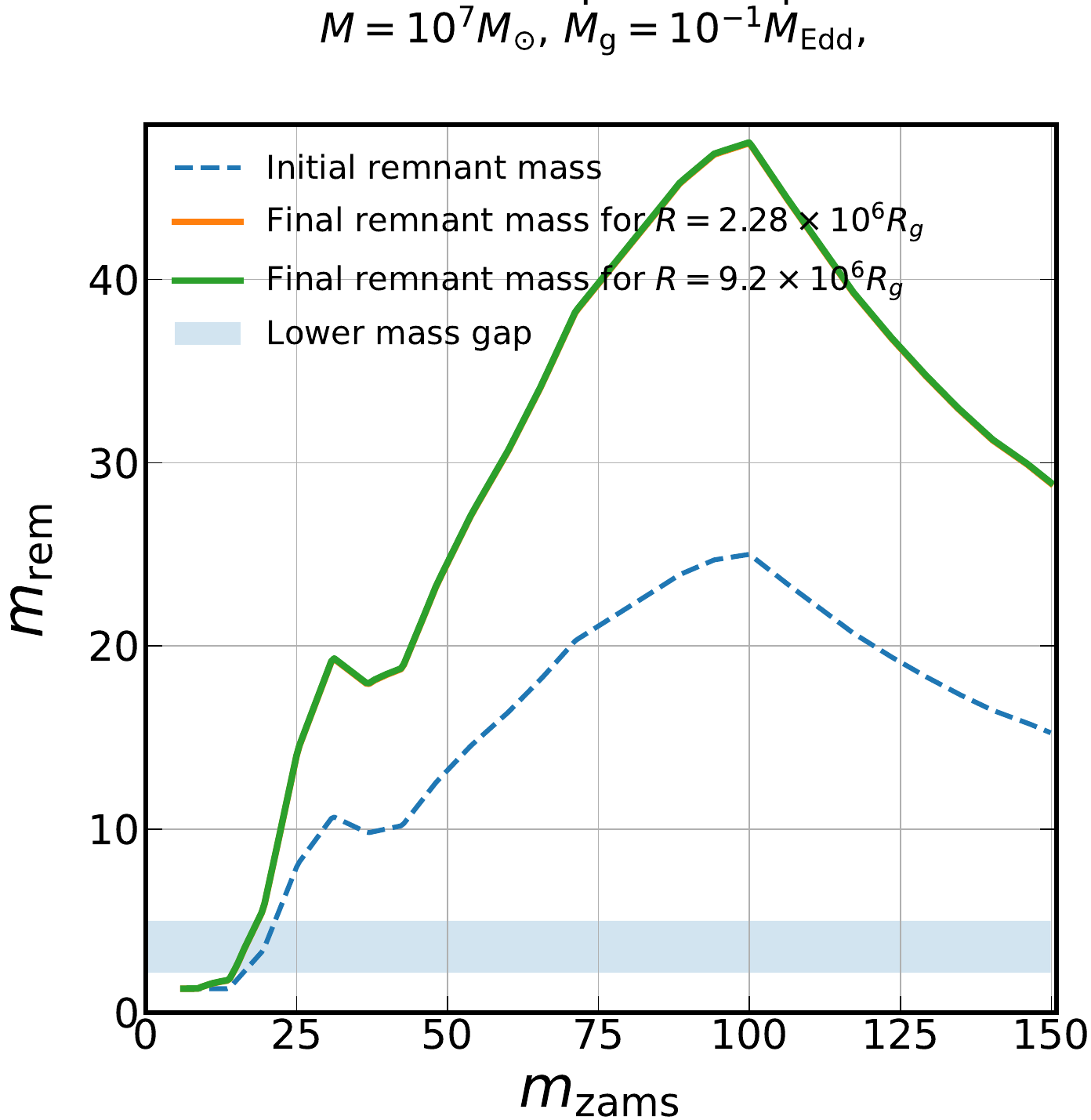}
 
 }
 
 \subfloat[\label{subfig: final mass BH 1e8 spin up}]{
 \centering
 \includegraphics[width= 0.305\linewidth]{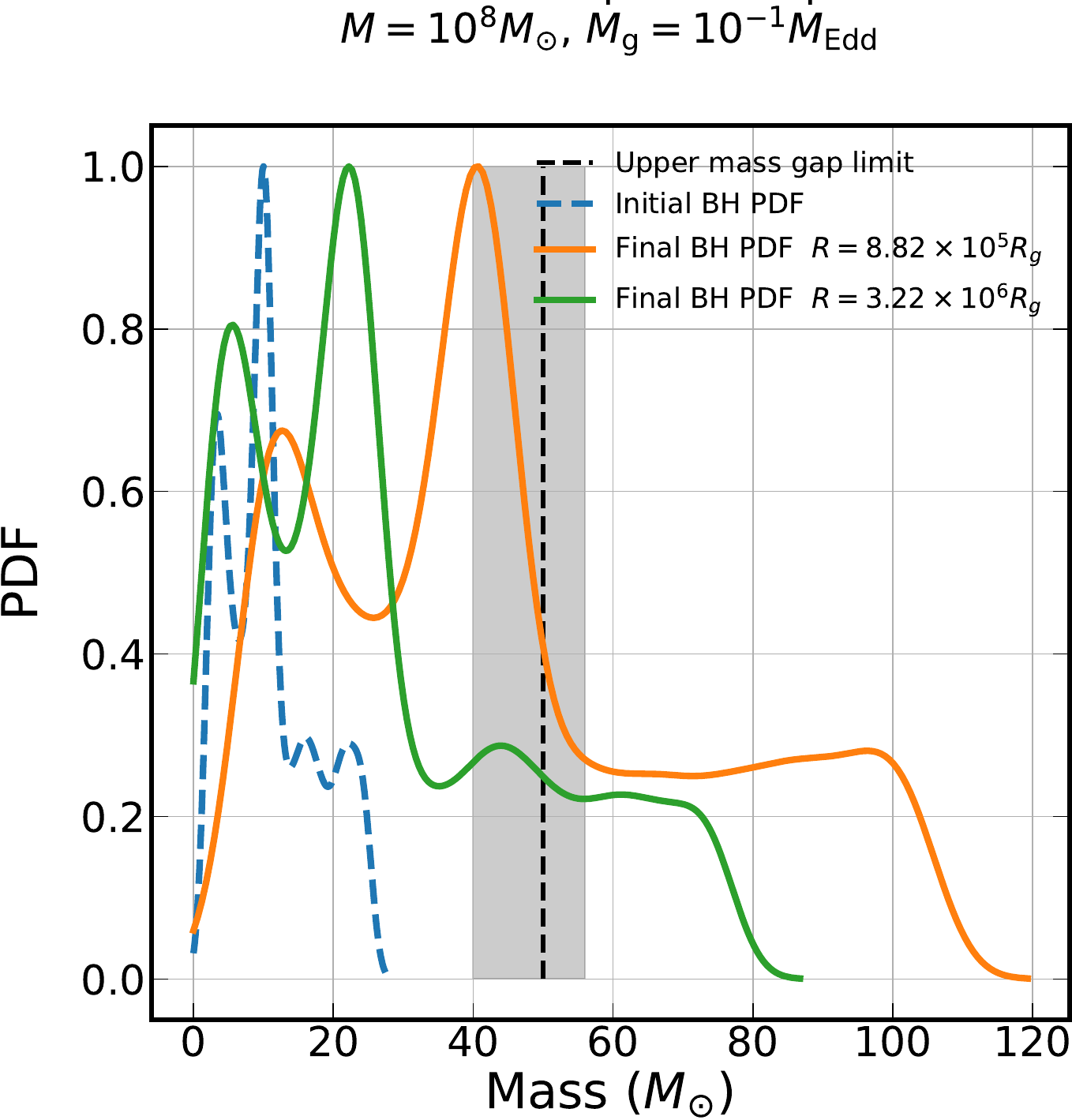}
 
 }
 \subfloat[\label{ubfig: final mass BH 1e8 max spin }]{
 \centering
 \includegraphics[width= 0.305\linewidth]{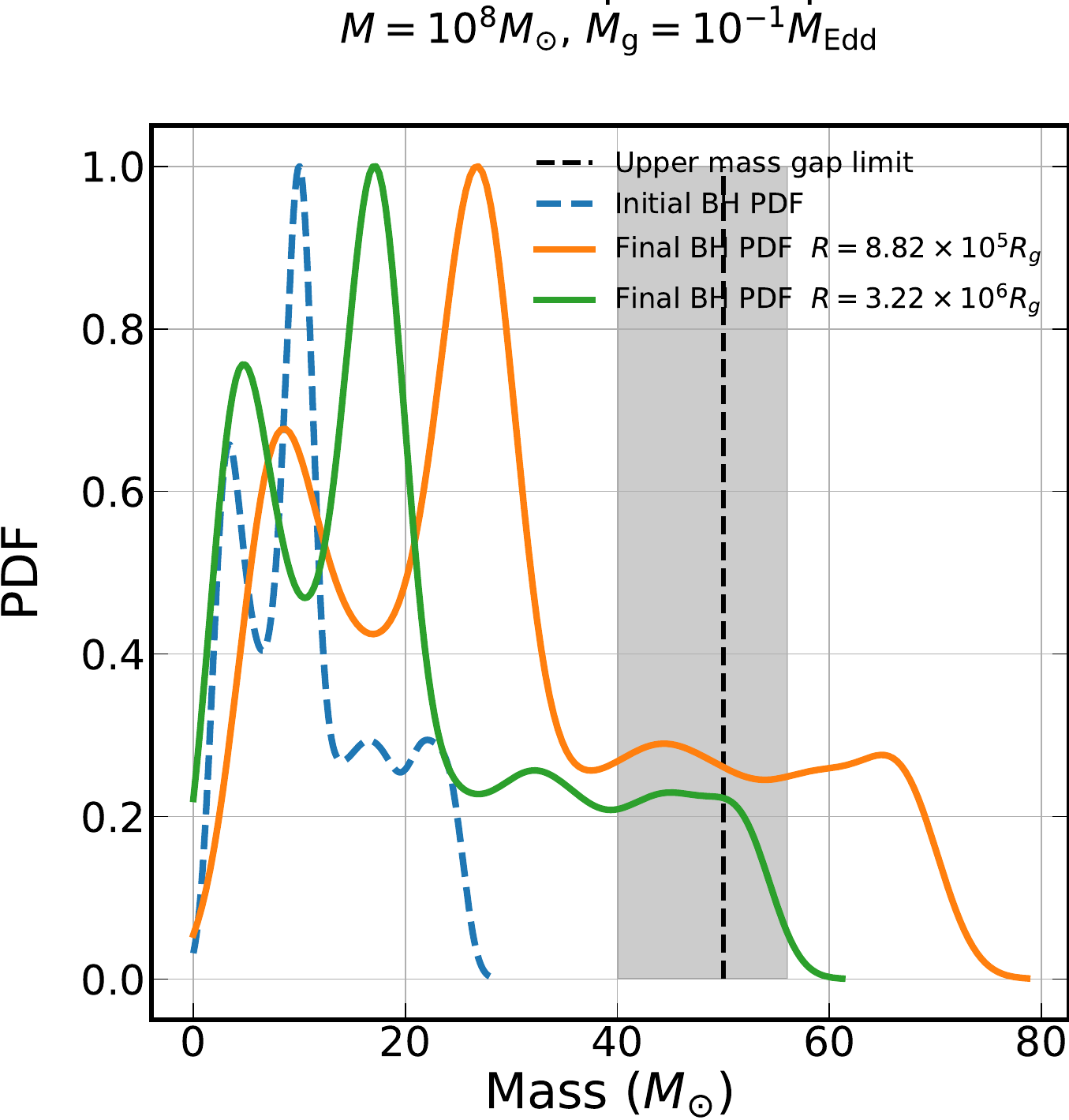}
 
 }
 \subfloat[\label{subfig: final mass BH 1e7}]{
 \centering
 \includegraphics[width= 0.305\linewidth]{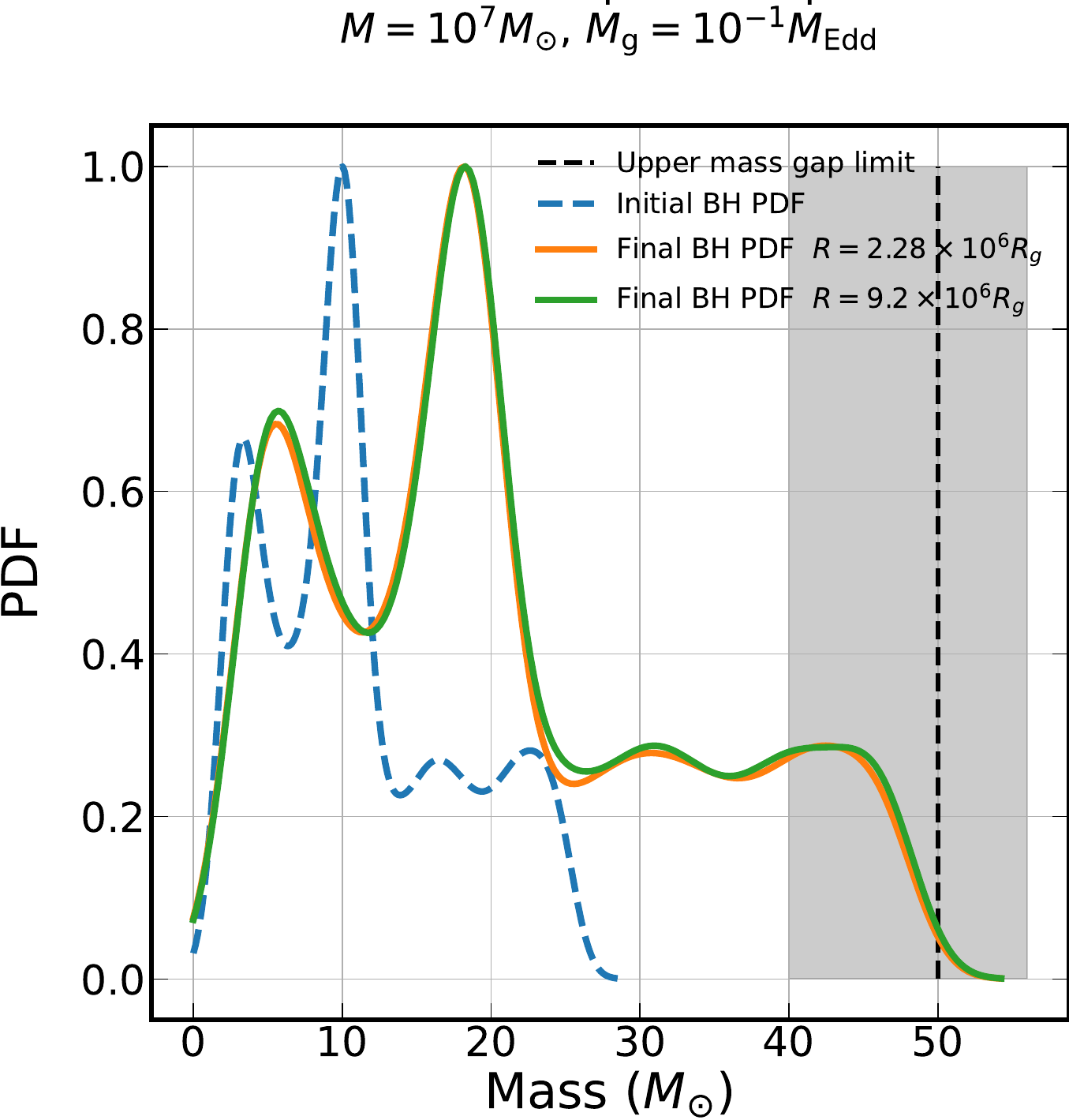}
 
 }
 
 \subfloat[\label{subfig: final mass NS 1e8 spin up}]{
 \centering
 \includegraphics[width= 0.305\linewidth]{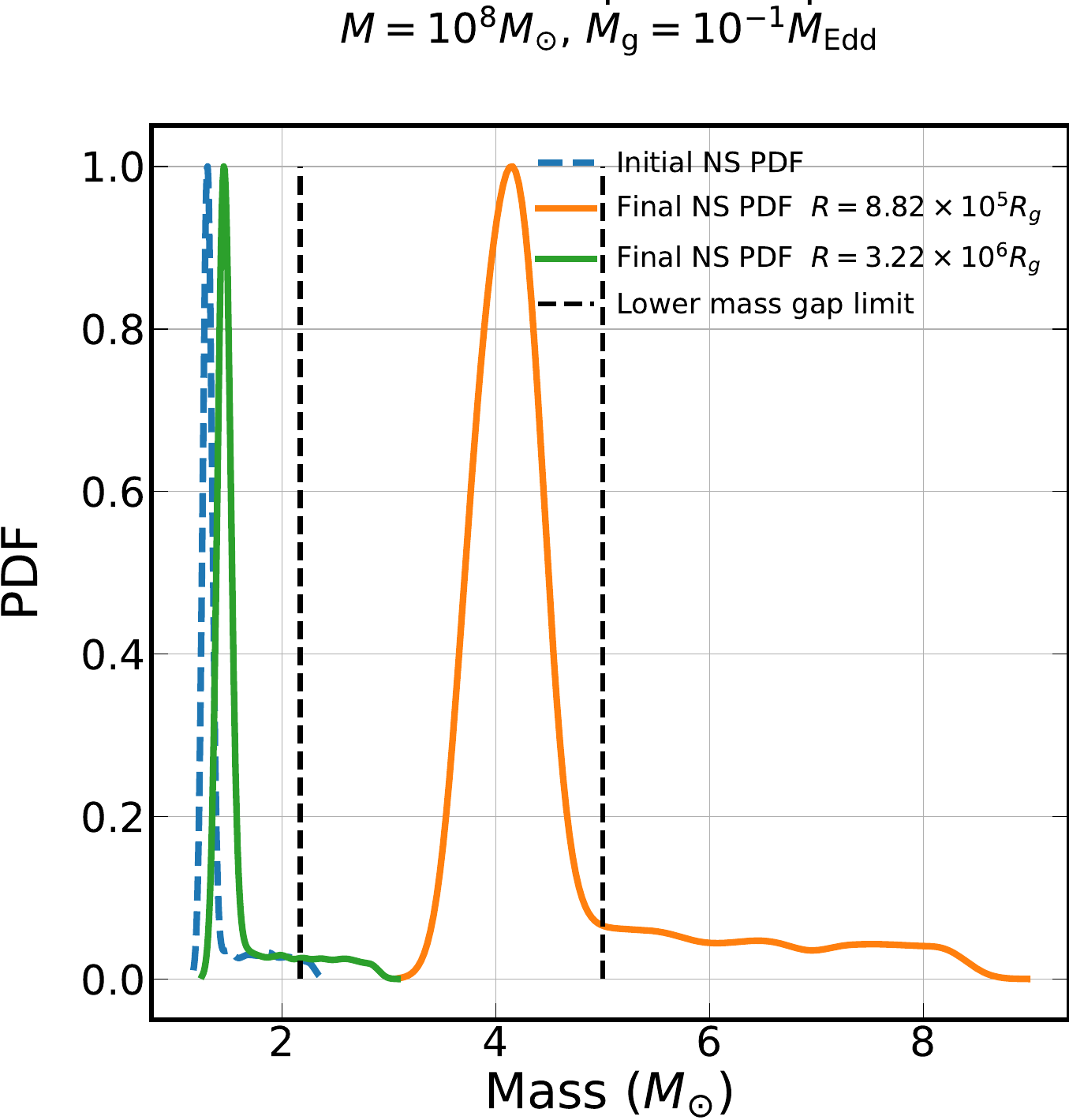}
 
 }
 \subfloat[\label{ubfig: final mass NS 1e8 max spin }]{
 \centering
 \includegraphics[width= 0.305\linewidth]{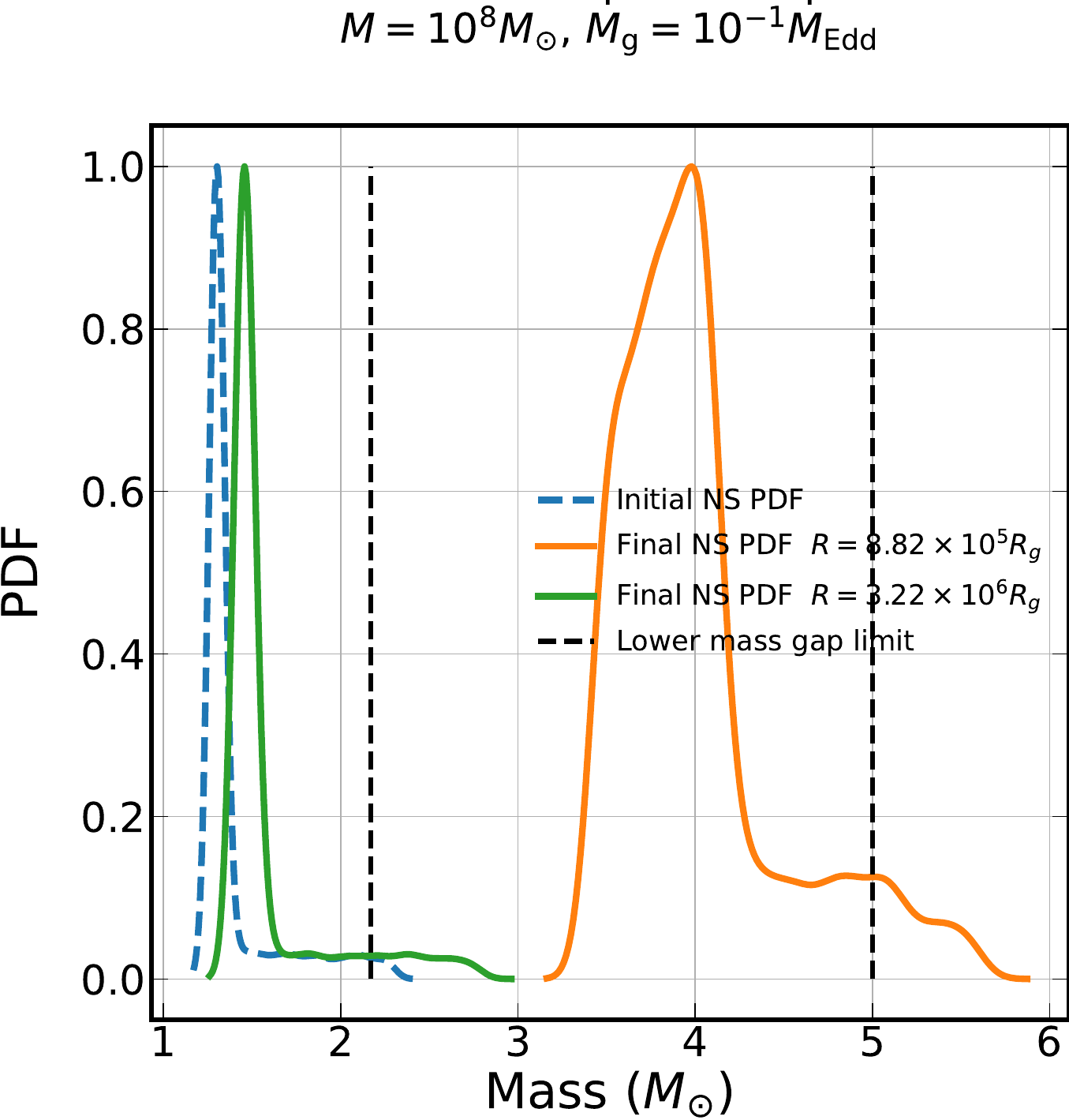}
 
 }
 \subfloat[\label{subfig: final mass NS 1e7}]{
 \centering
 \includegraphics[width= 0.305\linewidth]{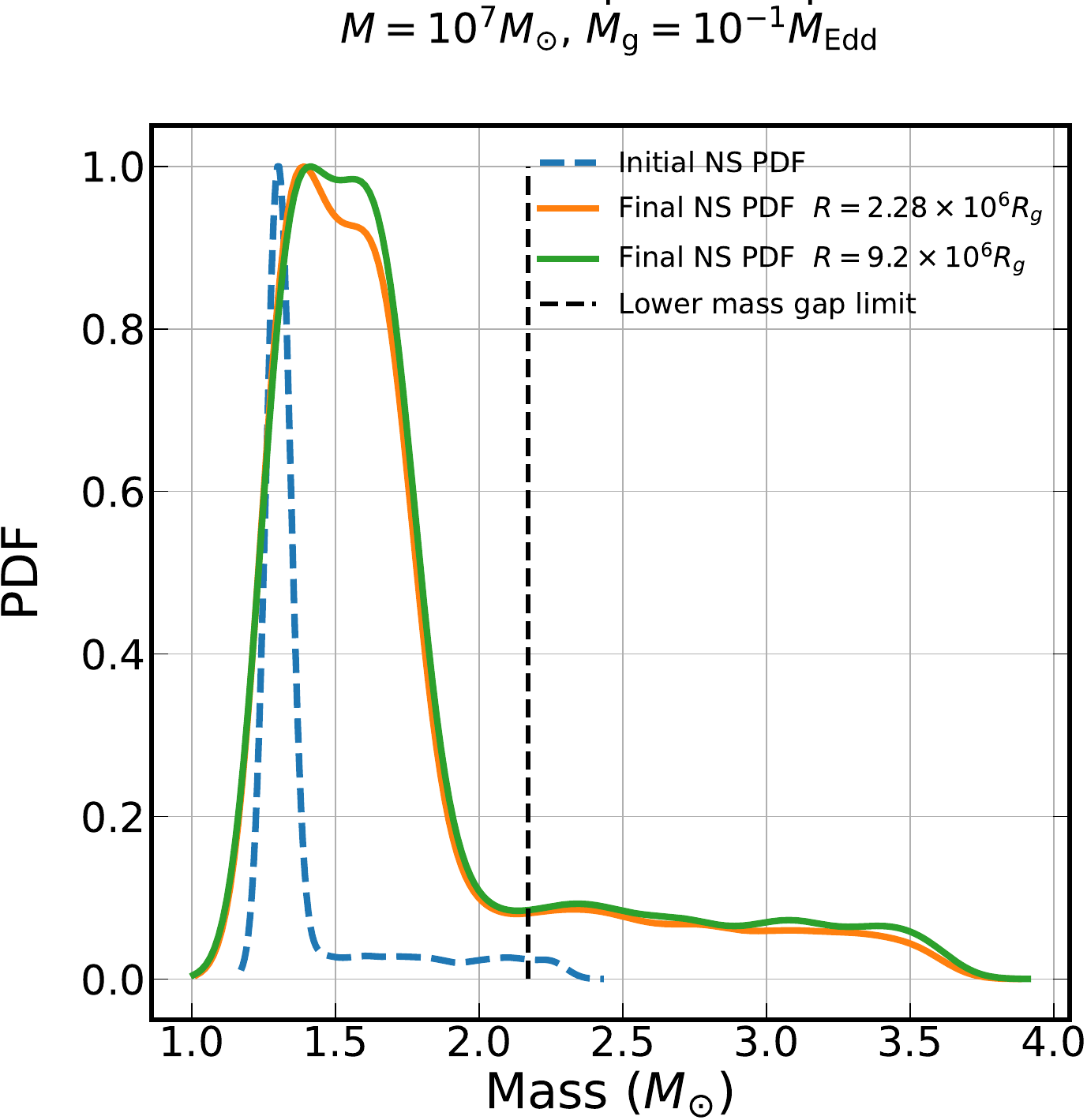}
 
 }

 \caption{Eddington-limited mass growth of ECOs in the AGN disk. The \textit{first row} plots show the final mass $m_{\rm rem}$ as a function of the ZAMS mass $m_{\rm ZAMS}$. These plots show both the ECO mass at formation (dashed blue line) and the ECO mass after a time $t_{\rm AGN}$ spent in the AGN disk at different radii (orange and green solid lines). The \textit{second row} plots convolve these results with a top-heavy initial mass function to show the mass distributions of ECOs born as stellar-mass BHs; the \textit{third row} is the same, but showing mass distributions of ECOs born as neutron stars. The first and third columns show results for ECOs with \textbf{no initial spin} that spin up during the accretion, where the \textit{first column} is for $10^8 M_{\odot}$ SMBH and the \textit{third} is for $10^7 M_{\odot}$ SMBH. The \textit{second column} shows results for accretion onto ECOs born with \textbf{maximum spin} for $10^8 M_{\odot}$. } 
 \label{fig : mass growth}
\end{figure*}

Embedded neutron stars and BHs will grow through accretion of AGN gas. If an embedded neutron star acquires significant mass, it will undergo an accretion-induced collapse (AIC) and produce a BH in the so-called lower-mass gap \citep{Yang+20, Perna+21}. Accretion onto embedded BHs, regardless of their origin, will (i) increase their mass and (ii) alter their spin \citep{Tagawa+20a}. If the accretion is coherent (i.e. has a persistent direction of angular momentum), which we assume to be the case, then the embedded BHs will spin up. Growth of embedded BHs will eventually produce objects in the ``upper,'' or pair-instability, mass gap, which cannot be populated through isolated stellar evolution \citep{SperaMapelli2017}. Because ECOs can undergo mergers and emit GWs {\it after} significant mass growth,\footnote{ECOs may potentially also increase their mass significantly while in an inspiraling binary state. Early works found that hydrodynamic hardening was too efficient to permit significant mass growth \citep{Stone+17}, but more recent results on circumbinary accretion \citep{Tang+17} led \citealt{ShuXu+18} to a different conclusion.} due to single-single captures inside \citep{Bellovary+2016} or outside \citep{Tagawa+20b} of migration traps, it is of great importance to understand their plausible mass and spin distributions, which may offer unique diagnostics of GWs produced by the AGN channel.

We used our disk structure results and an initial compact object mass distribution (see \sref{sec: numercial method}) in order to calculate the mass growth of ECOs during an AGN episode. 
More specifically, for particular radii in the AGN disk, we integrated \cref{eq: RBH accretion rate} over the AGN lifetime (but subtracting the main-sequence lifetime of the stars that turned into the initial ECOs\footnote{We note that this procedure only applies for ECOs that form {\it in situ} in the first generation of star formation; ECOs that form later will have less time to grow, whereas preexisting ECOs captured through gas drag will have more time to grow.}). We cap the reduced Bondi-Hoyle accretion rate at the Eddington accretion rate. For neutron stars, we set a constant radiative efficiency of $\eta = 0.1$, and pick a maximum mass for AIC of $2.17 M_{\odot}$ \citep{MargalitMetzger2017}. For BHs, we use a spin-dependent radiative efficiency 
assuming accretion disks that are prograde equatorial \citep{Bardeen+72}. Since typical AGN episodes last at most a few Salpeter times, the outcomes of ECO growth depend sensitively on the initial distribution of BH spins $a_\bullet$, which we consider two limiting cases of. First, we calculate mass growth for BHs that are born in the Schwarzschild limit ($a_\bullet=0$) and spin up to the Thorne limit ($a_\bullet = 0.998$) using the thin disk mass/spin growth relation \citep{Thorne1974}. We also consider the opposite limit, computing mass growth for BHs born with spin at the Thorne limit, and which do not spin up at all. 

We plot our results in \cref{fig : mass growth}, which shows both initial and final ECO mass distributions at two different radii (disk gas at the outer radius is less dense and therefore the mass growth there is not Eddington capped), for two different SMBH masses and two different assumptions about initial BH spins. 
We see that even in the maximum birth-spin case, which accretes less mass due to higher initial radiative efficiency, a nonnegligible fraction of BHs grow above the pair-instability mass gap ($m_\bullet \gtrsim 50 M_\odot$) for BHs formed through core collapse \citep{SperaMapelli2017}. This only fails to occur for the most short-lived AGN episodes. 
This mass growth gives a possible explanation for the mass gap BH(s) in the GW190521 event \citep{Abbott+20b}, separate from the more frequently discussed explanation of hierarchical mergers and growth. We also see a nonnegligible number of NSs that grow in mass and collapse into BHs in the lower-mass gap, although this gap is not as fundamental as the upper mass gap and may reflect small number statistics in the sample of BHs with mass measurements in X-ray binaries \citep{Kreidberg+2012}\footnote{In fact, recently \citet{Jayasinghe+2021} have presented a $3M_{\odot}$ lower-mass gap BH candidate.}. Considering the fact that our disk model is puffier (and thus features lower reduced Bondi--Hoyle accretion rates) than many other AGN models, rapid ECO growth will likely occur in most other AGN scenarios as well.

Finally, we note that if we had not capped the ECO accretion rate at the Eddington limit, then these BHs would have rapidly grown into IMBHs and begun consuming the vast majority of the total gas accretion rate $\dot{M}$ flowing through the AGN disk, rendering the entire two-fluid AGN picture inconsistent. An example of such results are shown in \aref{app: additional reults}, and provide an additional post hoc justification for our decision to cap ECO accretion at Eddington.

\section{Discussion and Conclusions}
\label{sec:conclusions}
We have formulated time-dependent equations for a multifluid, 1D thin AGN disk model, which over many decades in radius has an energy equation dominated by feedback from embedded BHs (i.e. an ``FDAF''). This model, which is based on the classic $\alpha$ prescription for angular momentum transport, produces stars and eventually embedded compact objects in Toomre-unstable zones or, alternatively, the disk captures stars and BHs from a preexisting nuclear star cluster. Feedback from embeds (particularly ECOs; we have shown that stellar feedback is usually negligible) eventually self-regulates much of the disk to be marginally stable against further fragmentation. We defer a full solution of the time-dependent equations to future work; in this paper, we have explored the structure of such a disk in two different steady-state limits. 

We argue that existing disk models that rely on feedback to self-regulate to marginal Toomre stability $(Q_{\rm{T}}=1)$ \citep{SirkoGoodman2003,Thompson+2005,DittmanMiller20} produce disks that are marginally stable {\it in the continuum limit} of feedback sources, but which can suffer greater instability due to discreteness effects. Specifically, the injection of stellar/ECO feedback into the disk is local and discrete. Even if an azimuthally averaged energy injection rate is capable of heating a radial annulus to a $Q_{\rm T}=1$ state, a sufficiently discrete distribution of embedded objects \newp{may} fail to mix heat uniformly through $2\pi$ in azimuth. In order to achieve a marginally stable result in all azimuthal sectors, we may therefore need to consider disks with azimuthally averaged $Q_{\rm{T}}\gg1$, a consideration we quantify via the heat-mixing parameter $\mathfrak{M}_{\varphi}$ (Eq. \ref{Azimuthal mixing parameter}). With these assumptions and conditions, we looked into two limiting, steady-state scenarios: 
\begin{enumerate}
 \item CMI regime: at smaller radii, Toomre instability does not occur in simple disk models and the dominant population of ECOs will be those migrating in from larger radii (where they enter the disk either from an unstable, star-forming zone, or alternatively through capture by gas drag). Here we assume a CMI of ECOs via (primarily) type I migration torques.
 
 \item Pileup regime: at larger radii, lower gas densities mean that the migration time is longer than typical AGN lifetimes, and ECOs exist in a zero-migration limit. In this regime, ECOs are formed and/or captured until the disk 
 stabilizes itself against further fragmentation,
 where the number density of ECOs is determined by the need to maintain at least marginal Toomre stability ($Q_T\geq 1$) with efficient mixing of feedback heat ($\mathfrak{M}_\varphi \geq 1$).

\end{enumerate}

From our results, it is clear that neither of these regimes can describe a full AGN accretion disk. 
The pileup solution is generally valid at the outer radii, and the CMI solution in the inner radii, with an intermediate gap covering roughly one order of magnitude in radius where neither solution is valid. 
Understanding this intermediate regime is beyond the scope of this paper, as it requires the solution of the full time-dependent set of disk equations. The results for the steady-state equations nevertheless lead us to the following main conclusions: 
\begin{enumerate}
 \item Stellar feedback is inefficient and self-limiting, as the end result of disk-mode star formation is the production of ECOs. Once the first generation of massive stars have exploded as core collapse SNe, accretion feedback from ECOs will dominate the AGN energy equation at most radii. In order to create a self-regulating disk via feedback from ECOs, we cannot consider azimuthally averaged Toomre stability parameters, as $Q_{\rm T}=1$ zones will usually fail to mix heat through all azimuthal sectors. Requiring that an effective heat-mixing parameter $\mathfrak{M}_\varphi \ge 1$ creates large zones with $Q_T\gg1$.
 
 \item Similar to past self-regulating AGN models, at large radii our disks are much puffier (greater aspect ratio $H/R$, lower density $\rho \propto Q_{\rm T}^{-1}$) than a Shakura--Sunyaev-type disk. This results in much shorter viscous timescales at large radii (in the pileup regime) and helps to solve the inflow timescale problem of simple Shakura--Sunyaev AGN disk models. In particular, FDAFs extend the range of validity of local $\alpha$ viscosity prescriptions by one to two orders of magnitude in radius.
 
 \item At small radii, inside the Shakura--Sunyaev Toomre instability point, FDAF solutions (i.e. the CMI regime) usually default to Shakura--Sunyaev-like ones, with the exception of the highest SMBH masses $M$ and the highest ECO inflow rates $\dot{M}_\bullet$. For $M\lesssim 10^8 M_\odot$, a Shakura--Sunyaev-like model provides a good approximation for disk structure for any realistic ECO inflow rate.
 
 \item The number of embedded BHs in AGN disks is substantial. The total mass of ECOs is dominated by the large radii that exist in the pileup regime. 
 The large number of corotating BHs we predict in the pileup radii may make such disk regions ``breeding grounds'' for BH--BH mergers, as low relative velocities will facilitate two-body gas capture \citep{Tagawa+20b}.  Our approximate estimates suggest that capture of BHs from the preexisting nuclear star cluster will not always suffice to stabilize outer regions into the pileup regime.
 
 \item ECO mass growth occurs across the parameter space we have examined, and especially at higher SMBH masses ($>10^7 M_\odot$), a nonnegligible fraction of embedded BHs will grow beyond the pair-instability mass gap through gas accretion alone. Similarly, embedded neutron stars can grow in mass and collapse into BHs \citep{Yang+20b, Perna+21}. Together with hierarchical BH mergers \citep{GerosaBerti19, Yang+19, GerosaFishbach2021}, AGN environments offer two different ways to produce compact objects in each of the mass gaps. 
 
 \item Migration traps can form in our model at small radii (the CMI regime). These include both the classic migration traps first identified in \citet{Bellovary+2016}, where the sign of the migratory torque flips, and also a different type, which we refer to as a ``Zeno trap,'' in which the migration time increases as the objects migrate inwards, until the migration time exceeds the AGN lifetime. Zeno traps appear to exist only for high-mass SMBHs $M \gtrsim 10^8 M_\odot$. The existence of either type of trap may change the disk structure at lower radii. These traps attract ECOs from larger radii and play an important role in forming BH--BH binaries that will later merge \citep{Secunda+19,Secunda+20}. 
\end{enumerate}

Our model is idealized in a number of ways, which we hope to improve on in future research. First, and most importantly, we have only investigated steady-state limits of a more general set of time-dependent, two-fluid disk equations. In the future, we will explore fully time-dependent solutions to these equations. Second, we have generally ignored (i) ECO--ECO scatterings and (ii) capture and scattering of stars from a preexisting, quasi-spherical nuclear star cluster (\citealt{Syer+91, Bartos+17a}; although see simple calculations in \S \ref{sec: Disc structure}). These interactions will create ECOs on somewhat inclined, not fully circular orbits, which may migrate and accrete at rates that are different than what is included in our calculations (which have generally assumed a dynamically cold disk of embeds on fully circular, noninclined orbits). Third, {\it in situ} formation of ECOs may be complicated (helped or hampered) by nonstandard stellar evolution in AGN environments \citep{Cantiello+21}, which should be accounted for in the future.  Fourth, our model is ultimately a 1D semi-analytic treatment of a complex, 3D problem. While simulating a population of thousands to millions of embedded compact objects is a formidable task, local hydrodynamical simulations could likewise improve our simplified treatment of radial/azimuthal heat mixing. 

To summarize, the self-regulation of an AGN disk will eventually be dominated by feedback from a population of embedded compact objects and especially stellar-mass BHs. We have modeled ``FDAF'' accretion disks with a second, collisionless ``fluid'' of ECOs that heat the ambient gas via accretion feedback. In the steady-state limit, we see large deviations from Shakura--Sunyaev-type solutions at large radii. We show that for a disk self-regulating to a point of marginal stability against fragmentation via feedback from local phenomena, $Q_{\rm T}=1$ is a necessary but not sufficient condition for marginal stability, and one needs to consider nonaxisymmetric mixing of heat from local point sources. The large number of embedded BHs undergoing mass growth and passing through migration traps strengthens the viability of the AGN channel for producing BH-BH mergers and mass gap BHs.  By coupling these FDAF solutions to a semianalytic model for BH and binary BH evolution (e.g. \citealt{Tagawa+20b}), we hope to make improved predictions for GW-source formation in the AGN channel.

\section*{Acknowledgements}

The authors gratefully acknowledge helpful scientific discussions with Evgeni Grishin, Zoltan Haiman, Brian Metzger, Tsvi Piran, and Re'em Sari.  S.G. received support from the Israel Science Foundation (Individual Research grant 2565/19).  N.C.S. received financial support from NASA, through both Einstein Postdoctoral Fellowship Award Number PF5-160145 and the NASA Astrophysics Theory Research Program (Grant NNX17AK43G; PI: B. Metzger). He also received support from the Israel Science Foundation (Individual Research grant 2565/19).This research was also partially supported with the BSF portion of a NSF-BSF joint research grant (NSF grant No. AST-2009255 / BSF grant No. 2019772 to N.C.S. and S.G.).

\begin{appendix}

\counterwithin{figure}{section}
\counterwithin{equation}{section}
\counterwithin{table}{section}

\section{ Reduced Equations with Keplerian Angular Frequency} \label{app: reduced equations }
We reorder Equations \ref{eq: Vertical structure}, \ref{eq: hydrostatic equilibrium}, \ref{eq: sound speed}, \ref{eq: alpha viscosity} and \ref{eq: steady viscous transport} in the steady-state limit. We find a reduced version of these equations (substituting in the Keplerian orbital frequency) as a function of radius and 3D density $\rho$:

\begin{align}
\begin{rcases}
 \begin{split}
 1. \; & H\left(\rho,R\right) =\left(\frac{1}{6\pi}\right)^{\frac{1}{3}} \cdot\left(GM\right)^{-\frac{1}{6}}\cdot R^{\frac{1}{2}}\cdot\left[\alpha^{-1}\dot{M}_{\rm g}\left(1-\left[\frac{R_{0}}{R}\right]^{1/2}\right)\right]^{\frac{1}{3}}\rho^{-\frac{1}{3}} \\
 2. \; & c_{s}\left(\rho,R\right) = \left(\frac{1}{6\pi}\right)^{\frac{1}{3}}\cdot\left(GM\right)^{\frac{1}{3}}\cdot R^{-1}\cdot\left[\alpha^{-1}\dot{M}_{\rm g}\left(1-\left[\frac{R_{0}}{R}\right]^{1/2}\right)\right]^{\frac{1}{3}}\rho^{-\frac{1}{3}} \\
 3. \; & P\left(\rho,R\right) = \gamma^{-1} \left(\frac{1}{6\pi}\right)^{\frac{2}{3}}\cdot\left(GM\right)^{\frac{2}{3}}\cdot R^{-2}\cdot\left[\alpha^{-1}\dot{M}_{\rm g}\left(1-\left[\frac{R_{0}}{R}\right]^{1/2}\right)\right]^{\frac{2}{3}}\rho^{\frac{1}{3}} \\
 4. \; & \Sigma\left(\rho,R\right) =2\left(\frac{1}{6\pi}\right)^{\frac{1}{3}}\cdot\left(GM\right)^{-\frac{1}{6}}\cdot R^{\frac{1}{2}}\cdot\left[\alpha^{-1}\dot{M}_{\rm g}\left(1-\left[\frac{R_{0}}{R}\right]^{1/2}\right)\right]^{\frac{1}{3}}\rho^{\frac{2}{3}} \\
 \end{split} 
\end{rcases}
\label{Appendix disk ss equations reduced}
\end{align}

Substituting these expressions into \cref{eq: pileup equation} for the pileup regime or \cref{eq: General black hole flow rate} for the CMI regime allows us to solve for $S_\bullet \left(\rho,T_c,R \right)$. Together with \cref{eq: capped Luminocity}, we substitute these back into \cref{eq: full energy conservation,eq: optical depth,eq: pressure equation} to get a set of 2 equations and 2 variables that we solve via numerical root-finding. Note that for the Shakura--Sunyaev solution (often shown as a comparison) we set $S_{\bullet}=0$ but otherwise proceed in the same way.

\section{CMI solution maps} \label{app: CMI profile}
\begin{figure} 
\centering
 \subfloat[]{
 \centering
 \includegraphics[width=0.46\linewidth]{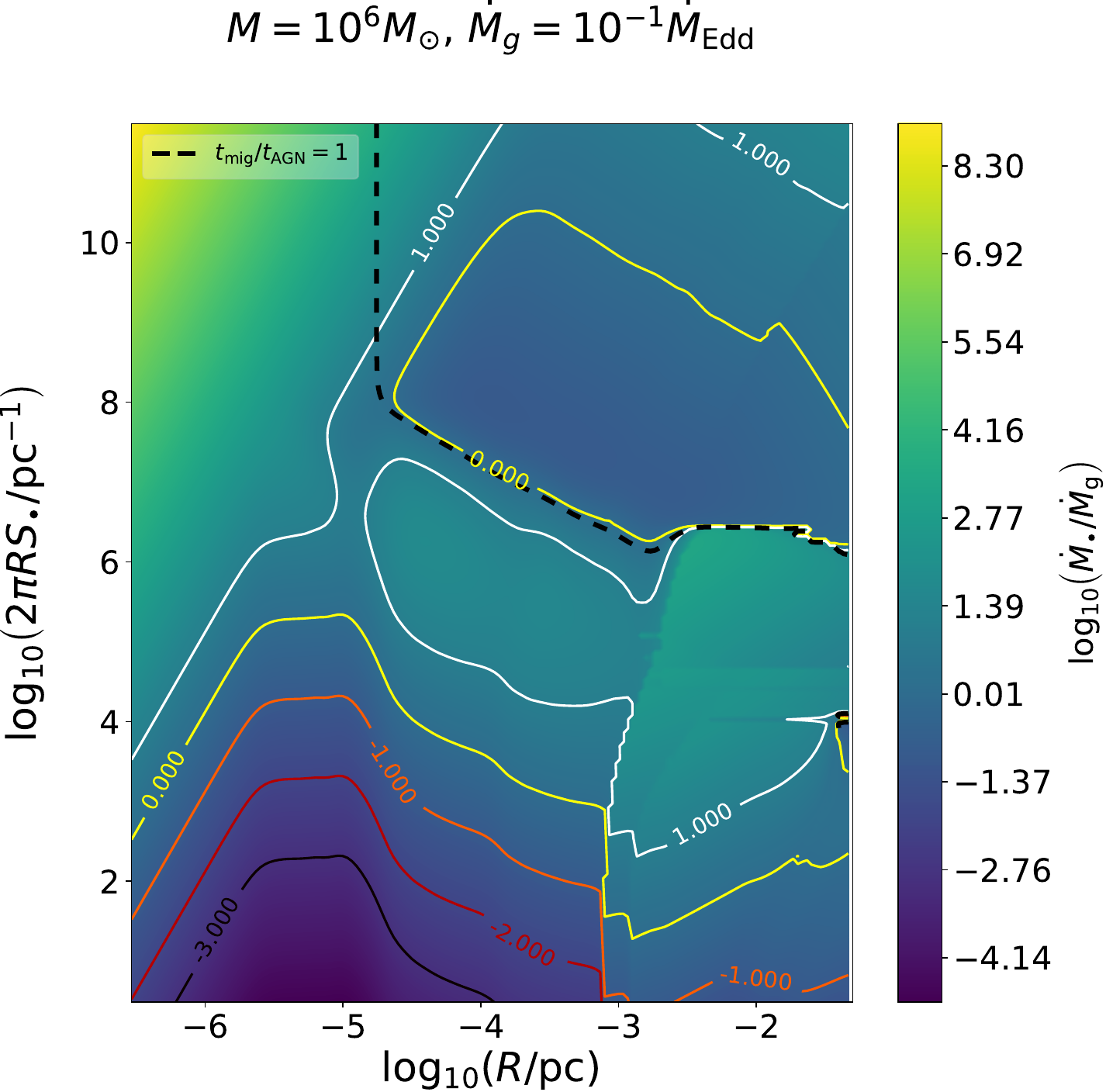}
 
 }
 \subfloat[]{
 \centering
 \includegraphics[width=0.46\linewidth]{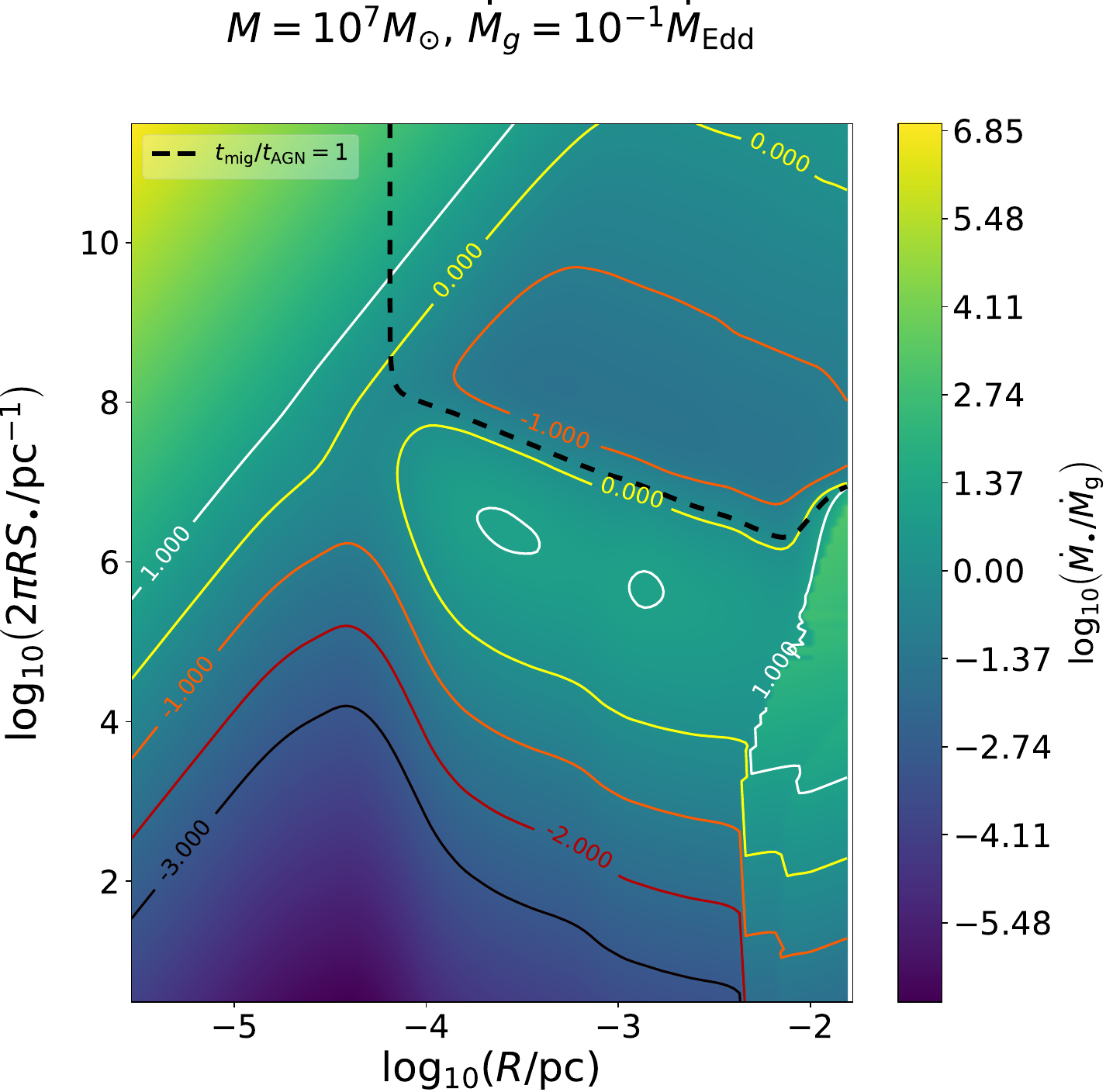}
 
 } 
 
 \subfloat[]{
 \centering
 \includegraphics[width=0.46\linewidth]{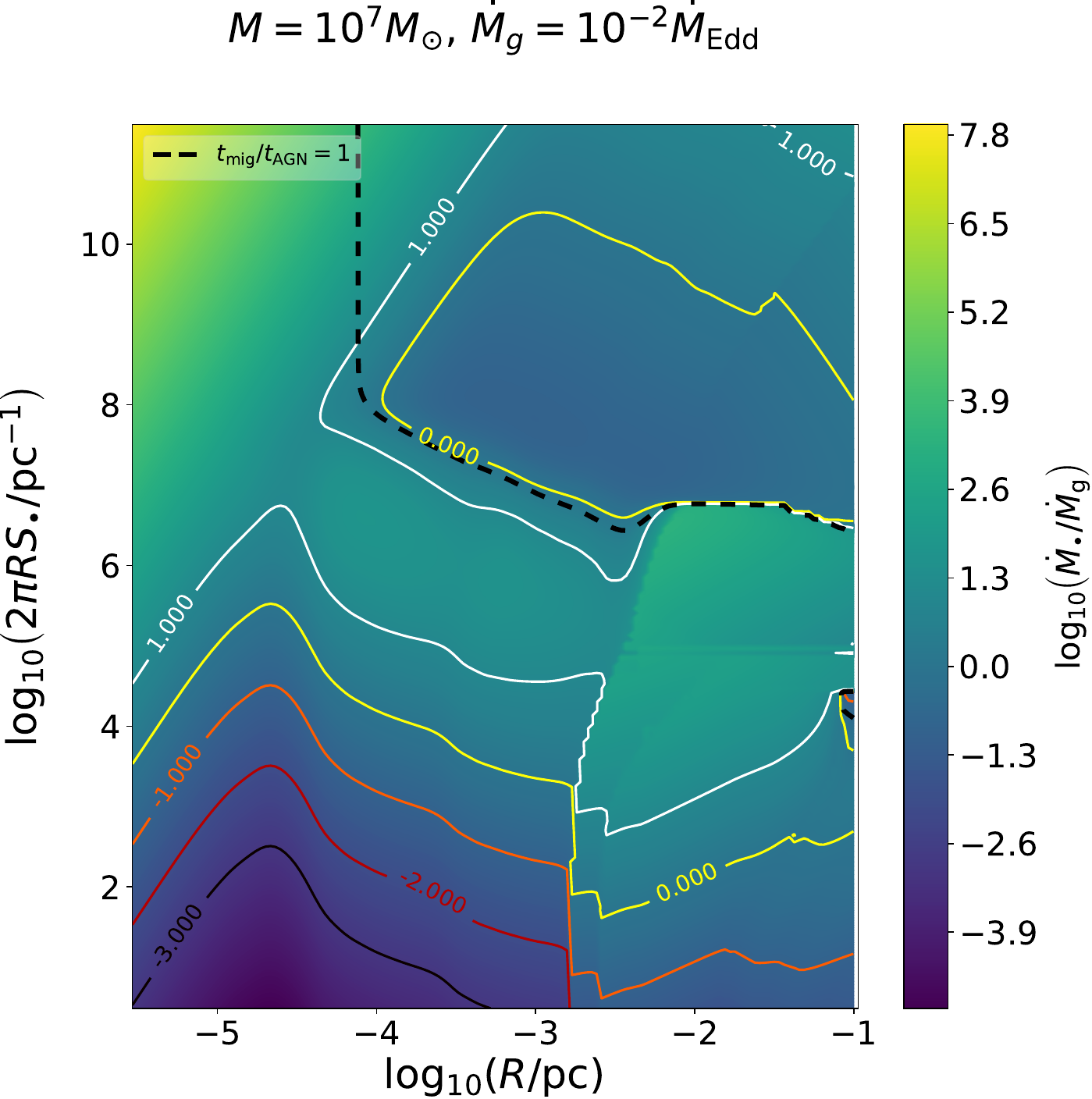}
 
 }
 \subfloat[]{
 \centering
 \includegraphics[width=0.46\linewidth]{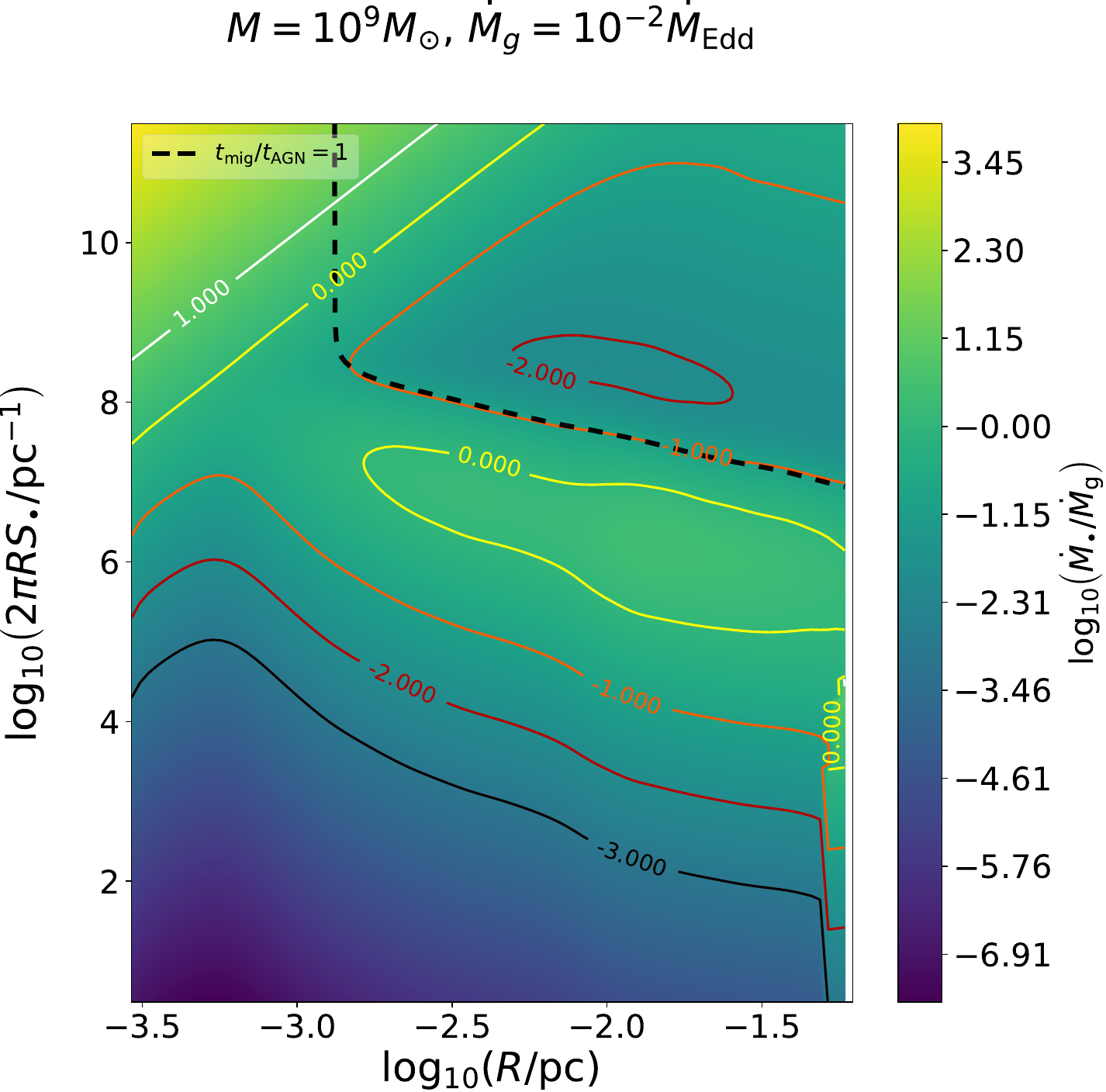}
 
 } 

\caption{$\dot{M}_{\bullet}/\dot{M}_{\rm g}$ heat maps as a function of radius and linear BH number density for different SMBH masses and gas accretion rates. The top panels are for $\dot{M}_{\rm g}=0.1 \dot{M}_{\rm Edd}$ with $M=10^6M_\odot$ ({\it top left}) and $M=10^7M_\odot$ ({\it top right}). The bottom panels are for $\dot{M}_{\rm g}=10^{-2} \dot{M}_{\rm Edd}$ with $M=10^7M_\odot$ ({\it bottom left}) and $M=10^9M_\odot$ ({\it bottom right}). In the area below (and to the left of) the black dashed line, the migration timescales are lower than the AGN lifetime. The different contour lines represent constant $\dot{M}_\bullet$. The lowest set of contour lines (which all roughly parallel each other) indicates solutions with Shakura--Sunyaev disk profiles.  At higher values of $\dot{M}_{\bullet}/\dot{M}_{\rm g}$, other branches of solutions emerge, although we do not expect these to be astrophysically relevant if $\dot{M}_{\bullet}/\dot{M}_{\rm g} \gtrsim 1$ or if $t_{\rm mig} \gtrsim t_{\rm AGN}$.} \label{fig: CMI appendix maps}
\end{figure}

In this appendix, we present heat maps (Fig. \ref{fig: CMI appendix maps}) in the style of Fig. \ref{fig: CMI constant S map}, but for different combinations of $M$ and $\dot{M}_{\rm g}$.  
As was the case in Fig. \ref{fig: CMI constant S map}, we see that for lower ratios of $\dot{M}_\bullet / \dot{M}_{\rm g}$, the ECOs do not affect the disk and we retrieve a Shakura--Sunyaev-type disk profile.  For larger values of $\dot{M}_\bullet / \dot{M}_{\rm g}$, other ``upper'' branches of solutions emerge that are characterized by much higher ECO surface densities $S_\bullet$, but these are sometimes unphysical, preferentially appearing for cases where $\dot{M}_\bullet \gtrsim \dot{M}_{\rm g}$ (a regime that breaks our assumption of negligible angular momentum transport through ECO migration and that is also in tension with the Soltan argument).  However, for larger SMBHs and accretion rates, the middle and upper branches of solutions become accessible for $\dot{M}_\bullet / \dot{M}_{\rm g} \lesssim 0.1$.  Even when the middle and upper branches of solutions do not require unphysical $\dot{M}_\bullet / \dot{M}_{\rm g}$ ratios, however, they are still often in tension with the CMI assumption of steady-state ECO mass flux, as $t_{\rm mig} \gtrsim t_{\rm AGN}$ is common.  In general, the uppermost solution branch is always characterized by $t_{\rm mig} \gg t_{\rm AGN}$, but for the largest SMBHs, the middle branch can have $t_{\rm mig} \lesssim t_{\rm AGN}$.

\section{Additional Results} \label{app: additional reults}


\begin{figure} 
 \centering
 \includegraphics[width=0.5 \linewidth]{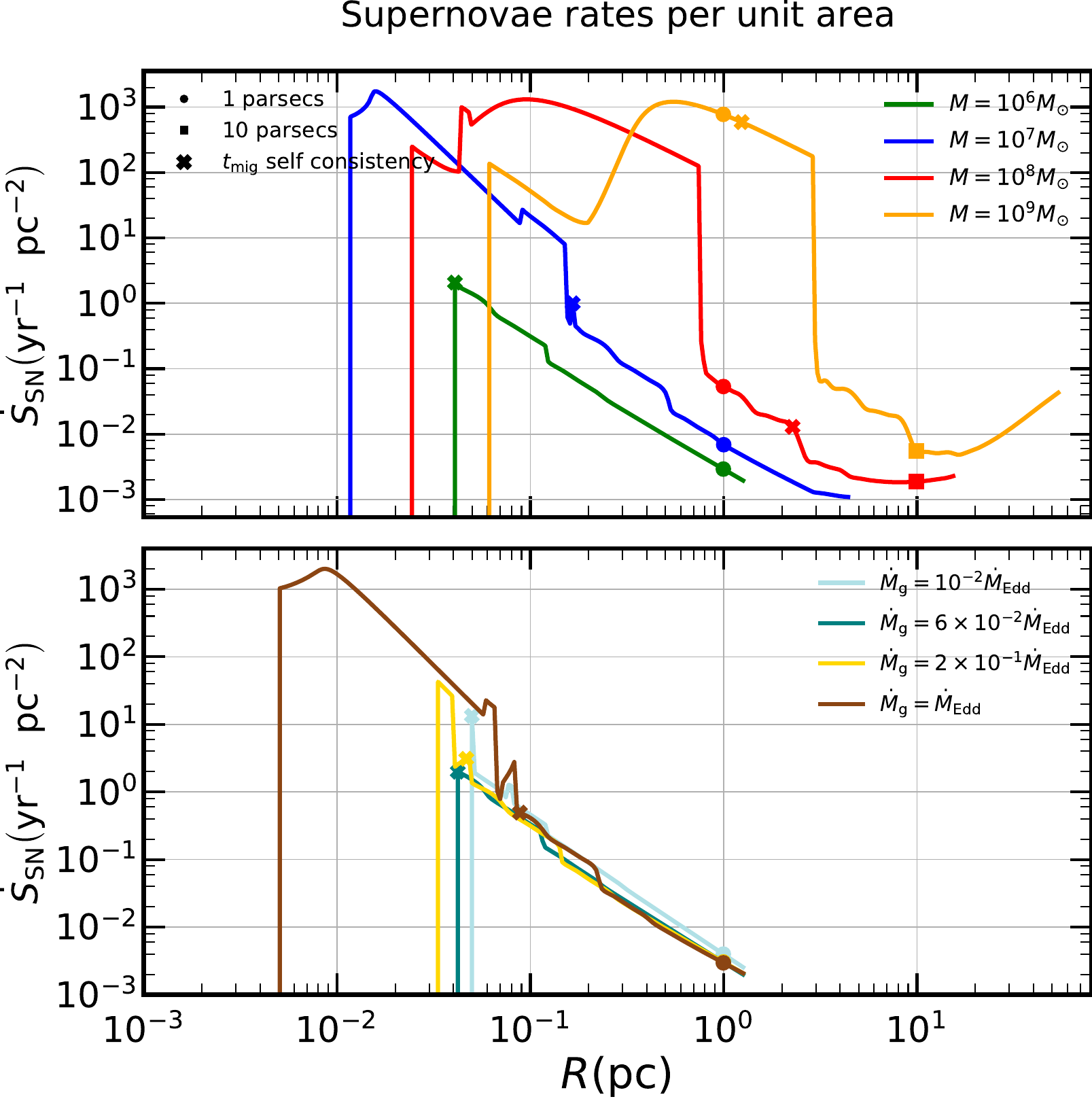}
 \caption{SN rates per unit area according to \cref{eq: SN rates} are plotted as a function of radius for the pileup solution. The {\it top panel} shows results for $\dot{M}_{\rm g}=0.1\dot{M}_{\rm Edd}$ for different SMBH masses. The pileup solution self-consistency point is marked by an ``x''. The {\it bottom panel} shows results for $M=10^6 M_\odot$ for four different accretion rates.  \newp{These SN rates are large enough to demonstrate the self-limiting nature of SN feedback: within a few Myr, enough ECOs will be produced in a SN-stabilized disk to shut off all further star formation, at least in the pileup regions.}}
 \label{fig:SN rates}
\end{figure}

\begin{figure} 
 \centering
 \includegraphics[width=0.5 \linewidth]{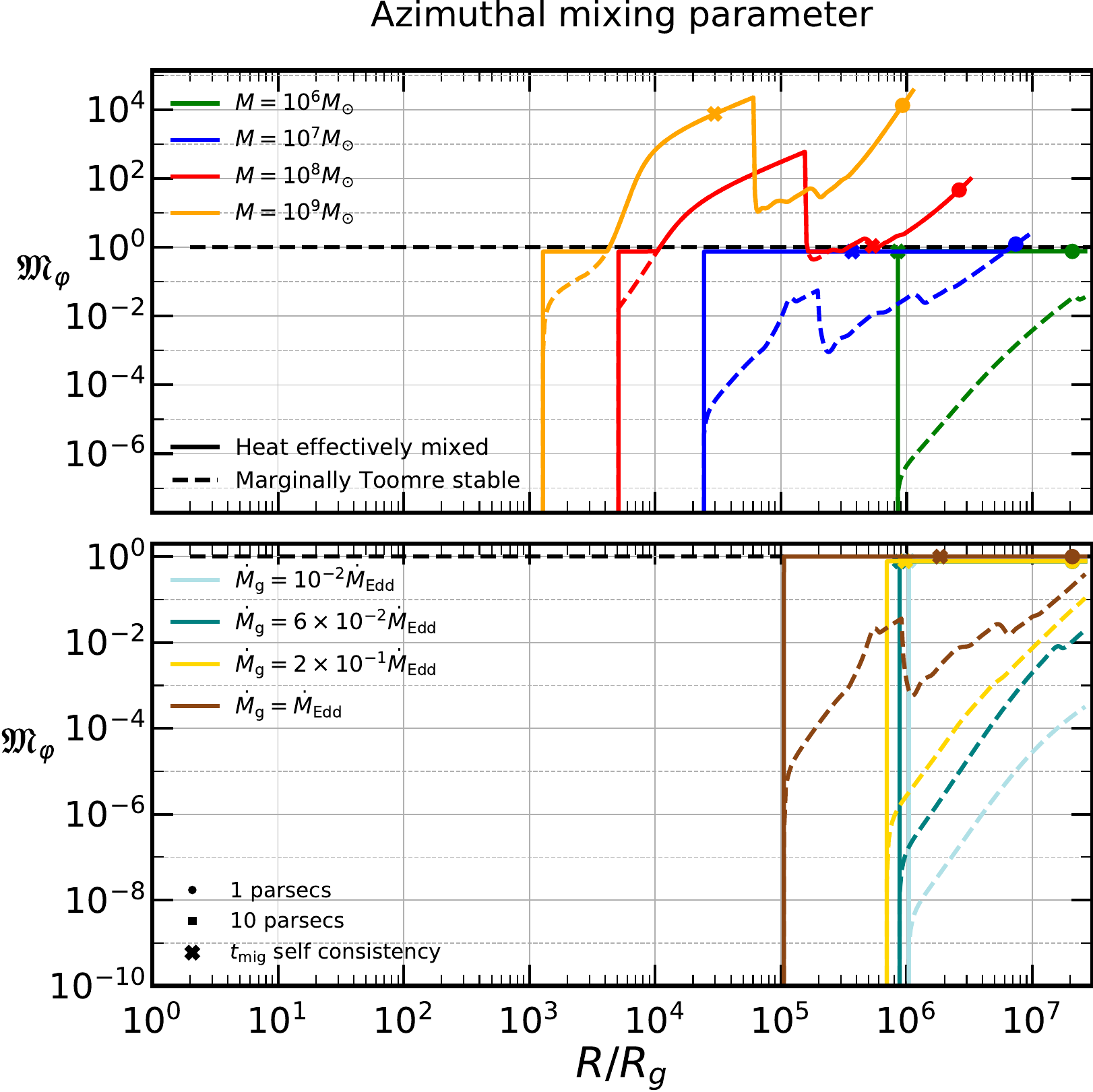}
 \caption{Azimuthal heat-mixing parameter as function of radius for the pileup solution. The pileup solution self-consistency point is marked by an ``x''. The {\it top panel} shows results for $\dot{M}_{\rm g}=0.1\dot{M}_{\rm Edd}$ for different masses. The {\it bottom panel} shows results for $M=10^6 M_\odot$ for four different accretion rates.}
 \label{fig:pileup m_phi}
\end{figure}

\begin{figure}
 \centering
 \includegraphics[width=0.5 \linewidth]{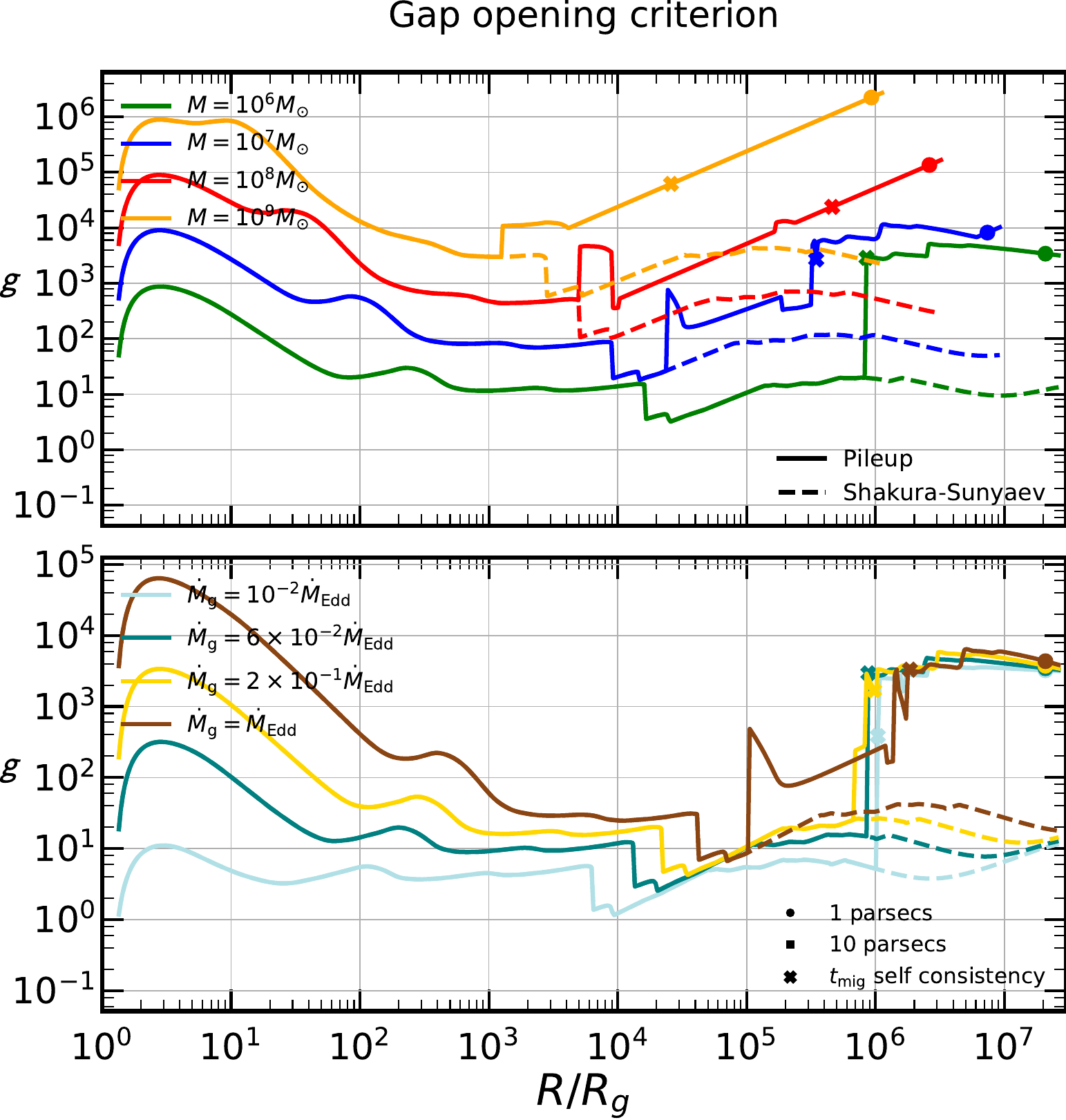}
 \caption{Gap-opening criterion according to \cref{eq: gap equation} \citep{Crida+06}. The pileup solution self-consistency point is marked by an ``x''. The {\it top panel} shows results for $\dot{M}_{\rm g}=0.1\dot{M}_{\rm Edd}$ for different masses. The {\it bottom panel} shows results for $M=10^6 M_\odot$ for four different accretion rates. }
 \label{fig: gap opening}
\end{figure}


\begin{figure} 
 \centering
 \includegraphics[width=0.5 \linewidth]{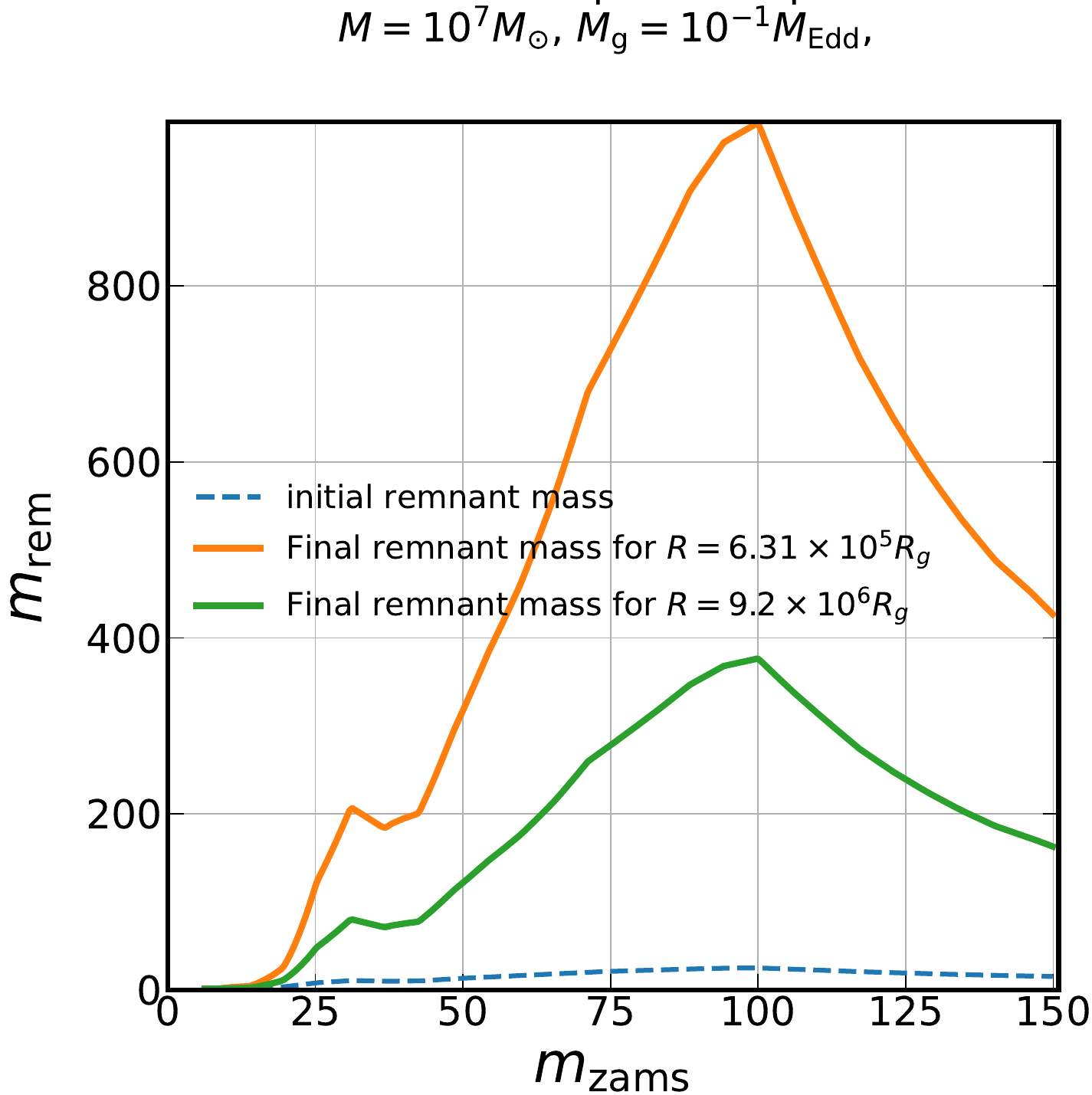}
 \caption{The final mass of ECOs after a time $t_{\rm AGN}$ is plotted as a function of the progenitor star ZAMS mass, consider accretion rates onto ECOs {\it that are not Eddington-limited}.  Here we assume that embedded BHs are born with spins at the Thorne limit, and we take pileup solutions around a $10^7M_\odot$ SMBH with $\dot{M}_{\rm g} = 0.1 \dot{M}_{\rm Edd}$. The two solid lines are for different radii with different densities, and the dashed line is the initial ECO mass.  Without the Eddington cap on the accretion rate, almost all ECOs will grow into intermediate mass BHs over the course of an AGN lifetime.}
 \label{fig: mass overgrowth}
\end{figure}


In this appendix, we show some self-consistency checks that are important for checking and motivating the different assumptions in our model. \paragraph{SN rates }
In \cref{fig:SN rates} we plot the SN rates (per unit area) that are required to equal the energy feedback from the BHs in the pileup regime (using \cref{eq: SN rates}). Integrating these results over the self-consistent pileup radii, the lowest rate we get is about one SN every $\sim 12$ years. These enormous SN rates can only proceed for a small fraction of typical AGN lifetimes before creating a large enough ECO population to completely shut off star formation (within the pileup region). 
\paragraph{Heat-mixing parameter}
In \cref{fig:pileup m_phi}, we see that across all of parameter space, the pileup zone always has radii where the azimuthal heat-mixing parameter $\mathfrak{M}_\varphi=1$, i.e. radii where applying marginal Toomre stability condition ($Q_{\rm T}=1$) to disk modeling would result in $\mathfrak{M}_\varphi \ll 1$.  This illustrates the self-consistency problems in traditional, $Q_{\rm T}=1$ disks with regards to heat mixing: they will possess large azimuthal sectors that cannot be effectively heated by BH feedback. 
\paragraph{Gap opening}
A massive enough object orbiting within a gaseous disk will open a gap in the gas, changing its subsequent interactions with the gas (e.g. it will no longer migrate in the type I manner). Using the gap-opening criterion parameter from \citet{Crida+06}, 
\begin{equation}
 \frac{3H}{4R} \left( \frac{m}{3M} \right)^{-1/3} + \frac{50 \nu M}{m R^2 \Omega} \lesssim 1, \label{eq: gap equation}
\end{equation}
we show in \cref{fig: gap opening} that the average ECO does not open a gap. This result is in tension with the findings of \citet{Tagawa+20b}, likely due to the different gas-disk model that they used.

\paragraph{Eddington limit on ECO accretion}
We have already shown in \cref{fig : mass growth} the mass growth of ECOs in our fiducial model, where growth rates are capped by the Eddington limit and ECOs grow by at most a factor of a few within the AGN lifetime.  In \cref{fig: mass overgrowth}, we show the mass growth of different ECOs without capping accretion rates at the Eddington limit.  Here we see that if the embed accretion rate is not capped at Eddington, then ECOs will swiftly grow into IMBHs.  This can pose substantial self-consistency problems for our (and likely other) AGN disk models, as a large population of IMBHs will throttle the gas supply to the central SMBH by opening gaps in the gas and also accreting most of the gas onto themselves.  This is a major consideration in our choice to cap ECO growth at Eddington.  Note that in this experiment, we have chosen the conditions that produced the least mass growth in \cref{fig : mass growth} (high initial BH spins, and $10^7M_\odot$ SMBH).
\clearpage

\end{appendix}

\bibliographystyle{aasjournal}

\bibliography{main}

\end{document}